\documentclass[a4paper,12pt]{article}
\pdfoutput=1
\usepackage{jcappub}
\usepackage{amsthm,graphicx}
\usepackage{epsfig}
\usepackage{latexsym, amssymb} 
\usepackage{amsmath}
\usepackage[normalem]{ulem}
\usepackage{xcolor}

\usepackage{rotating}
\hypersetup{
    colorlinks=true,
    linkcolor=blue,
    filecolor=magenta,      
    urlcolor=blue,
    breaklinks=true
 }
\def\Mpl{\mathrm{M}_{_{\mathrm{Pl}}}}

\def\beq{\begin{equation}}
\def\eeq{\end{equation}}
\def\br{\begin{eqnarray}}
\def\er{\end{eqnarray}}
\def\benu{\begin{enumerate}}
\def\efnu{\end{enumerate}}

\def\Mpc{\rm M_{\rm pc}}

\newcommand{\vtheta}{\boldsymbol{\theta}}
\newcommand{\bk}{\boldsymbol{k}}

\begin{document}
\title{Inflation Story: slow-roll and beyond}
\author[1,2]{Dhiraj Kumar Hazra,}
\author[2,3]{Daniela Paoletti,}
\author[4]{Ivan Debono,}
\author[5,6]{Arman Shafieloo,}
\author[4,7,8,9,10]{George F. Smoot,}
\author[11,12]{Alexei A. Starobinsky}
\affiliation[1]{The  Institute  of  Mathematical  Sciences,  HBNI,  CIT  Campus, Chennai  600113,  India}
\affiliation[2]{INAF/OAS Bologna, Osservatorio di Astrofisica e Scienza dello Spazio,
		Area della ricerca CNR-INAF, via Gobetti 101, I-40129 Bologna, Italy}
\affiliation[3]{INFN, Sezione di Bologna, via Irnerio 46, I-40127 Bologna, Italy}
\affiliation[4]{AstroParticule et Cosmologie (APC)/Paris Centre for Cosmological Physics, Universit\'e
Paris Diderot, CNRS, CEA, Observatoire de Paris, Sorbonne Paris Cit\'e University, 10, rue Alice Domon et Leonie Duquet, 75205 Paris Cedex 13, France}
\affiliation[5]{Korea Astronomy and Space Science Institute, Daejeon 34055, Korea}
\affiliation[6]{University of Science and Technology, Daejeon 34113, Korea}
\affiliation[7]{Institute for Advanced Study \& Physics Department, Hong Kong University of Science and Technology, Clear Water Bay, Kowloon, Hong Kong}
\affiliation[8]{Physics Department and Lawrence Berkeley National Laboratory, University of California, Berkeley, CA 94720, USA, Emeritus}
\affiliation[9]{Energetic Cosmos Laboratory, Nazarbayev University, Astana, Kazakhstan}
\affiliation[10]{Donostia International Physics Center,  University of the Basque  Country  UPV/EHU,  E-48080  San  Sebastian,  Spain}
\affiliation[11]{Landau Institute for Theoretical Physics RAS, Moscow, 119334, Russian Federation}
\affiliation[12]{Kazan Federal University, Kazan 420008, Republic of Tatarstan, Russian Federation}
\emailAdd{dhiraj@imsc.res.in, daniela.paoletti@inaf.it, ivan.debono@in2p3.fr, 
shafieloo@kasi.re.kr, gfsmoot@lbl.gov, alstar@landau.ac.ru} 

\abstract 
{We present constraints on inflationary dynamics and features in the primordial power spectrum of scalar perturbations using the Cosmic Microwave Background temperature, polarization data from Planck 2018 data release and updated likelihoods. We constrain the slow-roll dynamics using Hilltop Quartic Potential and Starobinsky $R+R^2$ model in the Einstein frame using the Planck 2018 binned \texttt{Plik} likelihood. Using the Hilltop as base potential, we construct Whipped Inflation potential to introduce suppression in the scalar power spectrum at large angular scales. We notice marginal (68\% C.L.) preference of suppression from the large scale temperature angular power spectrum. However, large-scale E-mode likelihood based on high frequency instrument cross spectrum, does not support this suppression and in the combined data the preference towards the suppression becomes negligible. Based on the Hilltop and Starobinsky model, we construct the Wiggly Whipped Inflation potentials to introduce oscillatory features along with the suppression. We use unbinned data from the recently released \texttt{CamSpec} v12.5 likelihood which updates Planck 2018 results. We compare the Bayesian evidences of the feature models with their baseline slow-roll potentials. We find that the complete slow-roll baseline potential is moderately preferred against potentials which generate features. Compared to Planck 2015 \texttt{PlikHM} bin1 likelihood, we find that the significance of sharp features has decreased owing to the updates in the data analysis pipeline. We also compute the bispectra for the best fit candidates obtained from our analysis.}

\maketitle

\section{Introduction}

Predictions of different models of inflation for primordial spectra of scalar and tensor perturbations~\cite{Starobinsky:1979ty,Starobinsky:1980te,Guth,Sato,Mukhanov,Linde:1981mu,Albrecht:1982wi,Hawking:1982cz,Starobinsky:1982ee,Guth:1982ec,Starobinsky:1983zz,Linde} are most effectively tested by Cosmic Microwave Background (CMB) surveys. Importantly, a space-based CMB survey can probe the signature of inflationary dynamics thanks to the full sky coverage that provides access to the largest angular scale information. Using CMB observations we can constrain the power spectra of the scalar and the tensor perturbations generated during inflation. If inflation is driven by a scalar field following strict slow roll dynamics (in a nearly flat potential), then both the scalar and tensor power spectra can be locally approximated by a power law form parametrized by the respective amplitudes ($A_S(k), A_T(k)$) and tilts ($n_S(k),n_T(k)$). The ratio $A_T/A_S$ relates to the scale of the inflaton potential while the other parameters refer to the overall shape of the potential. First measurements of the scalar amplitude $A_S$ was reported with COBE~\cite{COBE,COBEnormalization} and the spectral tilt $n_S$ with WMAP~\cite{Hinshaw_2013}. The current state of the art results are provided by the Planck 2018 data~\cite{Planck:2018Overview}. Planck has performed a cosmic variance limited measurement of the temperature anisotropies in the region where CMB is dominant. It has covered from the largest scales up to $\ell\sim 2500$, corresponding to a 5$'$ resolution. Observation of temperature and polarization anisotropies in the CMB sky enabled Planck to constrain the cosmological parameters such as baryon and cold dark matter densities, the angular diameter distance to the last scattering surface and the parameters for the spectrum of primordial fluctuations to the sub-percent level, with only the optical depth remaining above this uncertainty. Planck data has rejected the hypothesis of complete scale-invariant spectra $n_S=1$ with $8\sigma$ significance~\cite{Planck:2018Params}. This discovery confirms the overall shape of the inflationary potential. However, since the primordial B-modes are not yet detected, it can only put an upper bound on $A_T/A_S$~\cite{BK15} and therefore on the scale of inflation.

The Planck 2018 release brought two major improvements concerning our analysis. Firstly we refer to the improvement in the E-mode polarization on large angular scales due to the adoption of the cross power spectrum of $100\times143$ GHz channels from the High Frequency Instrument (HFI)~\cite{Planck:2018HFI}. The use of the cross spectrum has lowered the noise to the point of halving the uncertainty on the integrated optical depth together with moving it towards lower central values providing $\tau=0.054\pm0.007$~\cite{Planck:2018Params} at 68\% C.L. compared to $\tau=0.066\pm0.016$~\cite{Planck:2015Params} in Planck 2015. The CMB angular power spectra are proportional to $A_s e^{-2\tau}$ causing a full degeneracy among the scalar amplitude and the optical depth. The improvement in the E-mode polarization allows us to measure the optical depth from the reionization bump, reducing by consequence the error induced on the amplitude of scalar fluctuations. In second instance the improvement in the characterization of the temperature to polarization leakage and polar efficiency has allowed a cosmology-grade high-$\ell$ polarization which reduces the uncertainty on the scalar spectral index from $0.006$ in Planck 2015 (where only temperature could be used) to $0.004$ in Planck 2018 release. Therefore simply from the perspective of slow roll inflation, Planck 2018 brings new bounds on the potential.

Together with this overall shape of the potential, the Planck data, with accurate measurements of temperature and polarization anisotropies also hint at possible detailed structure of the inflationary potential. A closer look at the data subtracted from the best fit model with power law perturbation spectrum reveals localized outliers~\cite{Planck2018X} at different scales with some having up to 2-3$\sigma$ significance. If the outliers do not represent statistical fluctuations, then a simple solution can be addressed from a non-trivial inflationary dynamics. The characteristics of the outliers, and equivalently the reconstructions~\cite{Hazra:recon13broad,Hazra:reconP13,Hunt:2015iua,Obied:2018qdr} point towards two main types of features, namely large scale suppression and bursts of oscillations at certain scales~\footnote{for earlier works on reconstruction with the pre-Planck data, see~\cite{Hannestad:recon00,Shafieloo:recon03,Mukherjee:recon03,Bridle:recon03,Kogo:recon05,Leach:recon05,Tocchini-Valentini:recon05,Shafieloo:recon07,Paykari:recon09,Nicholson:recon09,Gauthier:recon12,Vazquez:recon12,Hazra:recon13,Hunt:recon13}}. Oscillating features are also present in the polarization data~\cite{Planck2018X}. Therefore it is important to explore whether these oscillations have a single primordial origin. 

In single field models of inflation, the presence of a (smoothed) step/discontinuity, kink or an inflection point~\cite{Starobinsky:Kink,Ivanov:1994pa,Adams:Step0,Covi:Step1,Ashoorioon:Step2,Joy:2007na,Joy:2008qd,Hazra:Step3,Miranda:Step4,Benetti:Step5,GallegoCadavid:Step6,Chluba:Step7,Bousso:Step8,Allahverdi:PI,Jain:PI} in a nearly flat potential, or its sinusoidal modification~\cite{Chen:Osc00,McAllister:Osc0,Flauger:Osc1,Pahud:Osc2,Flauger:2010Osc,Chen:2010Osc,Aich:Osc3,Peiris:Osc4,Meerburg:Osc5,Easther:Osc6,Motohashi:Osc7,Miranda:Osc8}, can lead to departures from strict slow roll dynamics that results in a burst of oscillations or persistent oscillations in the primordial power spectra (PPS). Two field models of inflation can also generate sharp feature signals~\cite{Cremonini:2fm1,Achucarro:2fm2,Braglia:2fm7,Braglia:2fm8} or resonant signals~\cite{Chen:2fm3,Chen:2fm4,Chen:2fm5,Chen:2fm6,Braglia:2fm9,Braglia:2021ckn,Braglia:2fm10,Braglia:long} in the PPS that help fitting the data better than standard slow roll models.
Change in the speed of sound during inflation also can source these types of features in the PPS~\cite{Canas-Herrera:2020mme}. Although none of these models is favoured with respect to the standard model by Bayesian indicators, the analyses of these features remain a crucial point in current and future data. In fact, some of the large and intermediate scale features have been consistently present in the data since WMAP~\cite{WMAP:1Peiris} with ever increasing precision. In some models with features, the fitting of these outliers in both temperature and polarization is robust against the increase of data points~\cite{Hazra:2016fkm,Braglia:2021ckn}. 

We compare the updated Planck likelihood testing it against slow roll inflation and a set of models that allow variation in the form of change in slope (low-$\ell$ power suppression in temperature anisotropy) and oscillatory behaviour to fit to observed excursions of the CMB anisotropy power spectrum. A framework that allows these kinds of behaviour is Wiggly Whipped Inflation (WWI) (see, ~\cite{Hazra:2014jka,Hazra:2014goa,Hazra:2016fkm,Hazra:2017joc,LHuillier:2017lgm,Debono20}). Scalar primordial spectra generated from WWI contain two characteristic features: at large scales scalar power spectra is suppressed, at the intermediate and small scales the power spectra exhibits oscillations. They include different features relevant to the CMB data and its baseline configuration allows us to compare with the strict slow roll scenarios. Therefore in this single framework we can address constraints on inflation with standard slow-roll and with initial or intermediate fast-roll phases. The baseline slow-roll part of the WWI framework is constrained with the Planck power spectrum data on temperature, E-mode polarization, lensing and B-mode data from BICEP-KECK-Planck joint analysis~\cite{BK15}. 

Since WWI framework generates a suppression of the scalar spectrum at the large scales, it is supported by the CMB temperature data~\cite{Hazra:2016fkm} for fitting the well known lack of power observed in the low multipole region since WMAP. The new increased accuracy of E-mode polarization data from HFI opens the possibility to investigate the possible presence of a suppression at large scales also in this channel (although likely more difficult due to the reionization bump). After the Planck official release in 2018 (followed by the likelihood release in 2019), an updated \texttt{CamSpec} {\tt 12.5HMcln} likelihood for temperature and polarization data was publicly released where the authors have used improved data analysis pipeline to produce an upgraded \texttt{CamSpec} likelihood. Since WWI framework had already been tested against {\tt Plik bin1} (unbinned) likelihood (from Planck 2015)~\cite{Hazra:2016fkm}, in this paper we have used the new \texttt{CamSpec} likelihood for the intermediate and small scale features. We also compare the samples with better likelihoods in different temperature and polarization combinations to examine their overlap. Our analysis is more conservative than Planck~\cite{Planck2018X} feature analysis where the nuisance parameters were fixed to the Planck baseline best fit values. Apart from estimating inflationary parameters, we carry out a Bayesian model comparison between the models within the WWI framework. We search for the best fits and identify a few representative spectra that provide improvement in fit to the data compared to standard slow roll models. 

Three point correlation functions of the primordial perturbations are known as the bispectra and are commonly quantified with a parameter $f_{\rm NL}$ which is a suitable dimensionless ratio of the bispectra and the product of the power spectra. Therefore $f_{\rm NL}$ is one of the quantifier of the primordial non-Gaussianity in a model. Using temperature and polarization data Planck reports constraints~\cite{Planck:2019NG} on $f_{\rm NL}$ as, $f_{\rm NL}^{\rm local}=-0.9\pm5.1$, $f_{\rm NL}^{\rm eq}=-26\pm47$, $f_{\rm NL}^{\rm orth}=-38\pm24$ at local, equilateral and orthogonal limits respectively. The 68\% uncertainties indicate that the results are consistent with Gaussian nature of primordial fluctuations. Note that these constraints are obtained assuming a scale independent bispectrum amplitude. Slow roll models of inflation in a canonical Lagrangian generate negligible non-Gaussianity~\cite{Maldacena:2002vr,Seery:2005wm}. However departure from slow rolls generate scale dependent oscillations in the $f_{\rm NL}(k)$~\cite{Chen:2006xjb,Chen:Osc00,Chen:2010xka,Flauger:2010Osc,Chen:2010Osc,Martin:2011sn,Hazra:2012BINGO,Hazra:2012kq,Adshead:2013zfa,Achucarro:2013cva,Sreenath:2014BINGO,Martin:2014kja,Achucarro:2014msa,Fergusson:2014hya,Meerburg:2015owa,Appleby:2015bpw,Dias:2016rjq}. Therefore apart from the power spectrum, features considered in our analysis generate oscillatory and non-negligible bispectra. In the context of WWI models, the bispectra were discussed in~\cite{Hazra:2014goa,Hazra:2016fkm}. With the updated constraints on the potential parameters with the final Planck data, in this paper we also compute the $f_{\rm NL}$ for the best fit candidates. 

The paper is organized as follows: in~\autoref{framework} we discuss the slow-roll and beyond slow-roll models we plan to compare against the data. In~\autoref{DataAnalysis} we provide a detailed discussion of the data and the likelihoods used in the analyses. We also discuss the choice of priors and the Bayesian model comparison methodology. Following that, in~\autoref{ResultsSection} we compare the models with the marginal likelihoods, present the posterior distribution of the parameters, discuss the best fit primordial power spectra and compute the bispectrum amplitude. Finally, we summarise our conclusions in~\autoref{DiscussionSection}. For readers who are familiar with modelling and Planck data analysis, we would like to suggest that they can skip~\autoref{DataAnalysis} and move to~\autoref{ResultsSection} directly after~\autoref{framework}.

\section{Framework}~\label{framework}

A power law primordial spectrum is largely consistent with Planck data at all scales~\cite{Planck13:inflation,Hazra:reconP13,Planck2018X}. The stringent constraint on the spectral tilt translates to a strong constraint on the slope of the inflationary potential \cite{Planck2018X}. Moreover, recent results from BICEP-KECK-Planck (BK15)~\cite{BK15} joint analysis have bounded the tensor-to-scalar ratio ($r$) to $r<0.056$ at 95\% C.L.~\cite{Planck2018X}. Therefore, within these data combinations, the slow roll inflationary models that we will consider as {\it baseline models} must have sufficient freedom to allow variations of the spectral tilt and produce a tensor-to-scalar ratio compatible with the current tight constraints~\footnote{BK18 data was released after the release of the paper. Although it provides stronger constraints with $r<0.036$ at 95\% C.L. we do not re-analyse our models with this data as the tensor-to-scalar ratio remains well within the BK18 bound.}.

For feature models we use the Wiggly Whipped Inflation framework which we introduced in~\cite{Hazra:2014jka,Hazra:2014goa}. This was also analyzed in~\cite{Hazra:2016fkm,Hazra:2017joc} for the constraints using Planck combined data from temperature and polarization and the forecasts for future CMB space-based observations. We note that since certain WWI best fit power spectra to the Planck 2015 data contain non-local oscillations extending to small scales, future large scale structure (LSS) surveys such as Euclid~\cite{Euclid}, LSST~\cite{LSST} are also expected to play an important role in constraining such models. We examined these constraints for WWI with Euclid in~\cite{Debono20}.  In this work, since our goal is to use the latest available Planck likelihood, we make some adjustments in the analysis of the models while keeping the main framework unchanged. 

We categorize the potential into 2 different classes of potentials wherein with different limits on the parameters we obtain different features in the primordial power spectrum.

 \subsection{WWI-potential:} The base slow roll potential used here is the Hilltop potential~\cite{Boubekeur:2005Hilltop}. The WWI-potential is expressed as,
 \begin{equation}
V({\phi})=V_{i} \left(1-\left(\frac{\phi}{\mu}\right)^{p}\right)+\Theta(\phi_{\rm T}-\phi) V_{i}\left(\gamma (\phi_{\rm T}-\phi)^{q}+\phi_{0}^q\right),~\label{eq:equation-WWI}
\end{equation}

In the limit of   $\gamma=\phi_0=0$ the potential reduces to,
    \beq
    V({\phi})=V_{i} \left(1-\left(\frac{\phi}{\mu}\right)^{p}\right),~\label{eq:hilltop}
    \eeq
    as the base slow roll potential for the WWI model~\cite{Hazra:2014goa}. We denote it as {\it WWI-base}. We fix $p=4$ for our analyses with which this model generates different spectral tilt and $r$ as discussed in~\cite{Efstathiou:2006ak}. We therefore allow $\mu$ to vary along with $V_i$.  Since this potential allows only strict slow roll, the primordial power spectrum is featureless.
 
    In the limit of $\phi_0=0$, the potential reduces to,
     \begin{equation}
V({\phi})=V_{i} \left(1-\left(\frac{\phi}{\mu}\right)^{p}\right)+\Theta(\phi_{\rm T}-\phi) V_{i}\left(\gamma (\phi_{\rm T}-\phi)^{q}\right),~\label{eq:equation-WI}
\end{equation}
  that allows the scalar field to start rolling down a steeper slope potential ending up on the WWI-base potential with a continuous transition. We will refer this potential as the Whipped Inflation potential~\cite{Hazra:2014jka}.
  
   Note that $\phi_{\rm T}$ is the value of the field where the transition occurs and $\gamma$ determines the departure from slow roll. We use $q=2$. However, in such cases, the slow roll parameter $\epsilon_{\rm H}=-\frac{\dot{H}}{H^2}$ remains much less than 1 and does not refer to the completely kinetic-dominated regime~\footnote{For inflation with initial kinetic domination see~\cite{Contaldi:2003KD,Hergt:2018KD,Ragavendra:2020KD}}. Such dynamics suppress the power at large and intermediate scales without affecting the smaller-scale part of the spectrum. This potential is useful to determine the significance of the power suppression of the CMB T-mode angular power spectrum at lower multipoles. We refer to this potential as {\it WI-potential}.
  
The general potential in~\autoref{eq:equation-WWI} allows the field to transit to the slow roll part through a discontinuity in the potential. Apart from the large scale power suppression, this potential also creates bursts of oscillations either localized to a particular cosmological scale or persistent at all scales depending on the sharpness of the transition $\Delta$ when modelled with a step function. $\phi_{0}^q$ determines the extent of the discontinuity. This general potential will be referred as the WWI-potential. 
  
\subsection{WWIP-potential:} The base slow roll potential used here is the $\alpha$-attractor~\cite{Kallosh:2013alpha1,Kallosh:2013alpha2,Kallosh:2013alpha3} model. WWIP-potential (Wiggly Whipped Inflation Prime potential that contains discontinuities in the derivatives of the potential) is expressed as:
\begin{equation}
V({\phi})=\Theta(\phi_{\rm T}-\phi) V_{i} 
\left(1-\exp\left[-\alpha\kappa\phi\right]\right)+\Theta(\phi-\phi_{\rm T}) 
V_{ii}\left(1-\exp\left[-\alpha\kappa(\phi-\phi_{0})\right]\right),~\label{eq:equation-WWIP}
\end{equation}
    which reproduces Starobinsky's $R+R^2$ model~\cite{Starobinsky:1980te} of inflation for $\alpha=\sqrt{2/3}$, $\phi_0\to \ln 2/(\alpha\kappa)$, $V_i=V_{ii}$ and $\kappa\phi\gg 1$. For $V_i=V_{ii}$, 
        \beq
   V({\phi})=V_{i} 
\left(1-\exp\left[-\alpha\kappa\phi\right]\right).~\label{eq:starobinskyRR2}
    \eeq
    
    We use it as a base potential for WWIP and we denote it as {\it WWIP-base}. In this potential the spectral tilt and the tensor-to-scalar ratio are fixed. Therefore in this potential we allow the $N_*$ of the pivot mode to vary. In our analysis we fix the initial scale factor by demanding the pivot mode $0.05~{\rm Mpc}^{-1}$ leaves the Hubble radius $N_*$ {\it e-folds} before the end-of-inflation.
 
 When $\phi_0\ne0$, note that we have two different slopes at two different field regions separated by $\phi_{\rm T}$. $V_i$ and $V_{ii}$ are related by demanding the potential to be continuous at the transition. While the potential is continuous, its derivatives are discontinuous. This potential also produces suppression of scalar power at the large and intermediate scales and wide oscillations at the intermediate to small scales.

 This framework, constructed by these potentials allows us to explore the constraints on slow roll inflation and different types of departure from slow-roll. We use the power spectrum calculator from the BI-spectra and Non-Gaussianity Operator (\texttt{BINGO})~\cite{Hazra:2012BINGO,Sreenath:2014BINGO} to solve the inflationary dynamics. First we solve the Klein-Gordon equation assuming initial slow roll with $3H\dot{\phi}=-V_\phi$. The initial value of the scalar field is solved using a shooting algorithm. We assume the inflation ends at 70 {\it e-folds} as it is necessary to have at least 70 {\it e-folds} to ensure that the largest scale participating in the convolution integral of power spectrum and transfer function stays well inside the Hubble radius at the onset of inflation. As mentioned before, we normalize the initial scale factor ($a(N)=a_i$ at $N=0$) by fixing the time when pivot mode $0.05~\mathrm{Mpc}^{-1}$ departs Hubble radius at $N=20$. Note that this is exactly 50 {\it e-folds} before the end of inflation as we have fixed $N_\mathrm{end}$. We integrate the curvature perturbation equations (and also the tensor perturbation equations) for each mode-$k$ from sub-Hubble $k=100aH$ using Bunch-Davies initial condition to super-Hubble $k=10^{-5}aH$ scales. For sharp change in the potential due to the presence of discontinuity, sub-Hubble evolution becomes important and therefore $k=100aH$ is not sufficient choice for the convergence of the integral. We therefore dynamically change this condition at smaller scales in such a way that the solutions to the differential equations converge. We use {\tt CAMB}~\cite{Lewis:1999CAMB} to compute the angular power spectrum. For WWI and WWIP models we compute the angular power spectra for temperature and polarization for all multipoles avoiding interpolation.

\section{Data and analysis}~\label{DataAnalysis}
In this section we will discuss the properties of the data we are going to use in our analysis. We also discuss the prior ranges on the parameters, the methodology of model comparison and the method to search the best fit primordial power spectra.  

\subsection{Planck data: likelihoods we use}

The Planck 2018 data release delivered the temperature and polarization likelihoods on both large and small angular scales. Large angular scales are provided by the lowT likelihood in temperature based on the {\tt{commander}} component-separated temperature map and the {\tt{Simall}} E-mode polarization likelihood based on the $100\times143\,\mathrm{GHz}$ cross-spectrum. As for the small angular scales, two different likelihoods have been delivered, based on different algorithms and data treatments, both based on Half-Mission data auto and cross power spectra for 100, 143 and 217 GHz \cite{Planck2018V}. For some cases we also add to the analysis the lensing likelihood which is derived from the four-point correlation function on Planck CMB maps \cite{PlanckVIII}. In~\autoref{tab:likelihood} we list all the likelihoods we use in our analysis. Note that depending on the nature of feature models we use different likelihood combinations. For all the models, we use lowT likelihood for TT and {\tt Simall} likelihood for EE at low-$\ell$. The both cover multipoles $2-29$. At higher multipoles the ranges depend on the channels considered (either TT or EE). These are provided in the table. Using the keys mentioned in the table we will denote the combinations of Planck temperature and polarization likelihoods used in our analysis. For the base models we only use binned likelihoods from \texttt{Plik}, since base models do not have features. 
For base models we add also the B-mode measurements from the BICEP-Keck 2015 data~\cite{BK15} at 95-150 and 220 GHz. The likelihood covers the intermediate multipoles between 50 and 300 and it includes also some of the WMAP and Planck channels and a mitigation of the foreground contamination.
For the base model constraints we use the \texttt{Plik}-TTTEEE+lowT+lowE+Bk15+lensing likelihood jointly. We set $\mu$ in WI and WWI and $N_*$ in WWIP case to their best-fit values obtained in base model analyses. Since WI-model produces only suppression at large scales, we use three combinations: \texttt{Plik}-TT+lowT, \texttt{Plik}-EE+lowE and \texttt{Plik}-TTTEEE+lowT+lowE. We keep one temperature-only likelihood separate where we do not use any low-$\ell$ polarization information to understand the extent of support towards low-multipole power suppression from large-scale polarization data. Since WWI and WWIP models have sharp oscillatory features, for these cases we use the unbinned likelihood. However, instead of the unbinned {\tt bin1} \texttt{Plik} likelihood used in the Planck inflationary feature analysis, we use the recently released \texttt{Camspec 12.5HMcl} likelihood~\cite{EG20} obtained from a reanalysis of Planck data. This likelihood is based on an alternative data pipeline that excludes the 100 GHz temperature autospectrum and cleans the 143 and 217 GHz channels using the 545 GHz one. It applies a different mask, and the version we use, which applies the most aggressive masks, enlarges the sky fraction up to 80\% in temperature and polarization. We denote this likelihood as~\texttt{Clean CamSpec}. In~\cite{Braglia:2021ckn,Braglia:2fm10} (where one of the author of this paper has co-authored) this likelihood is termed as EG20. For the WWI and WWIP feature models, we use this likelihood for high-$\ell$. In both these models we use TT+lowT+lowE, EE+lowE, TE+lowE and TTTEEE+lowT+lowE combinations for constraints. In the complete TTTEEE+lowT+lowE likelihood we have 9 nuisance parameters, where 6 are foreground and 3 are calibration parameters. 

\begin{table}[]
\begin{tabular}{|l|l|l|l|}
\hline
Likelihood type                                            & Key                                                      & \begin{tabular}[c]{@{}l@{}}Multipole\\  range ($\ell$)\end{tabular}     & \begin{tabular}[c]{@{}l@{}}Likelihood name \\ in official release\end{tabular}                                                                                  \\ \hline
\texttt{Commander}                                                 & lowT                                                     & 2-29                                                                            & $\tt{commander\_dx12\_v3\_2\_29}$                                                                                                                               \\ \hline
\texttt{Simall}                                                    & lowE                                                     & 2-29                                                                            & $\tt{simall\_100x143\_offlike5\_}$                                                                                                                \\ 
                                                    &                                                      &                                                                            & $\tt{EE\_Aplanck\_B}$                                                                                                                \\ \hline

\begin{tabular}[c]{@{}l@{}}\texttt{PlikHM}\\ binned\end{tabular}   & \begin{tabular}[c]{@{}l@{}}Plik-TT\\ Plik-EE\\ Plik-TTTEEE\end{tabular} & \begin{tabular}[c]{@{}l@{}}TT: 30-2508\\ EE: 30-1996\\ TE: 30-1996\end{tabular} & \begin{tabular}[c]{@{}l@{}}$\tt{plik\_rd12\_HM\_v22\_TT}$\\ $\tt{plik\_rd12\_HM\_v22\_EE}$\\ $\tt{plik\_rd12\_HM\_v22b\_TTTEEE}$\end{tabular}                   \\ \hline
 \texttt{Clean CamSpec~(EG20)}                                                   & \begin{tabular}[c]{@{}l@{}}TT\\ EE\\ TTTEEE\end{tabular} & \begin{tabular}[c]{@{}l@{}}TT: 30-2500\\ EE: 30-2000\\ TE: 30-2000\end{tabular} & \begin{tabular}[c]{@{}l@{}}$\tt{v12.5HMcln TT}$\\ $\tt{v12.5HMcln EE}$\\
$\tt{v12.5HMcln TE}$\\
$\tt{v12.5HMcln TTTEEE}$\end{tabular}       \\
\hline
 {\tt BICEP-KECK 2015} & BK15                                                    & 50-300                                                                            & $\tt{BK15\_bandpowers\_20180920}$                                                                                                                               \\ \hline

\end{tabular}
\caption{~\label{tab:likelihood}Likelihood codes based on Planck data that we use in this work. As explained in the text, we use different combinations of likelihoods for our analysis. Note that~\texttt{Clean CamSpec} here is the same as EG20 as termed in~\cite{Braglia:2021ckn,Braglia:2fm10}.}
\end{table}

\subsection{Priors on parameters}

In~\autoref{tab:PriorRanges} we tabulate the priors used for background and inflationary parameters. Since we use nested sampling, we use only relevant parameter ranges as uniform priors. For the parameters common to all the models, listed in the first row of the table, the priors ensure that the posterior distributions for the least constraining data (EE+lowE in most cases) are not prior-dominated in the baseline case. The round brackets show the priors adopted where lowE is not used. Since the reionization optical depth and therefore the amplitude of primordial spectrum is not well constrained without lowE, we use sufficiently larger priors to obtain converged two-tailed distributions. Inflationary parameters are divided into two classes: {\it standard} parameters that control the amplitude and the tilt and {\it feature} parameters that define the characteristics of the features. In the Hilltop and $R+R^2$ models, $V_0$ represents the amplitude of the primordial scalar power spectrum. In all feature models, the parameter $\phi_T$ is allowed to vary in the range that generates the features within the cosmological scales probed by Planck. In WI model the parameter $\gamma$ determines the extent of suppression, and a negative value implies amplification of scalar power at large scales. In this table, we provide the likelihood or likelihood combinations used in constraining each model in this paper.

\begin{table}[]
\centering
 \begin{tabular}{| l | l | l | l |}
 \hline
Model	&	Parameter	&	Prior range& Likelihood\\
\hline		
Common parameters	&	$\Omega_\mathrm{b}h^2$             	&	$[0.02, 0.0265]$&	\\
	&	$\Omega_\mathrm{c}h^2$    	&	$[0.1, 0.135]$&	\\
	&	$100\theta$ 	&	$[1.03, 1.05]$&	\\
	&	$\tau$     	&	$[0.03, 0.08] (0.03, 0.18)$&	\\
	\hline
Hilltop	&	$\ln(10^{10}V_i)$	&	$[0.8,3]$&Plik-TTTEEE + lowT 	\\
		&	$\mu$	&	[10,30]& +lowE + BK15	\\
	\hline

Starobinsky-$R+R^2$	&	$\ln(10^{10}V_i)$	&	$[-2,2]$& Plik-TTTEEE + lowT 	\\
$\alpha$-attractor	&	$N_*$	&	$[50,60]$& +lowE + BK15	\\
	&	$\alpha$	&	$\sqrt{2/3}$&	\\
\hline
WI	&	$\ln(10^{10}V_i)$	&	$[2.2,2.35] (2.2,2.5)$& Plik-TT+lowT		\\
	&	$\gamma$	&	$[-0.005,0.2]$&	Plik-EE+lowE\\
	&	$\phi_T$	&	$[13.05,13.3]$&	Plik-TTTEEE+lowT+lowE\\
	&	$\mu$	&	21&	\\

\hline

WWI	&	$\ln(10^{10}V_i)$	&	$[2.2,2.35]$& [{\tt Clean CamSpec} for high-$\ell$]  \\
	&	$\gamma$	&	$[0,0.07]$&	TT+lowT+lowE\\
	&	$\phi_{0}$	&	$[0,0.05]$&	TE+lowE\\
	&	$\phi_T$	&	$[13.05,13.3]$&	EE+lowE\\
	&	$\ln\Delta$	&	[-12,-3]&	TTTEEE+lowT+lowE\\
	&	$\mu$	&	21&	\\
	\hline

WWIP	&	$\ln(10^{10}V_i)$	&	$[0,0.3]$&[{\tt Clean CamSpec} for high-$\ell$]	\\
	&	$\phi_{0}$	&	$[0,1]$&TT+lowT+lowE	\\
	&	$\alpha$	&	$\sqrt{2/3}$&	TE+lowE	\\
	&	$\phi_T$	&	$[4.45,4.57]$&	EE+lowE\\
	&	$N_{*}$	&	51&		TTTEEE+lowT+lowE\\
	\hline
\end{tabular}
\caption{~\label{tab:PriorRanges}Prior ranges for the cosmological parameters. The numbers in the round parentheses denote the ranges used when lowE is not included in the analysis. We use uniform priors over the ranges mentioned in the table. Foreground and calibration nuisance parameters have the same ranges and priors as were used in the Planck 2018 and \texttt{Clean CamSpec} publications. The fourth column indicates the likelihood or combination of likelihoods used in constraining the model.}
\end{table}

The range of $\gamma$ is chosen to ensure two outcomes: first, that the potential remains positive, which may not be true below the chosen threshold, with a lower limit of $\gamma\sim-0.005$; second, that the upper limit implies a power suppression on large scales already ruled out by data to consider all possible scenarios allowed. The parameter $\phi_{0}$ is related to the amplitude of wiggles. In WWI and WWIP we keep the lower limit as 0 to include the no-wiggle scenario. Note that the upper limit of the $\phi_{0}$ for WWI should be selected in relation to $\Delta$ as they are correlated. Exploring this degeneracy prior to the data comparison, we select the upper limits that are expected to be ruled out for all values of $\Delta$. We use a logarithmic prior on the sharpness ($\Delta$) parameter of the step and allow $\ln\Delta$ to vary between -12 and -3. Large $\ln\Delta$ corresponds to wider features and is expected to be allowed only at large scales. Very small $\ln\Delta$ represents sharp features with higher frequencies. With small amplitudes, sharp features can persist over a large range of scales, but CMB data is not able to distinguish between sharp features beyond Planck's resolution. The wiggles in WWIP model are only parametrized by the location and amplitude. Similar to WWI model, the upper limit of amplitude ensures that the posterior probability becomes negligible well within the limit. We vary all the nuisance parameters required by the different likelihoods together with the cosmological and inflationary parameters. For calibration parameters and amplitude of galactic dust parameters we use Gaussian priors as discussed in~\cite{Planck2018V,EG20}. 

We would like to outline the logic of using different likelihoods at high-$\ell$s for different potential. For potentials with featureless primordial spectrum as described by~\autoref{eq:hilltop} and~\autoref{eq:starobinskyRR2} and for potentials with featureless spectrum at intermediate and small scales described by~\autoref{eq:equation-WI} we use Plik binned likelihoods for high-$\ell$s. In these cases, the use of unbinned likelihoods do not provide additional information but require too much resources. For potentials with features at intermediate and small scales ({\it e.g.}~\autoref{eq:equation-WWI} and~\autoref{eq:equation-WWIP}) we use {\tt Clean CamSpec} likelihood that is unbinned. Of course for marginal likelihoods we compare the feature models with their baseline for the same dataset combination. 

\subsection{Model selection}
We investigate whether the updated {\tt CamSpec 12.5HMcln} likelihood and the \texttt{PlikHM} likelihood based on Planck 2018 data support the variants of Wiggly Whipped Inflation, and we quantify the degree of support. This is a model selection or model comparison problem within the framework of Bayesian inference.

Parameter estimation methods involve the calculation of the maximum likelihood. This is the average of the likelihood over the prior distribution. The maximum likelihood is therefore larger for a model if more of its parameter space is likely, and smaller for a model with a large parameter space in which the likelihood is low (known as `wasted parameter volume'). 
This holds even if the likelihood function has very high peaks, because these peaks must compensate for volumes in parameter space where the likelihood is extremely low. A simpler theory with a more compact parameter space will have a larger maximum likelihood than a more complicated model, unless the latter is significantly better at explaining the data. 
A simple comparison of $\chi^2$ values obtained from parameter estimation techniques (such as MCMC) are not sufficient, since they do not take into account the prior volume of the models being compared. We must therefore calculate the model evidence for each model, and compare the model evidence using the Bayes factor.

 Bayesian inference is based on the logic of probability theory. The product rule gives us Bayes' theorem:
 \beq \label{eqn1}
 p(\vtheta | d,M)=\frac{p(d | \vtheta,M)p (\vtheta | M)}{p( d | M)}.
 \eeq

 The left-hand side is the posterior probability for the vector of unknown model parameters $\vtheta$ of length $n$ given the data $d$ under model $M$. The prior probability distribution function $p(\vtheta|M)$ is an expression of our state of knowledge before observing the data. This defines the prior available parameter space under the model $M$. The denominator $p (d | M)$ is the Bayesian evidence or model likelihood. This is the probability of observing the data $d$ given that the model $M$ is correct. 

 Model selection usually involves the calculation of the evidence or marginalised likelihood. This may be expressed as the multidimensional integral of the likelihood over the prior, or the parameter space under $M$:
 \beq
 Z \equiv p (d | M) = \int  p(d | \vtheta,M) p(\vtheta | M) \mathrm d \vtheta ,
 \eeq
 where $\vtheta$ is in general multidimensional, and $d$ is a collection of measurements (current or future).

 The Bayes factor is then the ratio of the evidence for two competing models $M_0$ and $M_1$, or the ratio of posterior odds. If the competing models have equal prior probability, then:
 \beq
 B_{01}\equiv\frac{p(d | M_0)} {p(d | M_1)}=\frac{p(M_0|d)}{p(M_1|d)}
 \eeq

We use {\tt PolyChord}\cite{Handley2015a,Handley:2015fda} to sample the parameter space and calculate the marginalized likelihoods for our models~\footnote{In particular, we use the {\tt PolyChord} integration of {\tt CosmoMC}~\cite{Lewis:2002COSMOMC} via {\tt CosmoChord}. {\tt Getdist}~\cite{Lewis:2019GETDIST} is used for post-processing.}. {\tt PolyChord} is a nested sampling algorithm, optimized for high-dimensional spaces. As such it is designed to calculate the marginalised likelihoods rather than the posterior, although posterior samples may still be obtained by using the dead and live points.

We compare the evidence for each model with the evidence for the respective featureless base models. In the base models, we fix the feature parameters to zero, and we keep the priors on all the other parameters unchanged. The comparison is made against the same dataset combination, as indicated earlier.

\section{Results}~\label{ResultsSection}
\subsection{Model comparison}

The results of our Bayesian model comparison are shown in~\autoref{tab:evidences}. The logarithm of the Bayes factor $\ln B$ is the difference of $\ln Z$ for the feature model and the featureless model.

The interpretation of the value of $\ln B$ depends to a large extent on the experiment being carried out. We use the Jeffreys scale to interpret the values in terms of the strength of evidence for the competing model (see \cite{Jeffreys:1961} and \cite{Trotta:2007}). An experiment for which $|\ln B|<1$ is usually deemed inconclusive, that is, the alternative model cannot be distinguished from the null hypothesis.

We note that $\ln B$ for all models falls between $0.12$ and $-2.8$. The only positive value is obtained for WI against the featureless model using TT+lowT. However, this is still within the inconclusive range. The TT+lowT+lowE, EE+lowE, and TTTEEE+lowT+lowE data sets give positive evidence against all three feature models. With TE+lowE, the evidence against WWIP is even stronger.


\begin{table}[]
    \centering
\renewcommand{\arraystretch}{1.2}
\begin{tabular}{l l l l}
\hline &  Model &&\\
Data set         & WI   & WWI  & WWIP  \\ \hline

TT+lowT          & 0.12 & NA    & NA     \\ 
TT+lowT+lowE     & NA    & -2.3 & -2.5  \\ 
TE+lowE          & NA    & -2.3 & -2.8  \\ 
EE+lowE          & -1.8 & -2.4 & -2.1  \\ 
TTTEEE+lowT+lowE & -1.4 & -2.5 & -1.61 \\ \hline
\end{tabular}
\caption{~\label{tab:evidences}Logarithm of the Bayes factor for the feature models compared to their respective slow-roll baseline model for the same dataset combinations. The data sets are as follows: for WI, high-$\ell$ TT, TE, EE and TTTEEE refer to \texttt{PlikHM} binned data, while for WWI and WWIP they refer to {\tt Clean CamSpec} data. NA for a model indicates that model has not been compared with the corresponding dataset.}
\end{table}

\subsection{Parameter estimation}
From the samples gathered using the nested sampling through {\tt PolyChord} we explore the posterior distributions of the parameters in all the models within the WWI family and for all dataset combinations considered in this analysis. 

\begin{figure*}[!htb]
\includegraphics[width=\columnwidth]{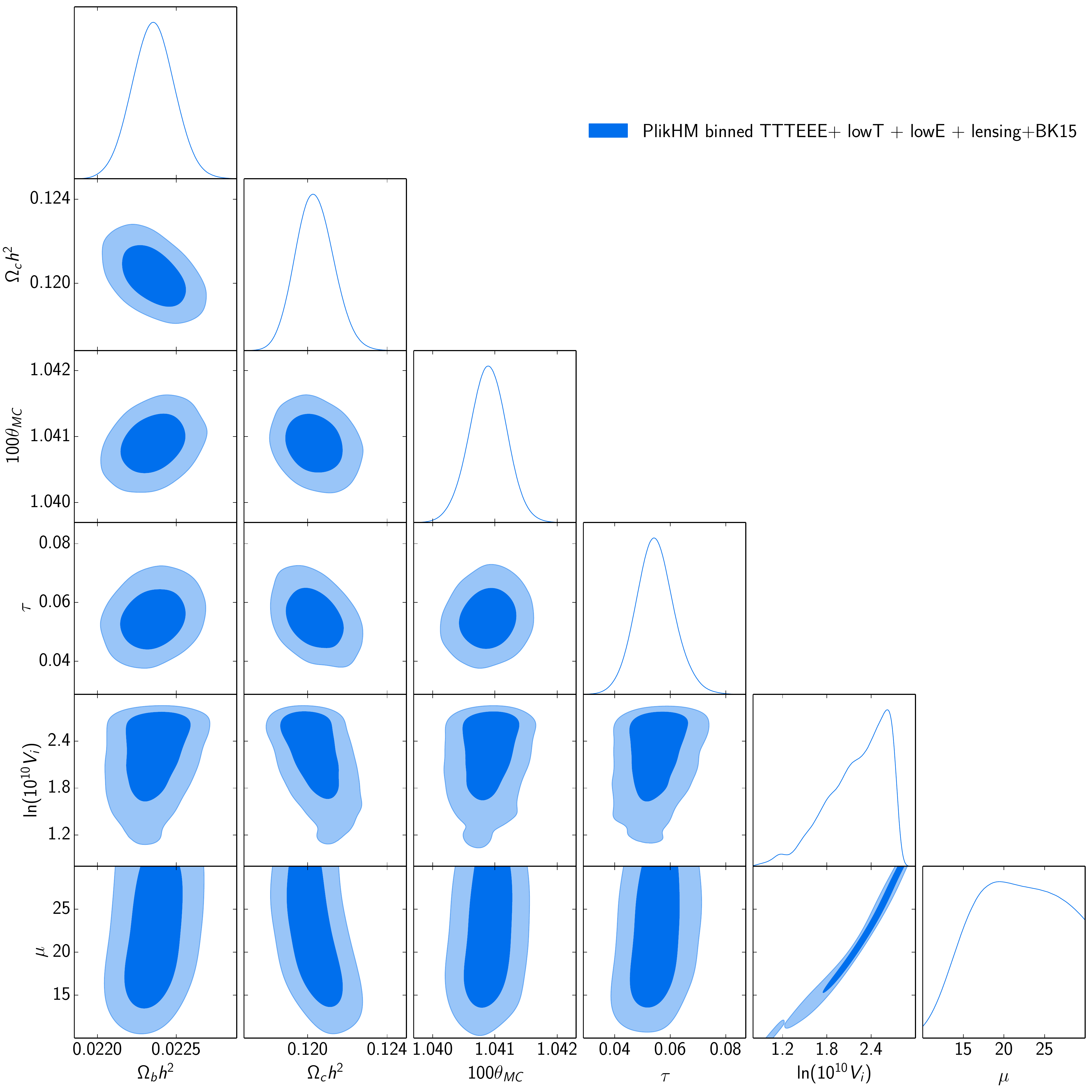}
\caption{\footnotesize\label{fig:hilltop}Constraints in Hilltop slow roll inflation using Planck 18 TTTEEE (\texttt{PlikHM} binned) + lowT + Simall + lensing + BK15 likelihood. We compare both primordial scalar and tensor power spectra with the data.}
\end{figure*}

\begin{figure*}[!htb]
\includegraphics[width=\columnwidth]{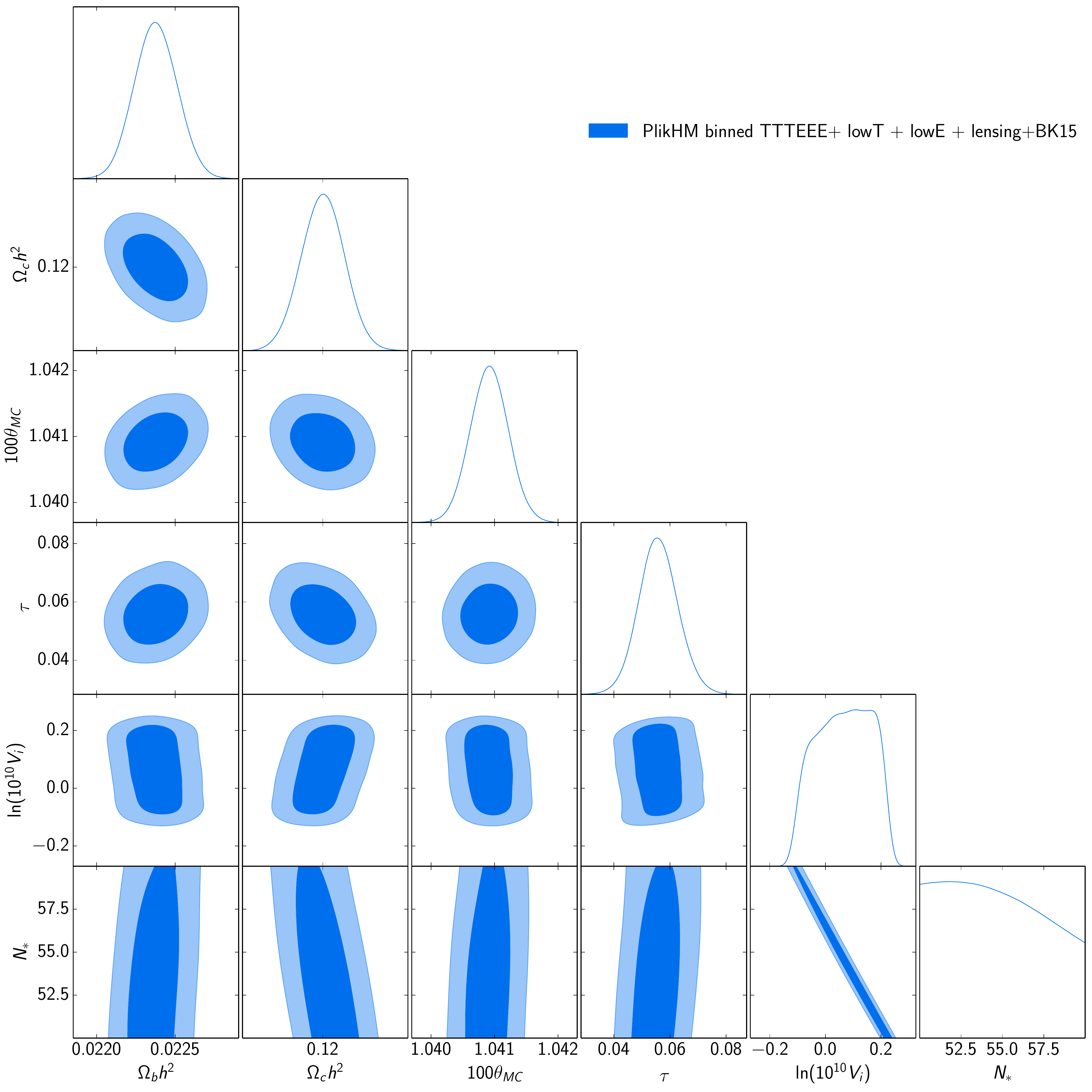}
\caption{\footnotesize\label{fig:starobinskyRR2}Triangle plot for Starobinsky ($R+R^2$) slow roll potential in Einstein frame. Here too, we compare both primordial scalar and tensor power spectra with the data.}
\end{figure*}

In~\autoref{fig:hilltop} we present the constraints on the background and inflationary potential parameters for Hilltop inflation discussed in~\autoref{eq:hilltop}. Here we use the complete Planck temperature, polarization and lensing datasets and BK15 likelihood. Compared to the Planck baseline results we do not notice changes in the background parameter constraints, as expected. The potential $V_i$ is strongly constrained by the overall normalization. However, compared to the constraints on $A_s$ in a power law primordial spectrum, the posterior distribution is found to be strongly skewed for $V_i$. This is an artefact of the degeneracy with $\mu$. In~\autoref{tab:HilltopRR2} the $1\sigma$ constraints and unidirectional 95\% limits (for distributions without two tails) are mentioned. Our analysis constrains $\mu>13.4~\Mpl{}$ at 95\% C.L. We use the best fit from this analysis and fix $\mu=21\Mpl{}$ for the WI and WWI case, where the Hilltop quartic represents the potential in the slow-roll phase. Constraints on $R+R^2$ are plotted in~\autoref{fig:starobinskyRR2} and tabulated in~\autoref{tab:HilltopRR2}. The data are not able to constrain the pivot scale $N_*$ beyond $1\sigma$ upper bound. This is expected as the change in the constraint in the spectral tilt is not substantial, and therefore the posterior remains prior-dominated. We find the best-fit value of $N_*\sim51$, which we use for the runs with features. Here too we find degeneracies between the potential $V_i$ and the pivot scale that substantially relax the constraint on $V_i$. 
\begin{table}[htbp]
    \centering
\setlength{\tabcolsep}{10pt} 
\renewcommand{\arraystretch}{1.5}
    \begin{tabular}{c c c}
    \hline
 & \texttt{PlikHM} binned TTTEEE & + lowT + lowE + lensing+BK15\\    
Param & Hilltop&Starobinsky $R+R^2$\\
\hline
$\Omega_bh^2$ & $0.0224\pm 0.0001$&$0.0224\pm 0.0001$\\
$\Omega_ch^2$ & $0.120\pm 0.001$&$0.120\pm 0.001$\\
$100\theta_{MC}$ & $1.0409\pm 0.0003$&$1.0409\pm 0.0003$\\
$\tau$ & $0.0546_{-0.0070}^{+0.0064}$&$0.0559_{-0.0068}^{+0.0069}$\\
$V_i$ & $2.21_{-0.21}^{+0.53}$& $0.0671_{-0.0802}^{+0.1297}$\\
$\mu$ & $>13.4$&NA\\
$N_{*}$ &NA& -\\
\hline
    \end{tabular}
    \caption{~\label{tab:HilltopRR2}Constraints on Hilltop quartic potential and Starobinsky $R+R^2$ model obtained using \texttt{PlikHM} binned TTTEEE + lowT + lowE, Planck lensing and BK15 likelihood. We compare both scalar and tensor power spectrum in these two models against the data. Note that in the Hilltop model, $\mu$ allows the variation of spectral tilt and the tensor-to-scalar ratios, while in the Starobinsky model $N_{*}$ determines the spectral tilt. In the runs for beyond slow-roll we use the best fit values of $\mu$ and $N_*$ from these analyses.}
\end{table}

\begin{table*}[htbp]
    \centering
\setlength{\tabcolsep}{10pt} 
\renewcommand{\arraystretch}{1.2}
    \begin{tabular}{c c c c}
\hline
 & & \texttt{PlikHM} binned& \\

Param & TTTEEE & TT & EE\\
 & lowT+lowE & +lowT & +lowE\\
\hline
$\Omega_bh^2$ & $0.0224\pm 0.0001$& $0.0223\pm 0.0002$& $0.0232\pm 0.0009$\\
$\Omega_ch^2$ & $0.120\pm 0.0010$& $0.120\pm 0.001$& $0.119\pm 0.003$\\
$100\theta_{MC}$ & $1.0409\pm 0.0003$& $1.0409\pm 0.0004$& $1.0397\pm 0.0008$\\
$\tau$ & $0.0590_{-0.0095}^{+0.0080}$& $0.1138_{-0.0261}^{+0.0259}$& $0.0571_{-0.0120}^{+0.0092}$\\
$\ln[10^{10} V_i]$ & $2.271_{-0.019}^{+0.017}$& $2.379\pm 0.052$& $2.278_{-0.0265}^{+0.0232}$\\
$\gamma$ & $<0.068$ & $<0.069$& $<0.166$\\
$\phi_T$ & $<13.19$ & $<13.22$& $<13.23$\\
\hline
    \end{tabular}
    \caption{~\label{tab:WI}Results for the WI potential. We report the $1\sigma$ uncertainties ($2\sigma$ upper bounds). Here we consider \texttt{PlikHM} binned likelihood for TT+lowT, EE+lowE and TTTEEE+lowT+lowE}
\end{table*}

\begin{figure*}[!htb]
\includegraphics[width=\columnwidth]{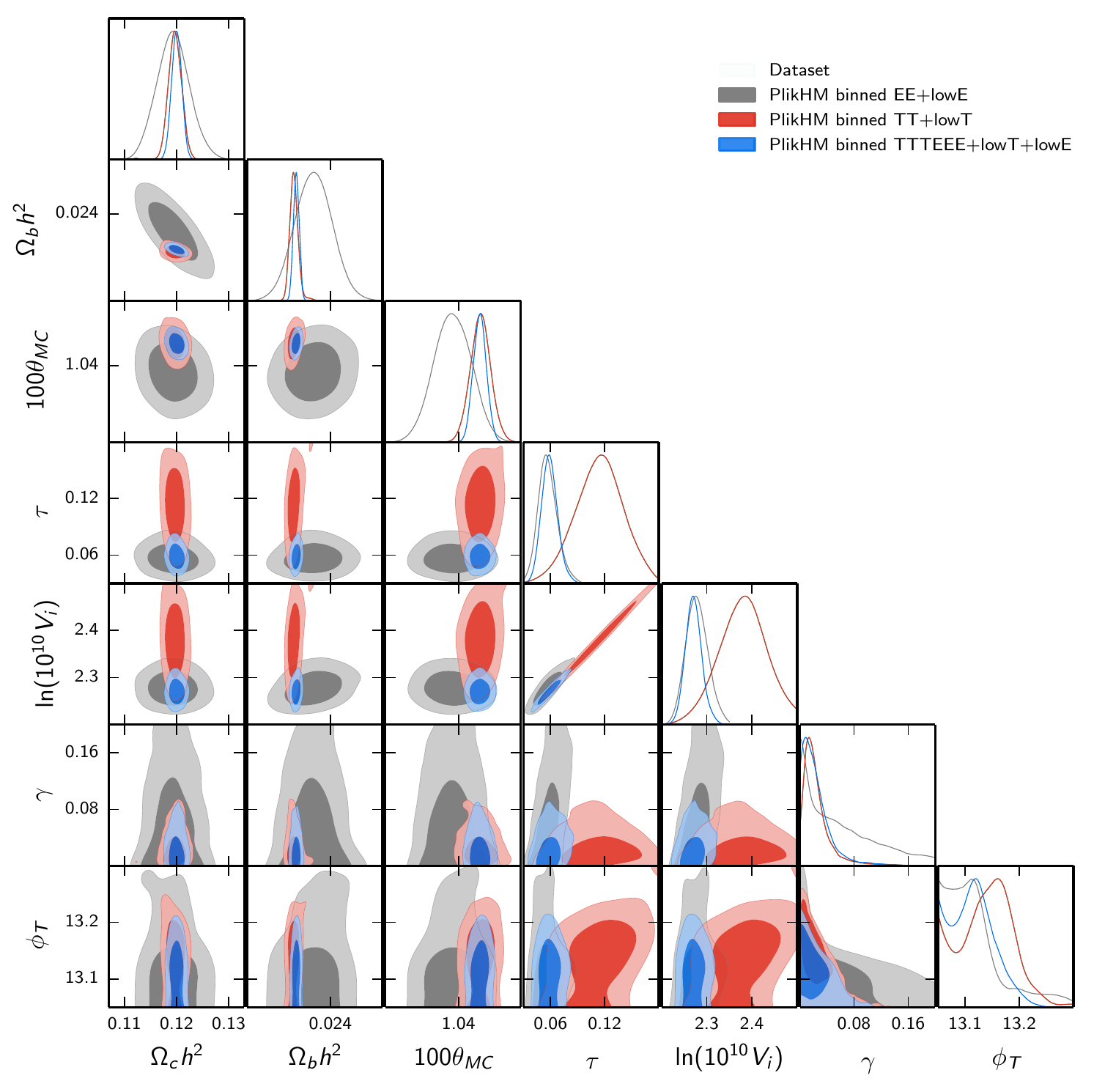}
\caption{\footnotesize\label{fig:WI}Constraints on Whipped Inflation from different data combinations \texttt{PlikHM+} binned TT + lowT, \texttt{PlikHM} binned EE + LowE and \texttt{PlikHM} binned TTTEEE + lowT + lowE. In temperature-only constraints we note marginal support for the large scale suppression. However since polarization-only likelihood (essentially LowE) does not support large-scale scalar suppression, in the complete likelihood we find decreased significance for the suppression. Large uncertainties in the determination of the amplitude and the reionization optical depth in the TT+lowT case occur owing to the absence of large-scale polarization data.}
\end{figure*}

Constraints on Whipped Inflation potential are shown in~\autoref{fig:WI}. Here we plot constraints from different datasets. Since we have fixed $\mu$ to the best-fit value in the base slow roll potential, and the best fit is obtained using B-mode observation data from BK15, we do not evaluate and compare the B-mode anisotropy spectrum. Since the present upper bound on the tensor power spectrum indicates no noticeable effect in the large-scale temperature anisotropy spectrum, we do not compute the tensor spectra either. The constraints plotted here are from \texttt{PlikHM} binned TT+lowT, EE+lowE and TTTEEE+lowT+lowE. Therefore, for the WI model, all the likelihoods that we discuss will belong to \texttt{PlikHM}. Here we do not consider TE data separately, since the suppression in the scalar power spectrum is constrained at large scales. In the combination of temperature data (TT+lowT) we do not include lowE to differentiate the support for the suppression from different available data at the largest scales. In this model $\gamma,\phi_T$ represent the extent and the location of the suppression of power. From the marginalized posterior distributions, a support between 1 and $2\sigma$ for the suppression can be noticed in TT+lowT. Since the low multipole temperature power spectrum data is consistently lower than the baseline model prediction, we find that a suppression provides improvement in fit compared to the baseline model, making it marginally favoured. As discussed before, we find positive but inconclusive Bayesian evidence for this model only in TT+lowT case. When we consider EE+lowE datasets, we do not find any support for large-scale features, which suggests that lowE data does not allow any broad features in the power spectrum. Due to the lack of support, additional prior volume from ${\gamma,\phi_T}$ is penalized in the marginal likelihood and we find inconclusive preference for baseline model over the WI model. When we combine all datasets, we find the preference for the WI model reduced due to the addition of lowE.
As expected, the value of $\ln B$ is found to be between the value obtained with TT+lowT and EE+lowE, which means that the evidence for the baseline over WI is slightly weakened with the combined datasets. It is also important to note the role of the lowE likelihood in constraining the scale of the potential $V_i$. Since in the constraints from temperature likelihood the amplitude and the optical depth are highly degenerate, the constraints on the potential and therefore the optical depth are weak. The lowE data breaks the degeneracy by putting stringent bounds on the optical depth, which in turn constrains $V_i$ significantly better than TT+lowT. The uncertainties on the parameters are tabulated in~\autoref{tab:WI}, where the drastic improvement in the bounds on the potential can be noticed. A comparison of $\tau$-constraints between~\autoref{tab:WI} and~\autoref{tab:HilltopRR2} or as obtained in Planck analysis~\footnote{see Planck Params table:~\href{https://wiki.cosmos.esa.int/planck-legacy-archive/images/4/43/Baseline_params_table_2018_68pc_v2.pdf}{PLA: Planck parameters table}} ($\tau=0.0544^{+0.0070}_{-0.0081}$) indicates a 10\% increase in uncertainties in both direction around an increased mean optical depth. A broad suppression in the power spectrum at large scales is marginally degenerate with the optical depth. This degeneracy increases the uncertainties and a preference of suppression in turn prefers an increase in the optical depth in the lowE data. Note that to compensate for the suppression in the lowE data, $\tau$ must increase to match the reionization bump in EE.

\begin{figure*}[!htb]
\centering
\includegraphics[width=0.49\columnwidth]{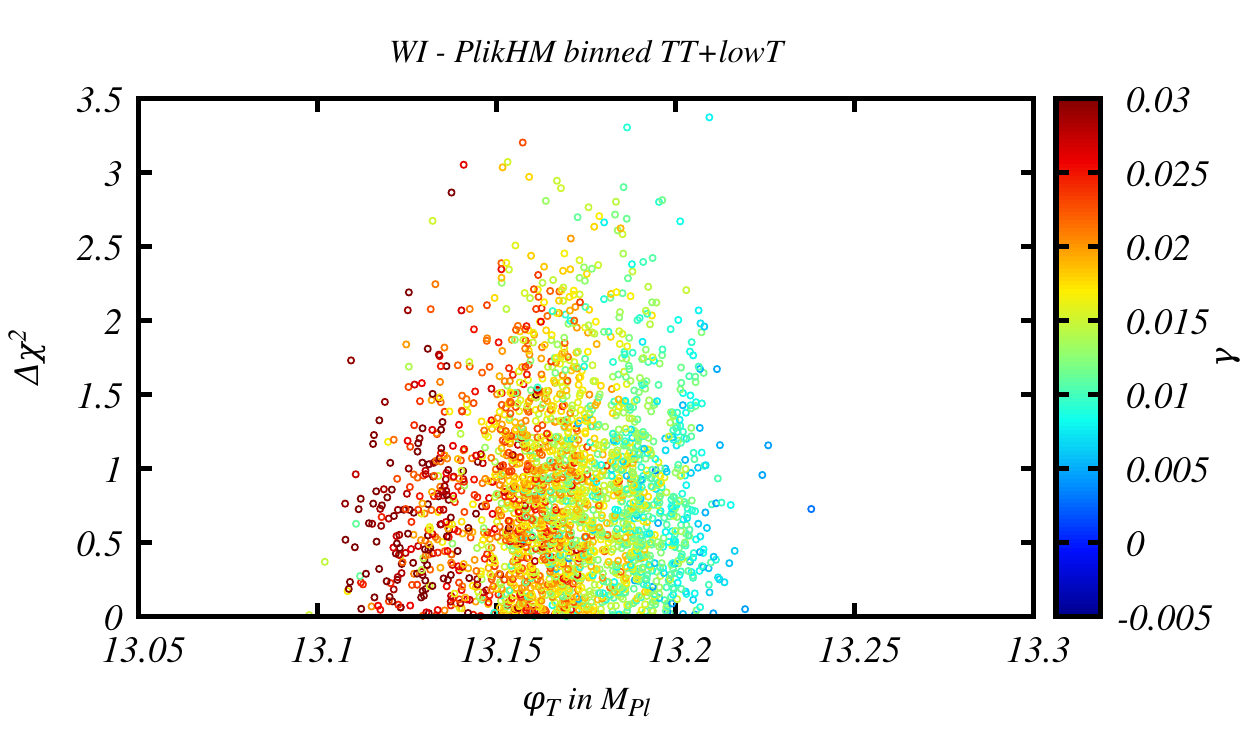}
\includegraphics[width=0.49\columnwidth]{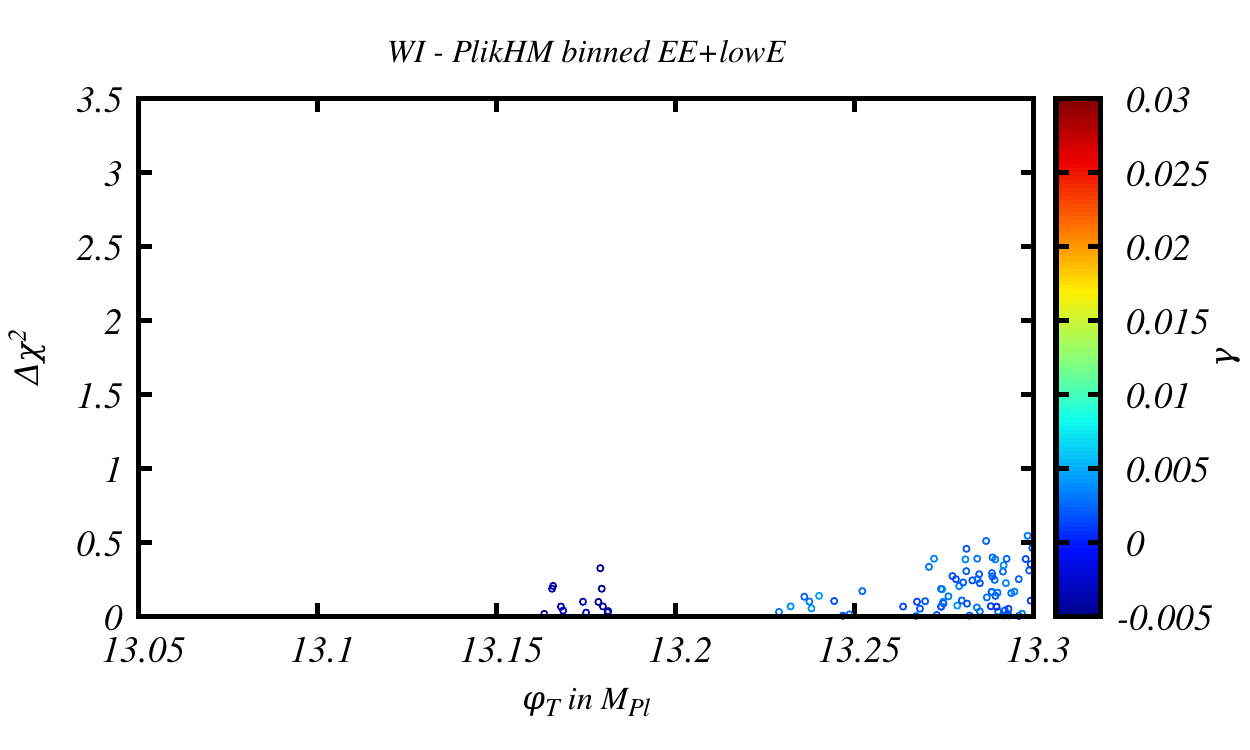}
\includegraphics[width=0.49\columnwidth]{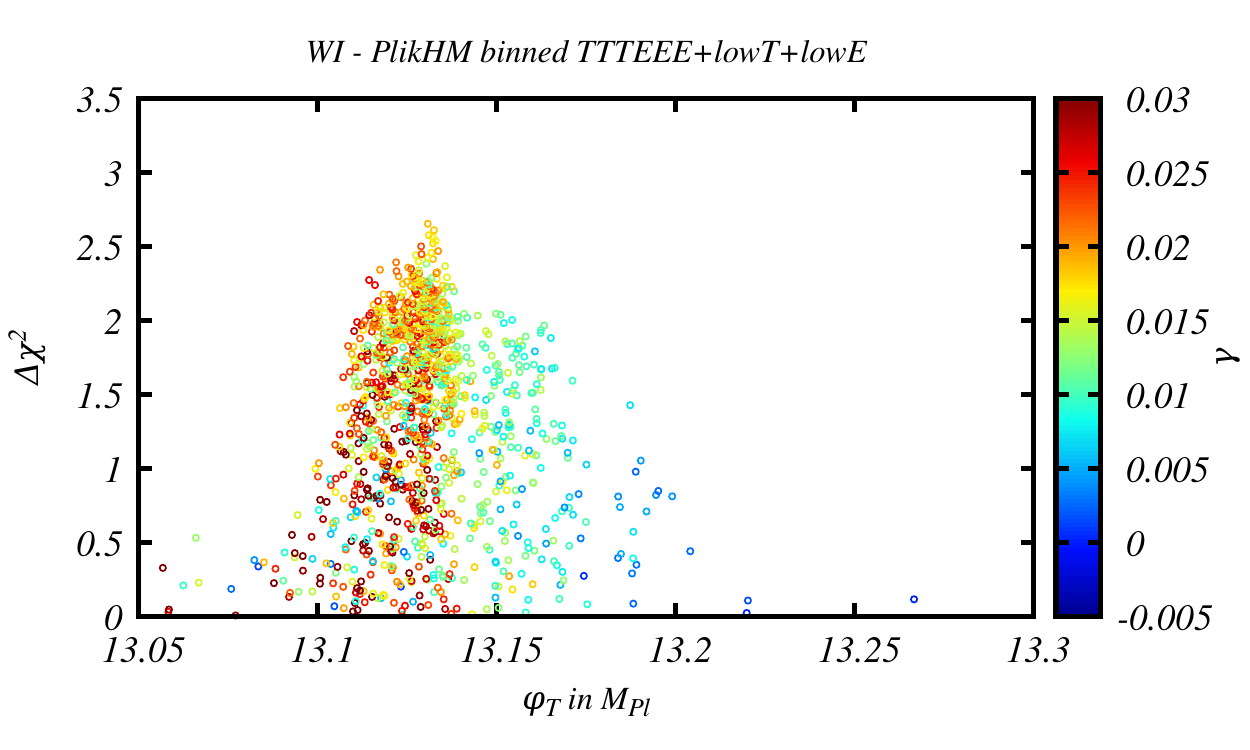}
\caption{\footnotesize\label{fig:WI-Delchi2} The samples with improvement in $\chi^2$ in the Whipped Inflation compared to its featureless baseline model. Note that we have compared the $\chi^2$ from the \texttt{PolyChord} analysis and therefore the maximum value does not necessarily mean global best-fit sample. We have plotted the samples against $\phi_T$ and the $\gamma$ parameter representing the location and magnitude of the large scale suppression in the scalar power spectrum. While TT+lowT data supports such transition with certain improvement in the $\chi^2$ compared to the baseline model we notice no such improvement in the EE+lowE data. Finally in the combined datasets we find less support towards large scale suppression compared to temperature only data.}
\end{figure*}

In~\autoref{fig:WI-Delchi2} we plot the improvement in $\chi^2$ {\it w.r.t.} the baseline best fit from the {\tt PolyChord} samples for different data combinations. The samples are plotted as a functions of the transition field value ($\phi_T$) which represents the location of the feature, and they are colored according to the value of $\gamma$ representing the extent of suppression. Here the higher $\phi_T$ represents a power spectrum with the suppression extending to smaller scales. Top right, left plots and the plot at the bottom correspond to TT+lowT, EE+lowE and TTTEEE+lowT+lowE respectively. For TT+lowT we find several samples providing better fit to the data compared to the baseline. A clear correlation is visible here. With higher $\phi_T$ Planck data only allows smaller $\gamma$ in order to provide the improved fit. For EE+lowE we do not obtain samples that improve the fit to the data. A few samples are located with negligible improvement ($\Delta\chi^2<0.5$). For the complete datasets we find marginally lower improvement compared to the temperature only data. The samples with higher $\gamma$ relocate to larger scales (higher $\phi_T$) as well. 

In our analysis, wiggles in the primordial spectrum are compared against the \texttt{Clean CamSpec} likelihood. Therefore in the context of WWI and WWIP potentials, mention of high-$\ell$ dataset combinations refer to \texttt{Clean CamSpec} 12.5 likelihood.

\begin{figure*}[!htb]
\includegraphics[width=\columnwidth]{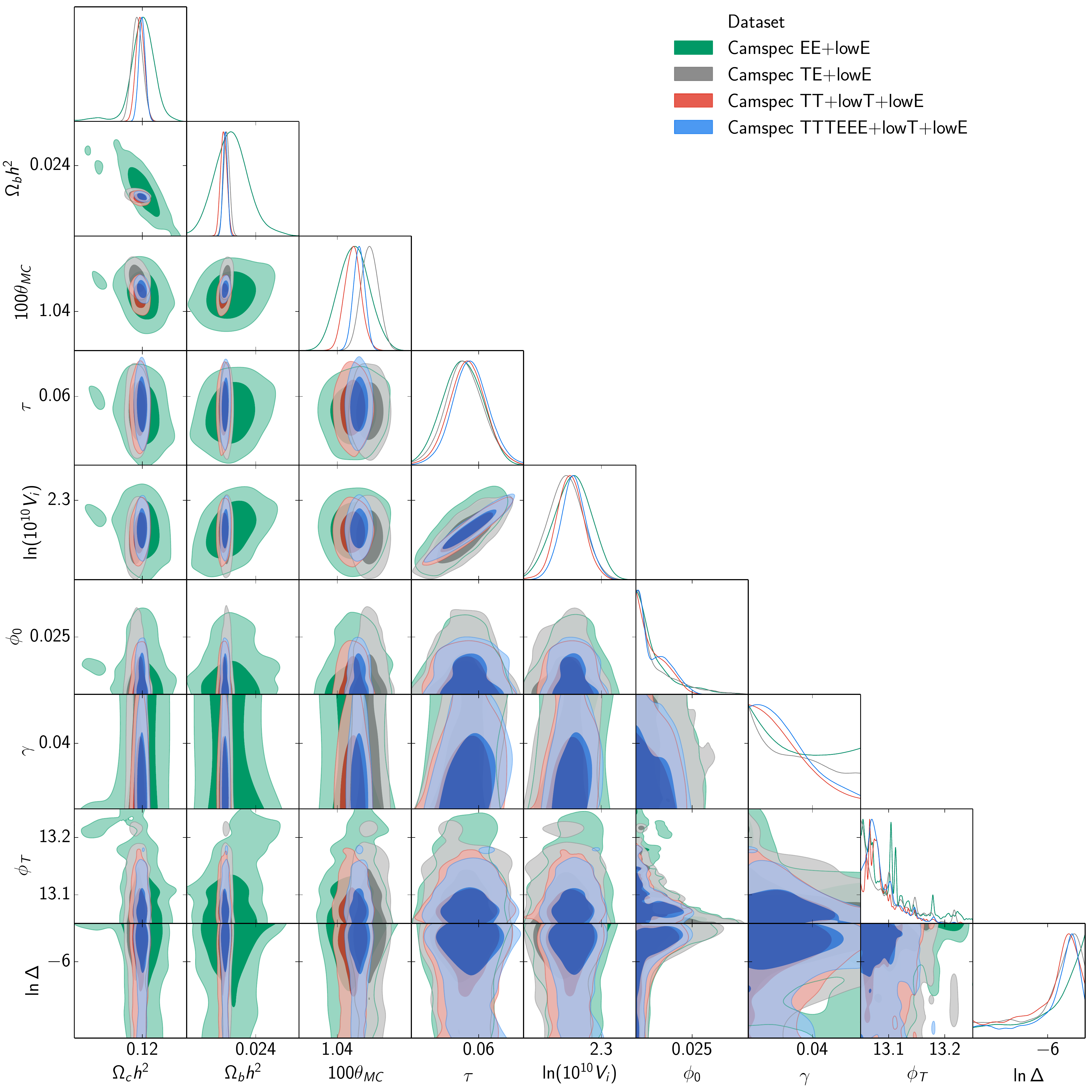}
\caption{\footnotesize\label{fig:WWI}Constraints on WWI potential for four different combination of datasets, namely \texttt{Clean CamSpec} TT+lowT+lowE, EE+lowE, TE+lowE and TTTEEE+lowT+lowE.}
\end{figure*}

\autoref{fig:WWI} plots the constraints on the WWI model. Given the increased degeneracy with 4 extra parameters compared to its slow roll base Hilltop quartic potential, we do not get significant determination of any feature parameters.~\autoref{tab:evidences} containing evidence also indicates preference towards slow roll baseline model compared to the WWI model. Marginalized posteriors on the suppression parameter $\gamma$ and the wiggle amplitude $\phi_0$ peak at zero values. We get a secondary plateau for $\phi_0$ parameter around $0.01\Mpl{}$ for TT+lowT+lowE and TTTEEE+lowT+lowE. Also for these two datasets, we find marginal preference on the sharpness of the transition parameter $\ln\Delta\sim-5$. The EE+lowE and TE+lowE datasets however do not show any preference on feature parameters in the marginalized posteriors. Constraints on the parameter values are listed in~\autoref{tab:WWI}. While for TE+lowE and EE+lowE we are unable to provide any significant bounds on $\gamma$ and $\ln\Delta$, we obtain upper bounds in these parameters when adding the low-$\ell$ temperature likelihood for the TT+lowT+lowE and TTTEEE+lowT+lowE combinations. We find the mild preference in the $\ln\Delta$ is not statistically significant at all in any combination of data as we are not able to constrain the parameter at 95\% in either direction around the mean. Compared to WI potential, if we compare~\autoref{tab:WI} and~\autoref{tab:WWI} in WWI model the bounds on $\gamma$ and $\phi_T$ are tighter. Note that here we can only compare the EE+lowE and TTTEEE+lowT+lowE case as in WI model we have not considered TT+lowT+lowE. The tighter constraints even with extra degrees of freedom are due to the occurrence of wiggles along with the suppression in this model, {\it i.e.} a non-zero $\gamma$ and $\phi_0$ imply both suppression and wiggles which are not allowed to extend below certain cosmological scales unless the features are sharp and sufficiently small in amplitude (as can be noticed in the correlation between $\gamma$, $\phi_0$ and $\ln\Delta$). Comparison of the $\Omega_bh^2$ posteriors in~\autoref{fig:WI} and~\autoref{fig:WWI} and the mean values in~\autoref{tab:WI} and~\autoref{tab:WWI} reveals that for WWI, the $\Omega_bh^2$ posterior from EE+lowE shifts towards TT+lowT+lowE. This does not occur for WI.
This is expected because the wiggles in the WWI are marginally degenerate with the baryon acoustic oscillations~\cite{Hazra:2016fkm} causing a shift in WWI of the polarization-only results towards temperature data, increasing the agreement between the two datasets.

\begin{table}[htbp]
    \centering
\setlength{\tabcolsep}{10pt} 
\renewcommand{\arraystretch}{1.2}
\begin{tabular}{c c c c c}
\hline
 & & Clean CamSpec& \\
Param & TTTEEE& TT& TE& EE\\
& lowT+lowE & lowT+lowE & lowE & lowE\\
\hline
$\Omega_bh^2$ & $0.0222\pm 0.0001$& $0.0221\pm 0.0002$& $0.0223\pm 0.0001$& $0.0226\pm 0.0010$\\
$\Omega_ch^2$ & $0.120\pm 0.001$& $0.119\pm 0.001$& $0.118\pm 0.002$& $0.120_{-0.003}^{+0.004}$\\
$100\theta_{MC}$ & $1.0410\pm 0.0003$& $1.0407\pm 0.0004$& $1.0415\pm 0.0004$& $1.0408\pm 0.0008$\\
$\tau$ & $0.0564_{-0.0077}^{+0.0079}$& $0.0550_{-0.0078}^{+0.0079}$& $0.0537_{-0.0088}^{+0.0083}$& $0.0530\pm 0.0091$\\
$\ln[10^{10} V_i]$& $2.264\pm 0.016$& $2.260\pm 0.017$& $2.255\pm 0.022$& $2.264\pm 0.025$\\
$\phi_{0}$ & $<0.021$& $<0.019$& $<0.029$& $<0.028$\\
$\gamma$ & $<0.059$& $<0.056$& $-$& $-$\\
$\phi_T$ & $<13.15$& $<13.15$& $<13.19$& $<13.21$\\
$\ln\Delta$ & $-$& $-$& $-$& $-$\\
\hline
    \end{tabular}
    \caption{\label{tab:WWI}Results for WWI potential. Here for high-$\ell$ likelihood we have used the \texttt{Clean CamSpec} unbinned angular power spectrum data. We note that no feature parameter is detected and we only obtain upper bounds in these parameters.}
\end{table}

\begin{figure*}[!htb]
\centering
\includegraphics[width=0.495\columnwidth]{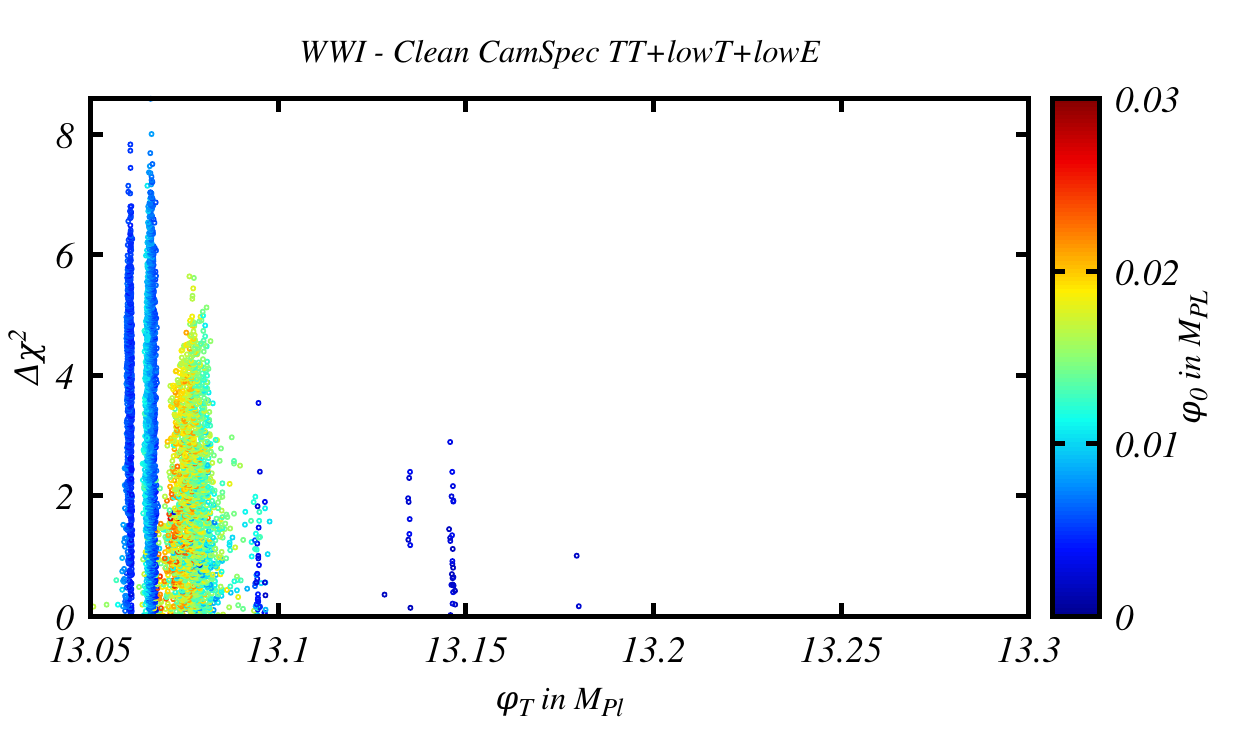}
\includegraphics[width=0.495\columnwidth]{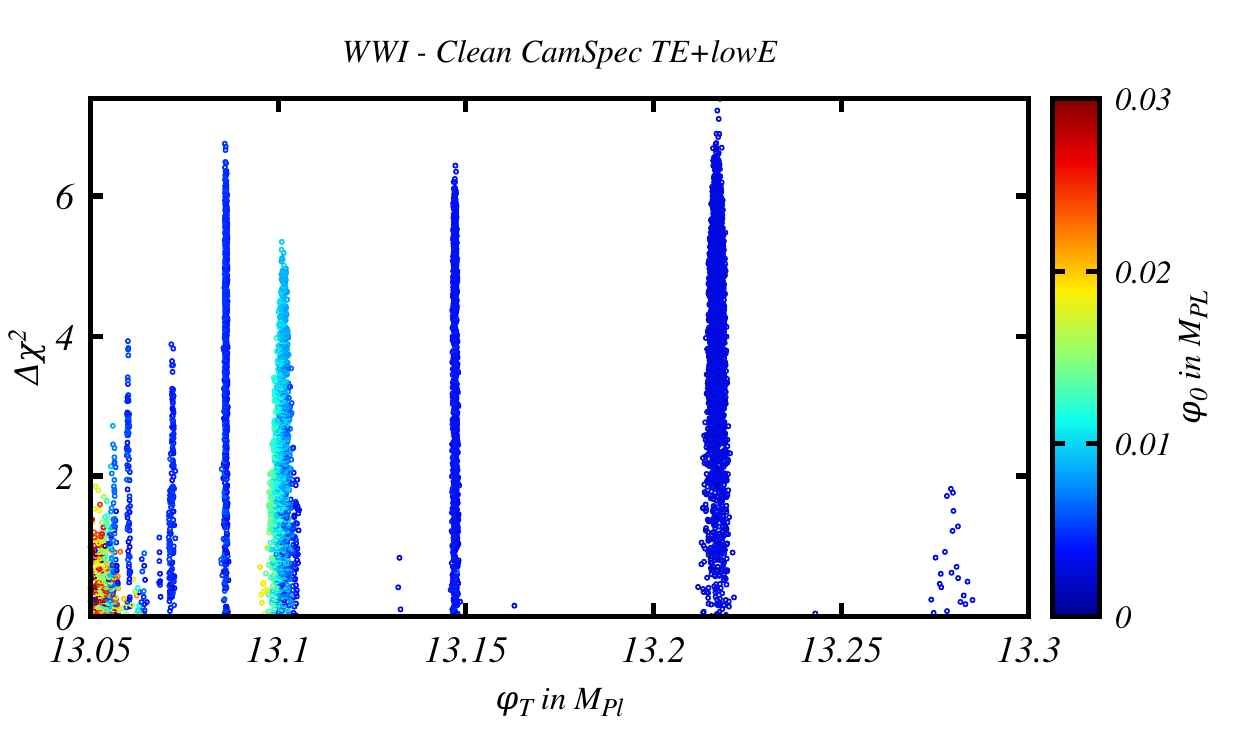}

\includegraphics[width=0.495\columnwidth]{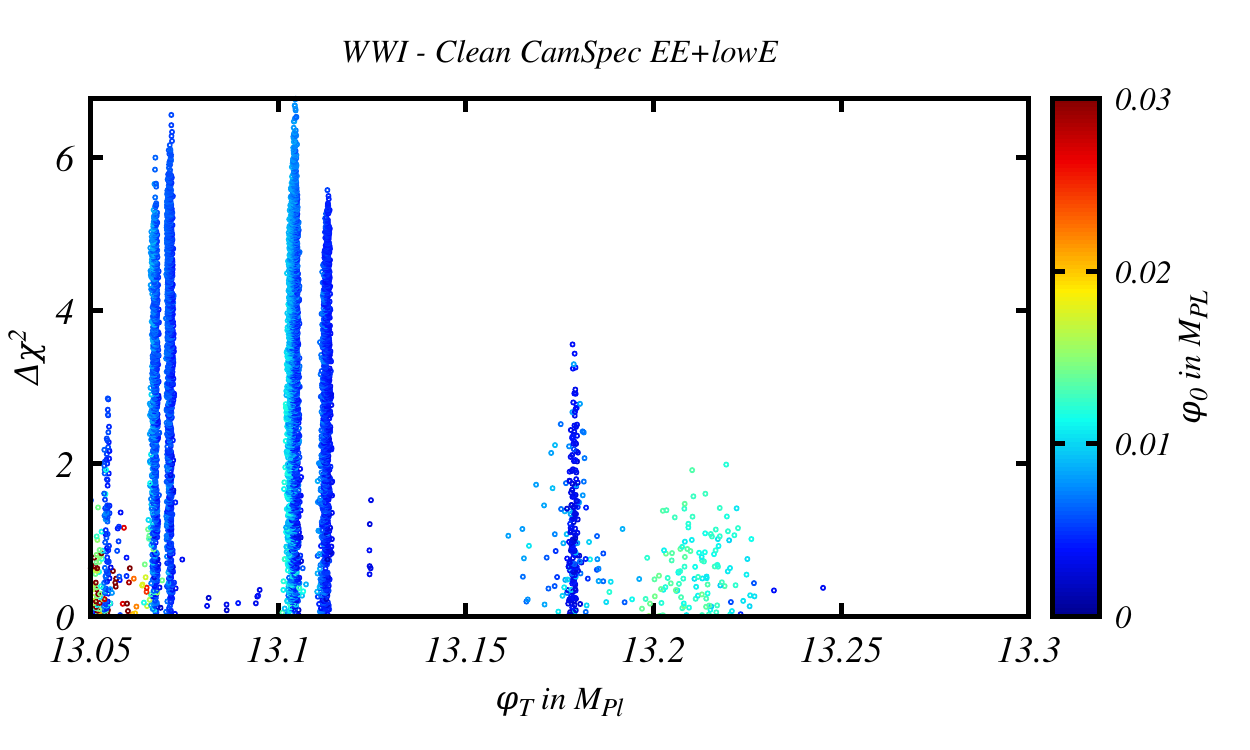}
\includegraphics[width=0.495\columnwidth]{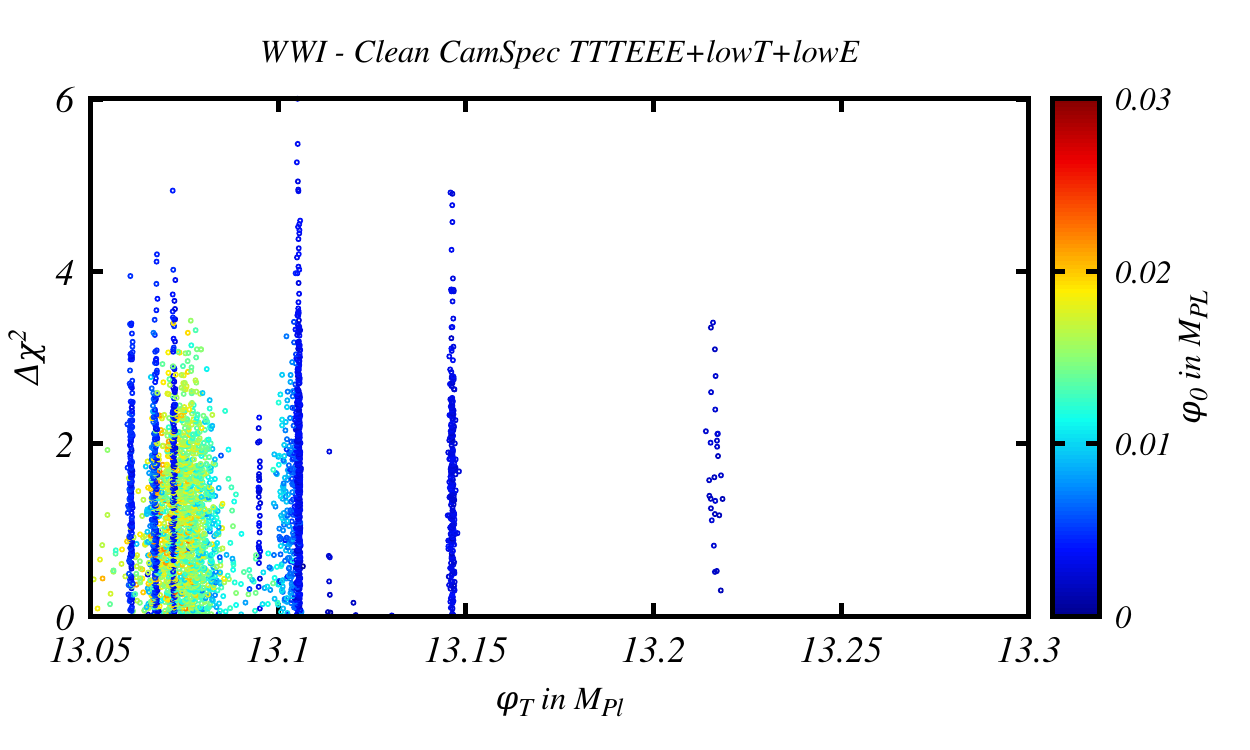}
\caption{\footnotesize\label{fig:WWI-Delchi2} The improvements in $\chi^2$ (from \texttt{PolyChord}) when WWI potential is used compared to its featureless baseline model are plotted against the position ($\phi_T$) and extent ($\phi_0$) of the discontinuity in the potential. Different locations and colors in the plot denotes different types of features from the WWI potential.}
\end{figure*}

\begin{figure*}[!htb]
\centering
\includegraphics[width=0.495\columnwidth]{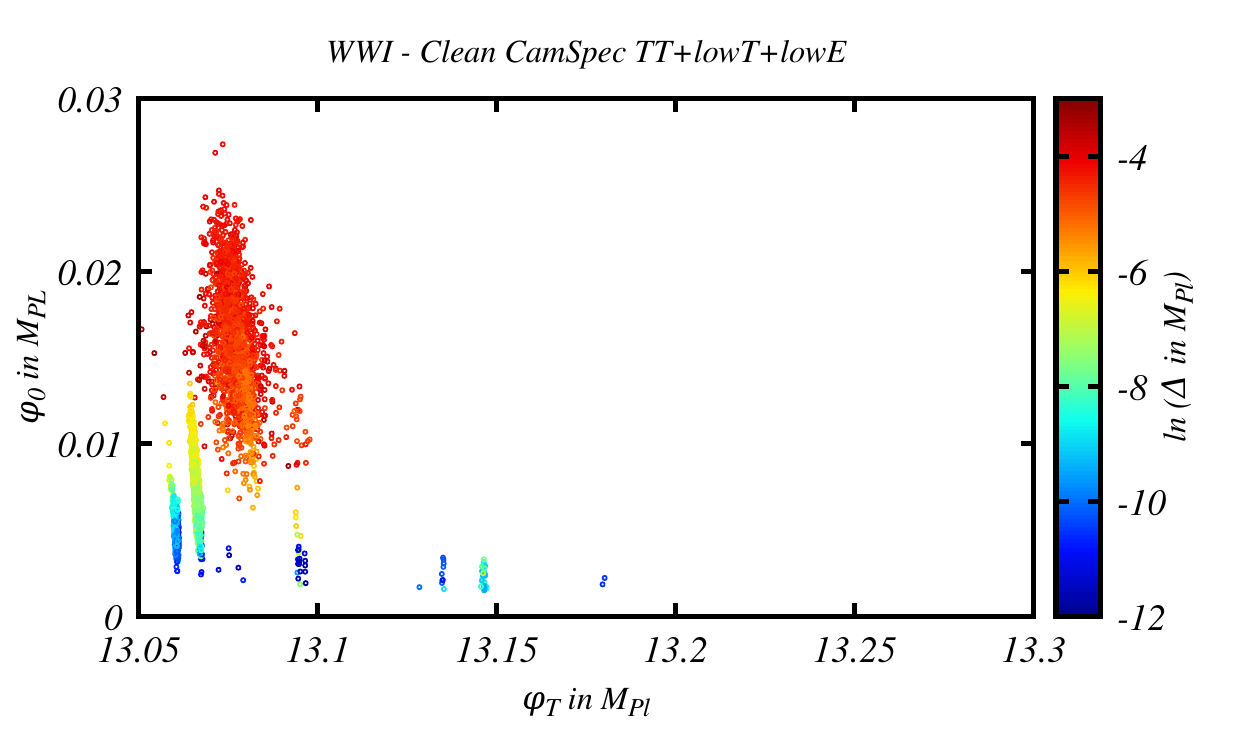}
\includegraphics[width=0.495\columnwidth]{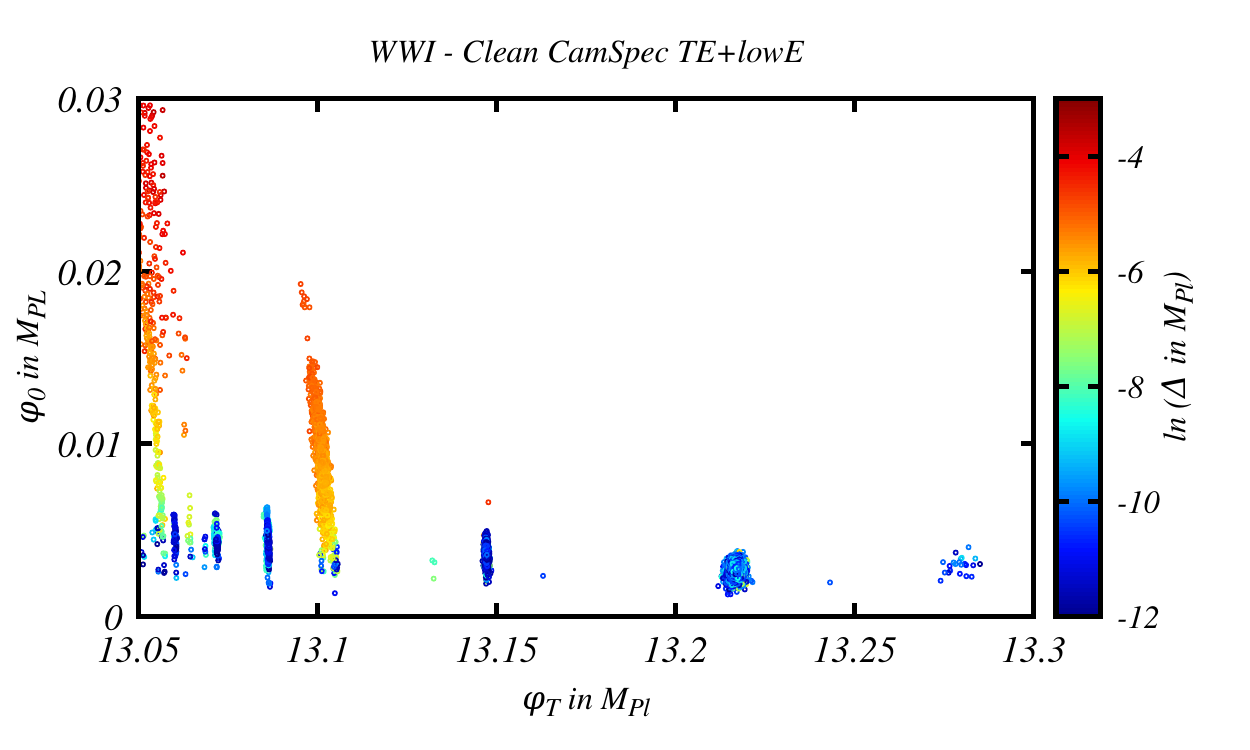}

\includegraphics[width=0.495\columnwidth]{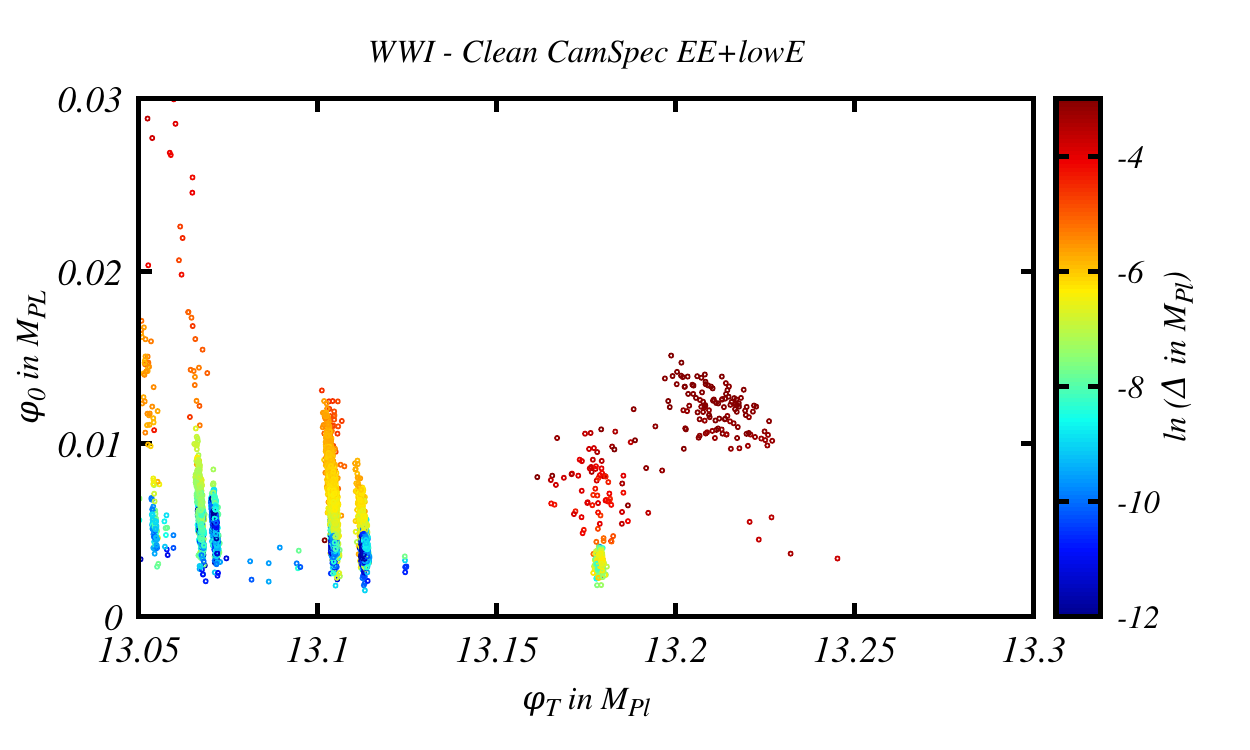}
\includegraphics[width=0.495\columnwidth]{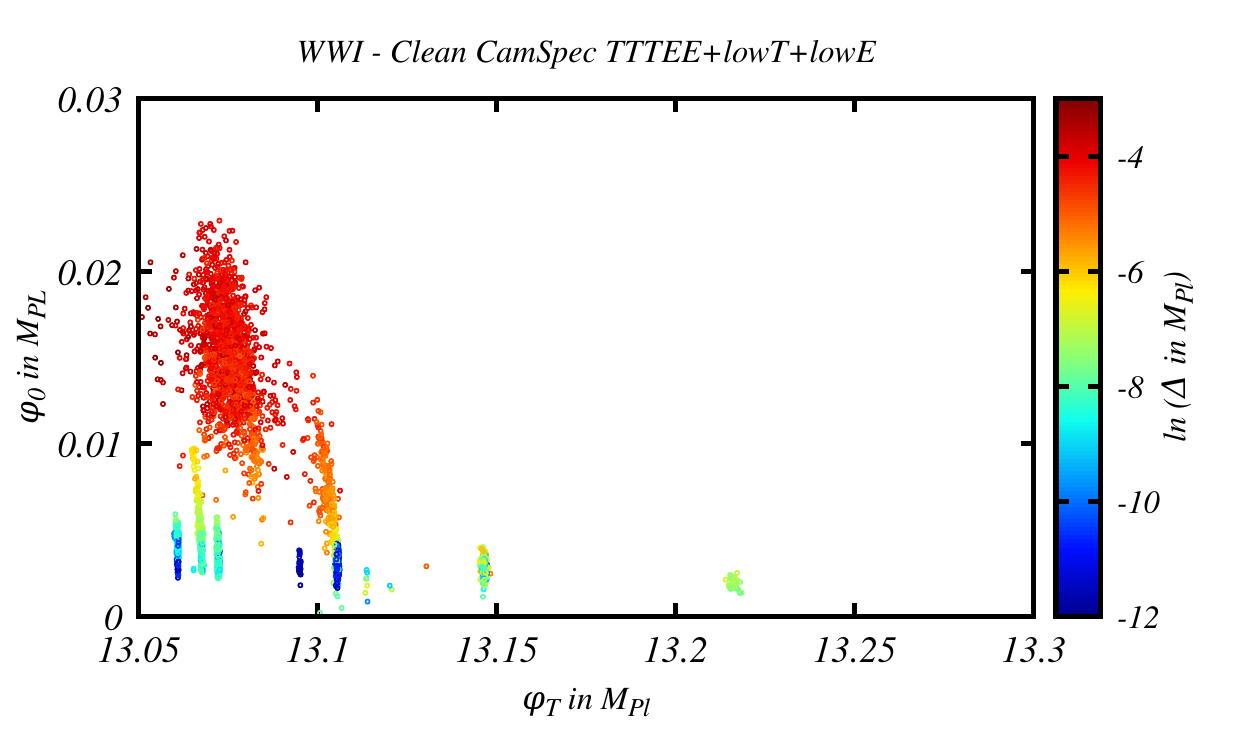}
\caption{\footnotesize\label{fig:WWI-map}Same samples are plotted as in~\autoref{fig:WWI-Delchi2}. Here all points represent samples that fit the data better than the baseline slow-roll potential and therefore they provide the map of the {\it favored} feature parameters. The plots for different datasets do not reveal any wide overlap within the parameter space. While in~\autoref{fig:WWI-Delchi2} for TT+lowT+lowE and TE+lowE show we find samples at $\phi_T\sim13.1\Mpl$ the same samples in this map reveal that these features do have different frequencies determined by $\Delta\phi$.}
\end{figure*}

While the marginal likelihood prefers the slow roll model over WWI model, we find several local best fit primordial features that fit the data better the baseline slow roll model. These features are important for discussion as future cosmic variance-limited polarization data or joint analysis with LSS can potentially detect one of these features with high significance if that represents the {\it true} model of the Universe. In order to identify the patches in the parameter space with better likelihood, we plot the samples in~\autoref{fig:WWI-Delchi2} where the improvement in $\chi^2$ from \texttt{PolyChord} is plotted as a function of $\phi_T$. The samples are colored according to the wiggle amplitude $\phi_0$. In samples from TT+lowT+lowE we notice warmer colored samples (higher amplitude wiggles) are located at large scales (low $\phi_T$). For TE+lowE and EE+lowE, the better likelihood samples are distributed nearly at all scales (as we have discussed in the marginalized posteriors before). TTTEEE+lowT+lowE has samples with better likelihood that are dominated by TT+lowT+lowE and TE+lowE. Importantly the combined data analysis has samples located sharply around $\phi_T=13.15\Mpl{}$, which is also visible in the samples from temperature and in the cross-correlation data. We can investigate the possibility of overlap of favored parameters spaces between different datasets in~\autoref{fig:WWI-map}. In this plot we show all samples that provide features with improvement to the data compared to strict slow roll model. The samples indicate the location and amplitude of the features that are colored by the sharpness/frequency of the features.
The extent of improvement in this plot cannot be visualized here but can be compared with~\autoref{fig:WWI-Delchi2}. We note that the red samples near $\phi_T=13.075\Mpl,\phi_0=[0.01-0.02]\Mpl$ in temperature panel (top left) also appear in the combined data panel (bottom right). We notice samples overlap in the top left and top right panel around $\phi_T=13.1\Mpl$ and as expected they appear in the combined dataset too. While there are several parameter spaces {\it favored} by different datasets, substantial overlap between parameter spaces in different panels can not be noticed. Evidently a feature supported by multiple datasets has the highest chance of representing the {\it true} model of the inflationary dynamics. 


\begin{table}[htbp]
    \centering
\setlength{\tabcolsep}{10pt} 
\renewcommand{\arraystretch}{1.5}
    \begin{tabular}{c c c c c}
\hline
 & & Clean CamSpec& \\
Param & TTTEEE& TT& TE& EE\\
& lowT+lowE & lowT+lowE & +lowE & +lowE\\

\hline
$\Omega_bh^2$ & $0.0222\pm 0.0001$& $0.0221\pm 0.0002$& $0.0223\pm 0.0002$& $0.0224_{-0.0010}^{+0.0008}$\\
$\Omega_ch^2$ & $0.120\pm 0.001$& $0.120\pm 0.001$& $0.119\pm 0.002$& $0.121\pm 0.003$\\
$100\theta_{MC}$ & $1.0410\pm 0.0003$& $1.0407\pm 0.0004$& $1.0415\pm 0.0004$& $1.0407\pm 0.0008$\\
$\tau$ & $0.0556_{-0.0085}^{+0.0071}$& $0.0537_{-0.0080}^{+0.0079}$& $0.0533_{-0.0086}^{+0.0088}$& $0.0548_{-0.0104}^{+0.0085}$\\
$\ln[10^{10} V_i]$& $0.190_{-0.017}^{+0.015}$& $0.185_{-0.017}^{+0.018}$& $0.179_{-0.022}^{+0.020}$& $0.199_{-0.028}^{+0.023}$\\
$\phi_{0}$ & $<0.144$& $<0.136$& $<0.270$& $<0.415$\\
$\phi_T$ & $-$& $-$& $-$& $-$\\

\hline
    \end{tabular}
    \caption{~\label{tab:WWIP}Results for WWIP potential. Here for high-$\ell$ likelihood we have used the \texttt{Clean CamSpec} unbinned angular power spectrum data.}
\end{table}

\begin{figure*}[!htb]
\includegraphics[width=\columnwidth]{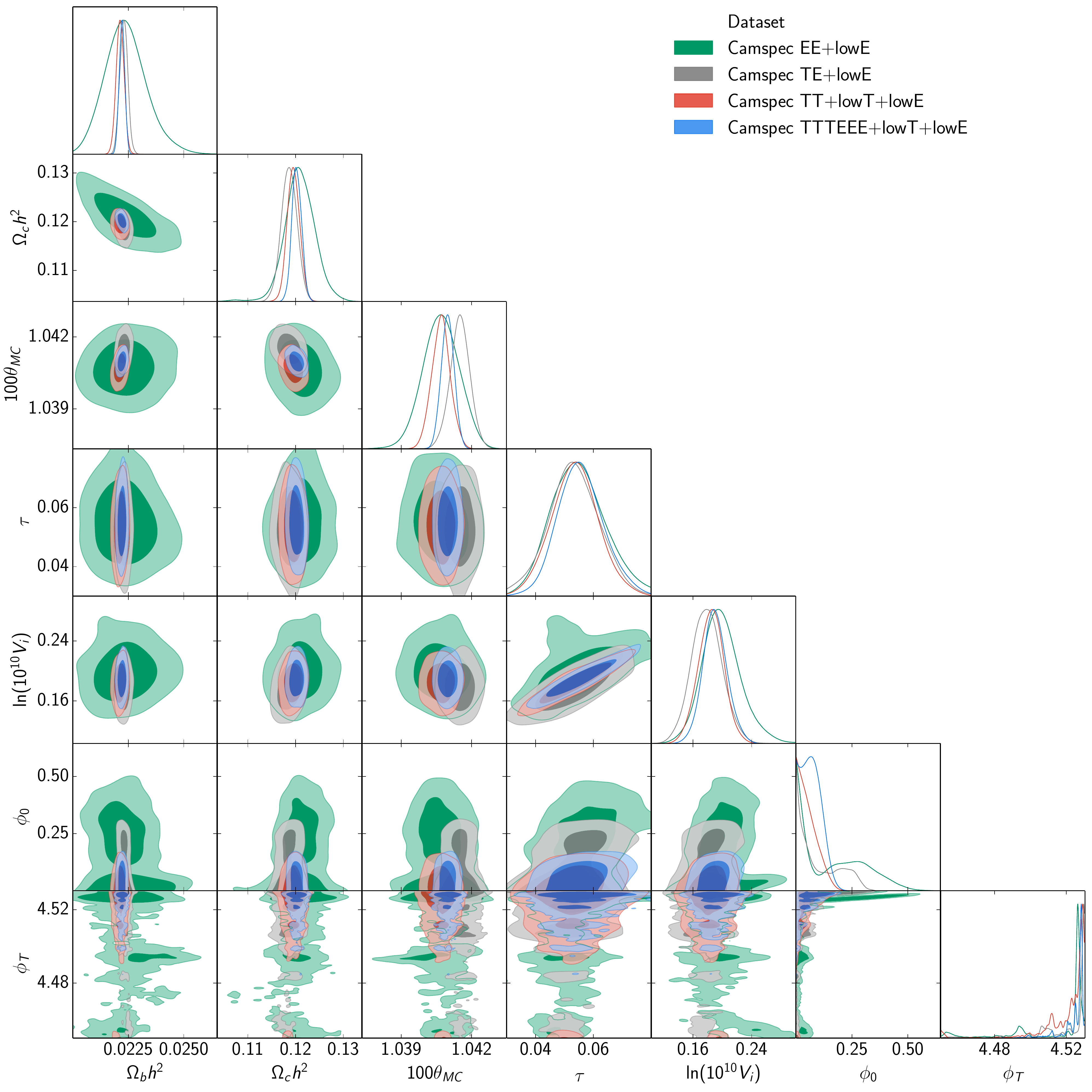}
\caption{\footnotesize\label{fig:WWIP}Constraints on WWIP potential for four different combination of datasets, namely \texttt{Clean CamSpec} TT+lowT+lowE, EE+lowE, TE+lowE and TTTEEE+lowT+lowE.}
\end{figure*}

\begin{figure*}[!htb]
\centering
\includegraphics[width=0.495\columnwidth]{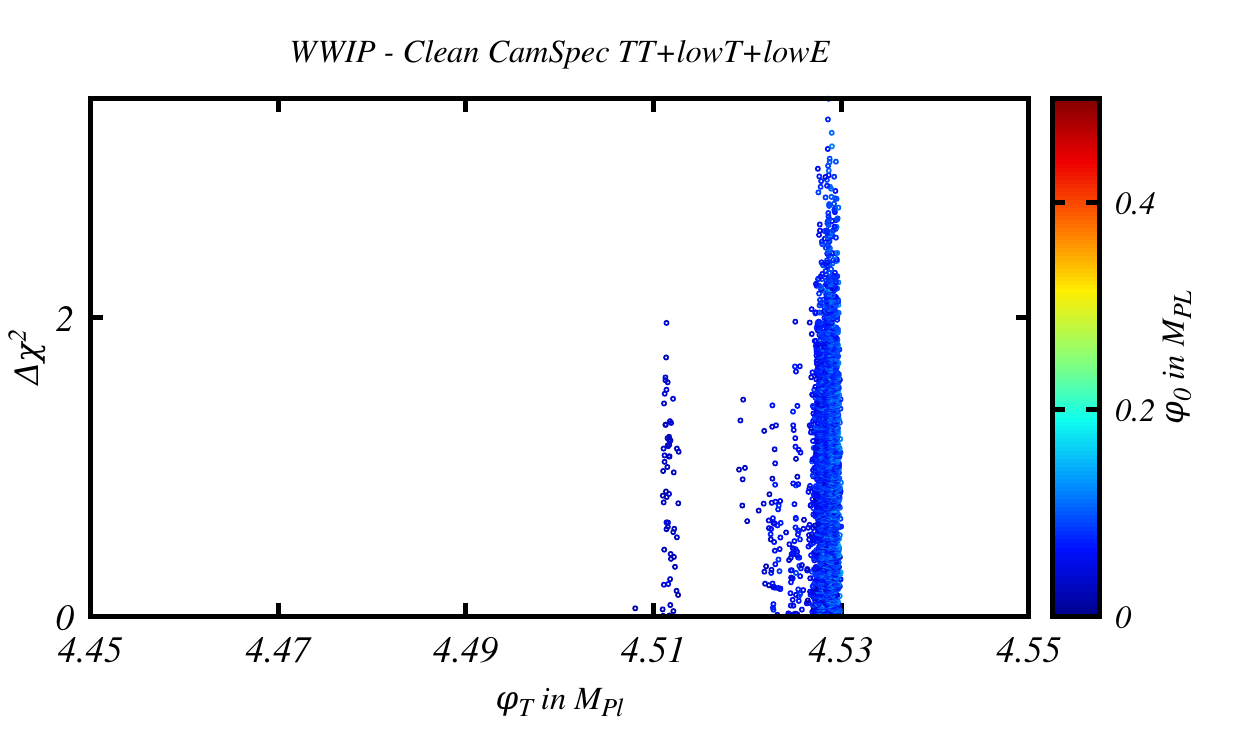}
\includegraphics[width=0.495\columnwidth]{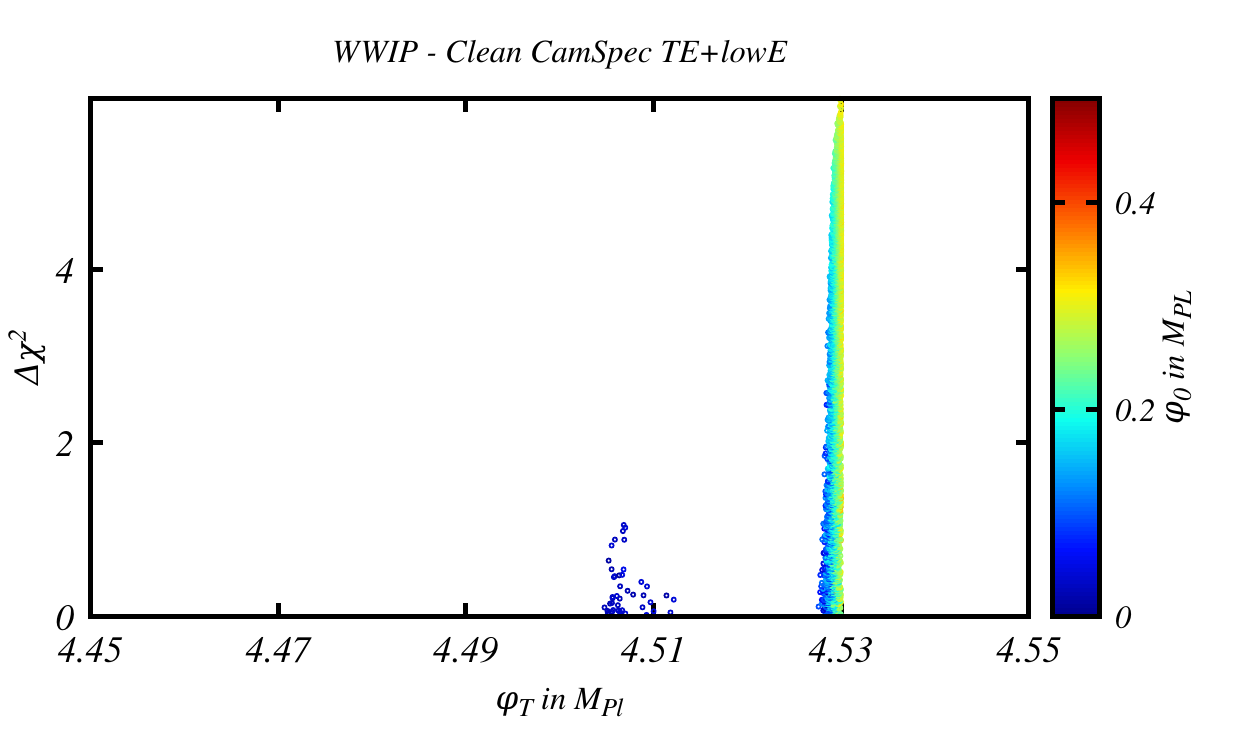}

\includegraphics[width=0.495\columnwidth]{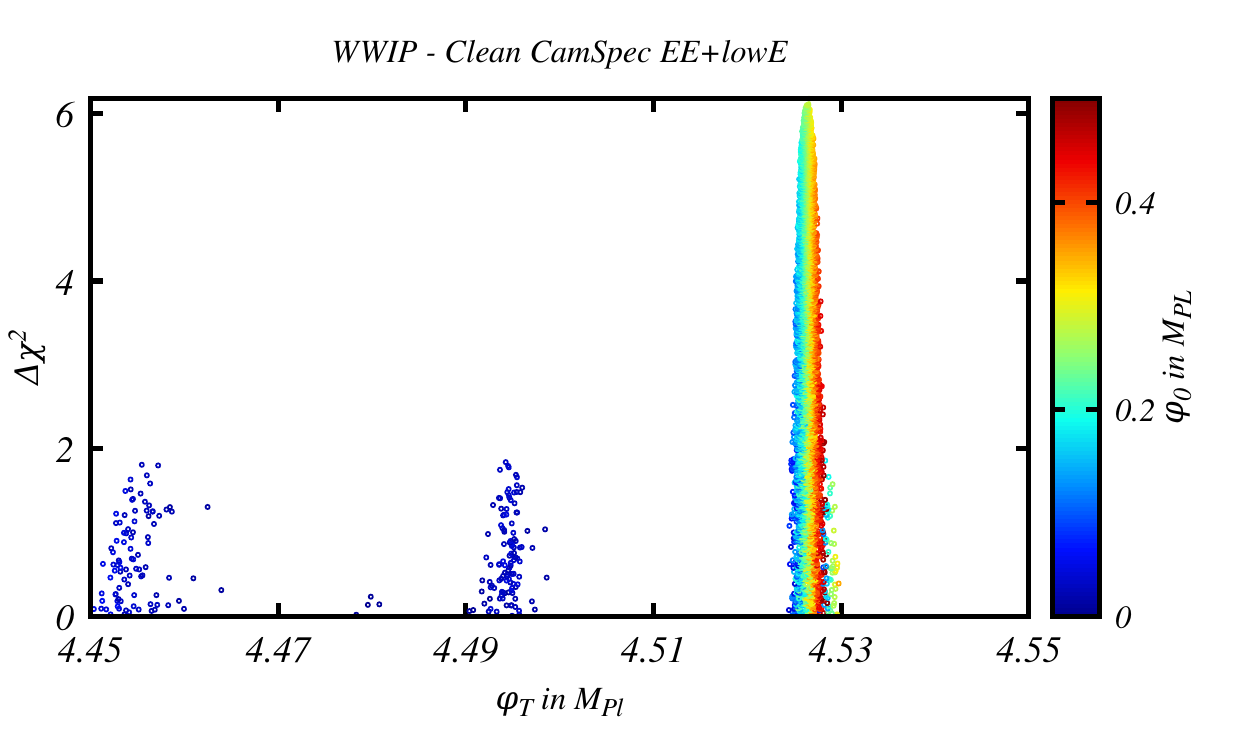}
\includegraphics[width=0.495\columnwidth]{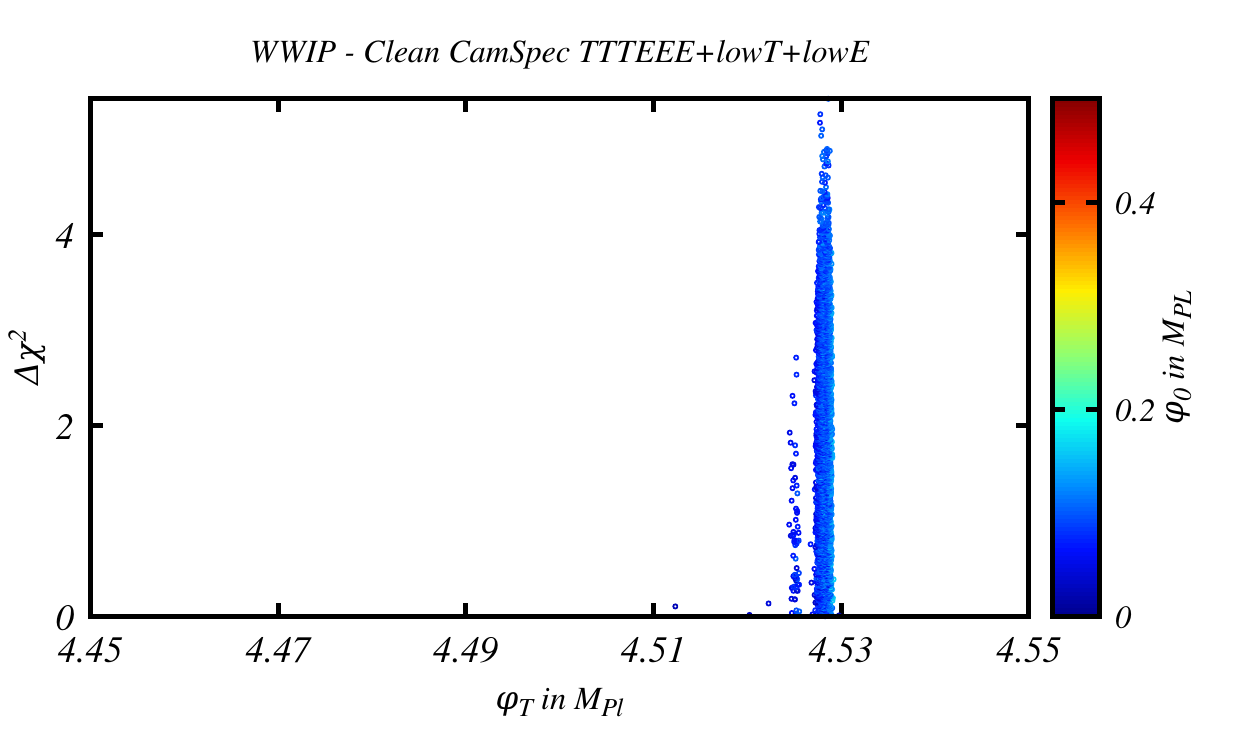}
\caption{\footnotesize\label{fig:WWIP-Delchi2}Same as in~\autoref{fig:WWI-Delchi2} but for WWIP potential. Here the colorplots explore the complete parameter space of WWIP potential with improvements in fit as the potential has two extra parameters {\it w.r.t.} the baseline model. We notice samples near $\phi_T=4.53\Mpl$ that are preferred by both TT+lowT+lowE and TE+lowE data. We find similar samples in the EE+lowE data but with higher $\phi_0$ values. Finally TTTEEE+lowT+lowE selects $\phi_T=4.53\Mpl$ samples as the region of maximum likelihood.} 
\end{figure*}

\begin{figure*}[!htb]
\centering
\includegraphics[width=\columnwidth]{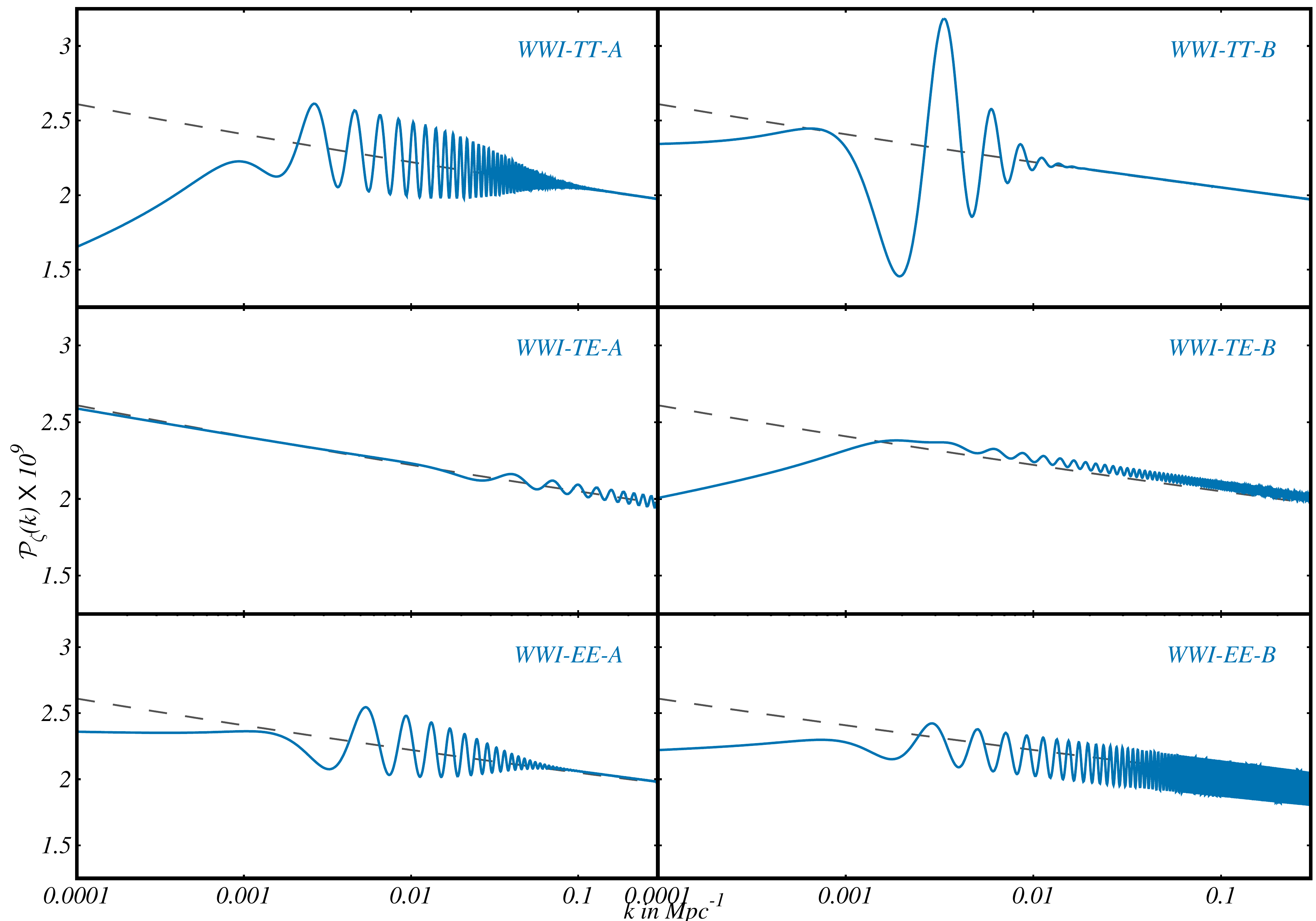}
\caption{\footnotesize\label{fig:PPS-WWI-TEX} Primordial power spectra that are local best fit to different combinations of datasets in WWI potential. We have plotted spectra that are visually different, though they provide similar improvement in fit to the data. We plot two such local best fits to TT+lowT+lowE (top panel), TE+lowE (middle panel) and EE+lowE (bottom panel). Note that though there are other local best fits to the data as shown in~\autoref{fig:WWI-map} here we show only two for each of the combinations.} 
\end{figure*}

\begin{figure*}[!htb]
\centering
\includegraphics[width=\columnwidth]{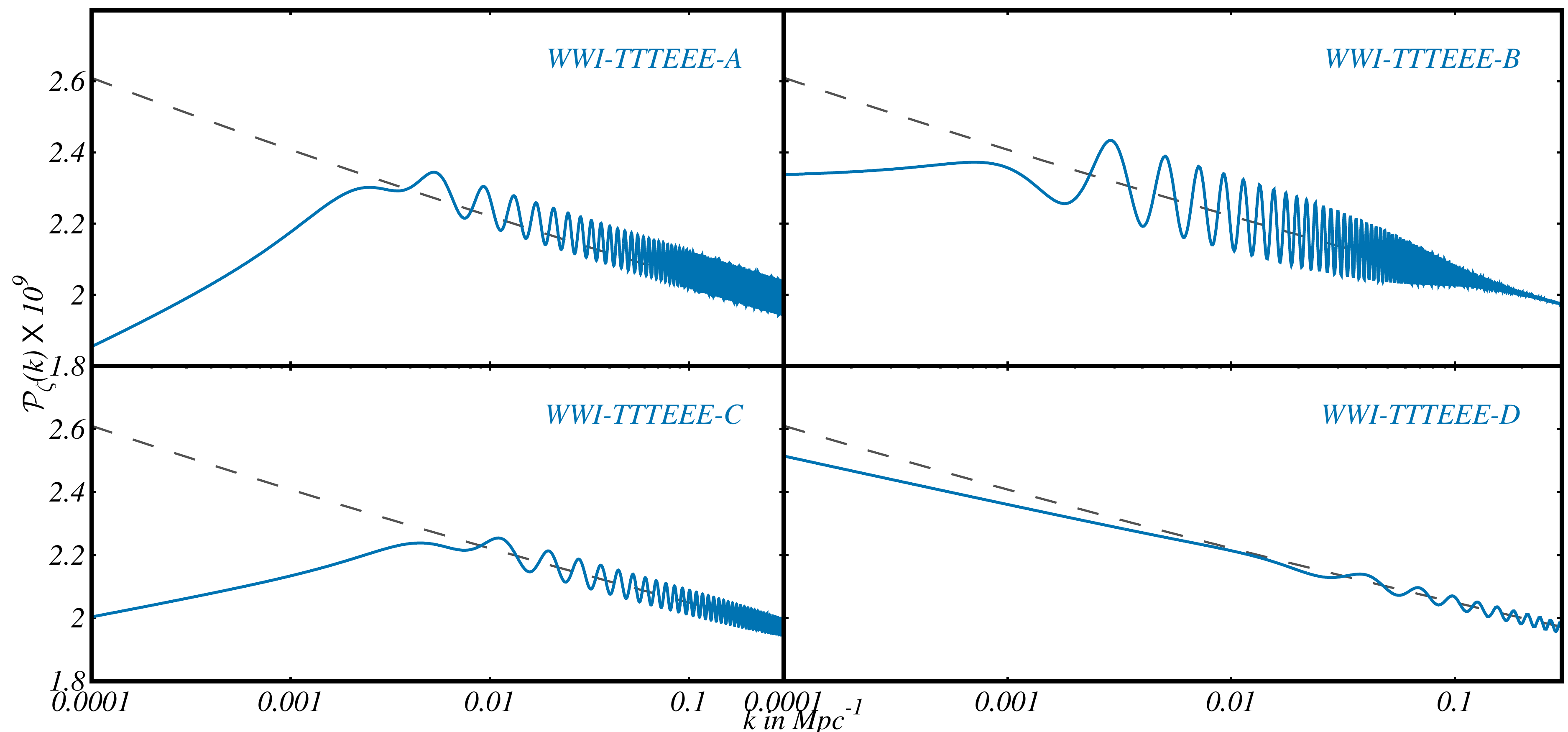}
\caption{\footnotesize\label{fig:PPS-WWI-TTTEEE} Primordial power spectra that are local best fit to the complete Clean \texttt{CamSpec} TTTEEE+lowT+lowE dataset in WWI potential. We have plotted four spectra that are visually different.} 
\end{figure*}

\begin{table}[]
\centering
\begin{tabular}{l|l|l|l|l}
\hline
Parameters         & {\scriptsize WWI-TTTEEE-A} & {\scriptsize WWI-TTTEEE-B} & {\scriptsize WWI-TTTEEE-C} & {\scriptsize WWI-TTTEEE-D} \\ \hline
$\ln[10^{10} V_i]$ & 2.272        & 2.265        & 2.263        & 2.262        \\ 
$\phi_{0}$         & 0.0033       & 0.0045       & 0.0030        & 0.0021       \\ 
$\gamma$           & 0.0210        & 0.0077       & 0.0129       & 0.0012       \\ 
$\phi_T$           & 13.10         & 13.07        & 13.15        & 13.22        \\ 
$\ln\Delta$        & -11.09       & -7.98        & -8.25        & -7.31        \\ \hline
$\Delta\chi^2$     & 7.7          & 6.9          & 9.25         & 6.3          \\ \hline
\end{tabular}\caption{~\label{tab:Bestfits} Parameter values for the four different bestfits from WWI model. The final row represents the improvement in $\chi^2$ compared to the baseline model.}
\end{table}
\begin{figure*}[!htb]
\centering
\advance\leftskip-1.5cm
\includegraphics[width=\columnwidth]{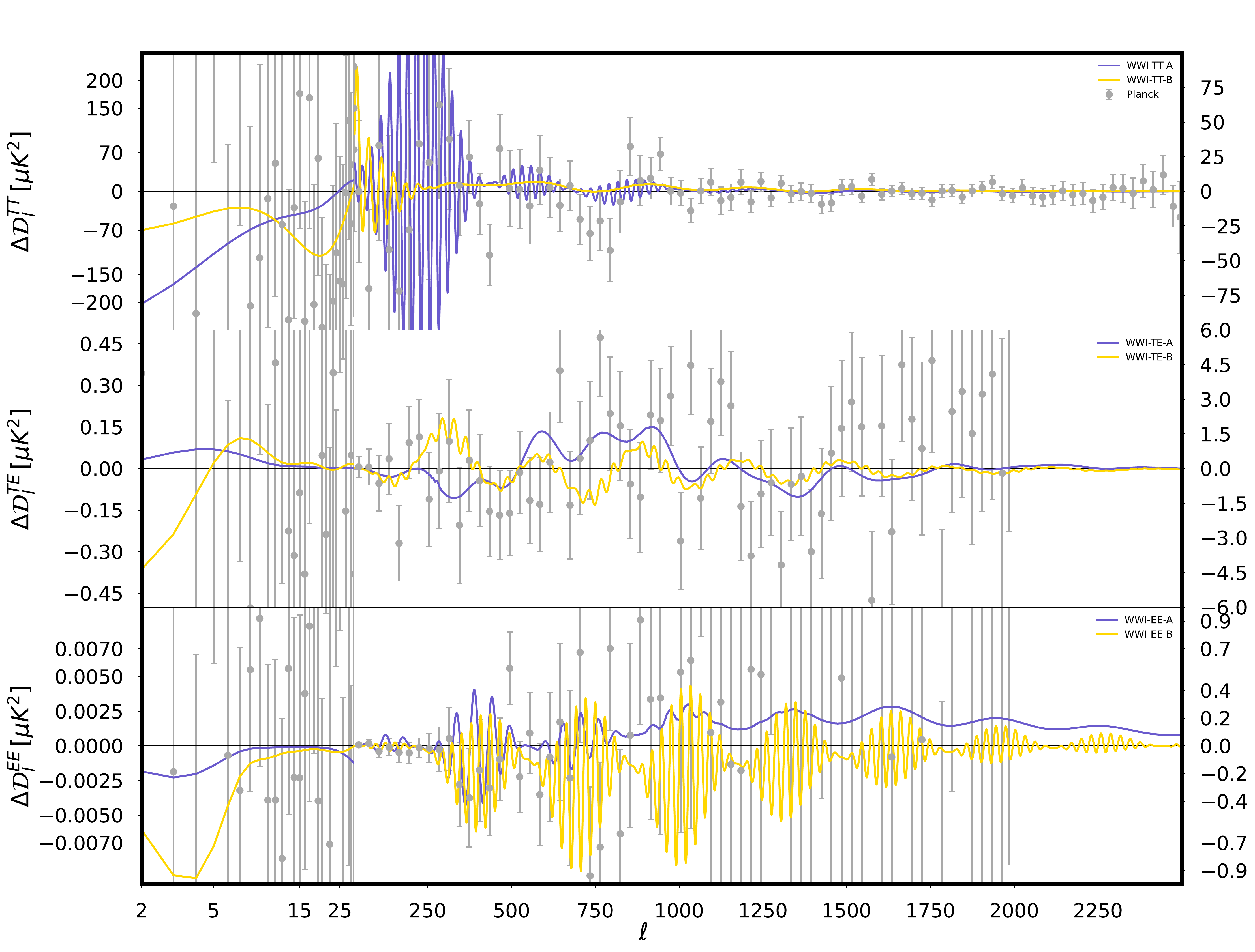}
\caption{\footnotesize\label{fig:resid-WWI-TEX} Best fit angular power spectra residual to the power law baseline best fit in WWI potential. These spectra correspond to the PPS plotted in~\autoref{fig:PPS-WWI-TEX}.} 
\end{figure*}

\begin{figure*}[!htb]
\centering
\advance\leftskip-1.5cm
\includegraphics[width=\columnwidth]{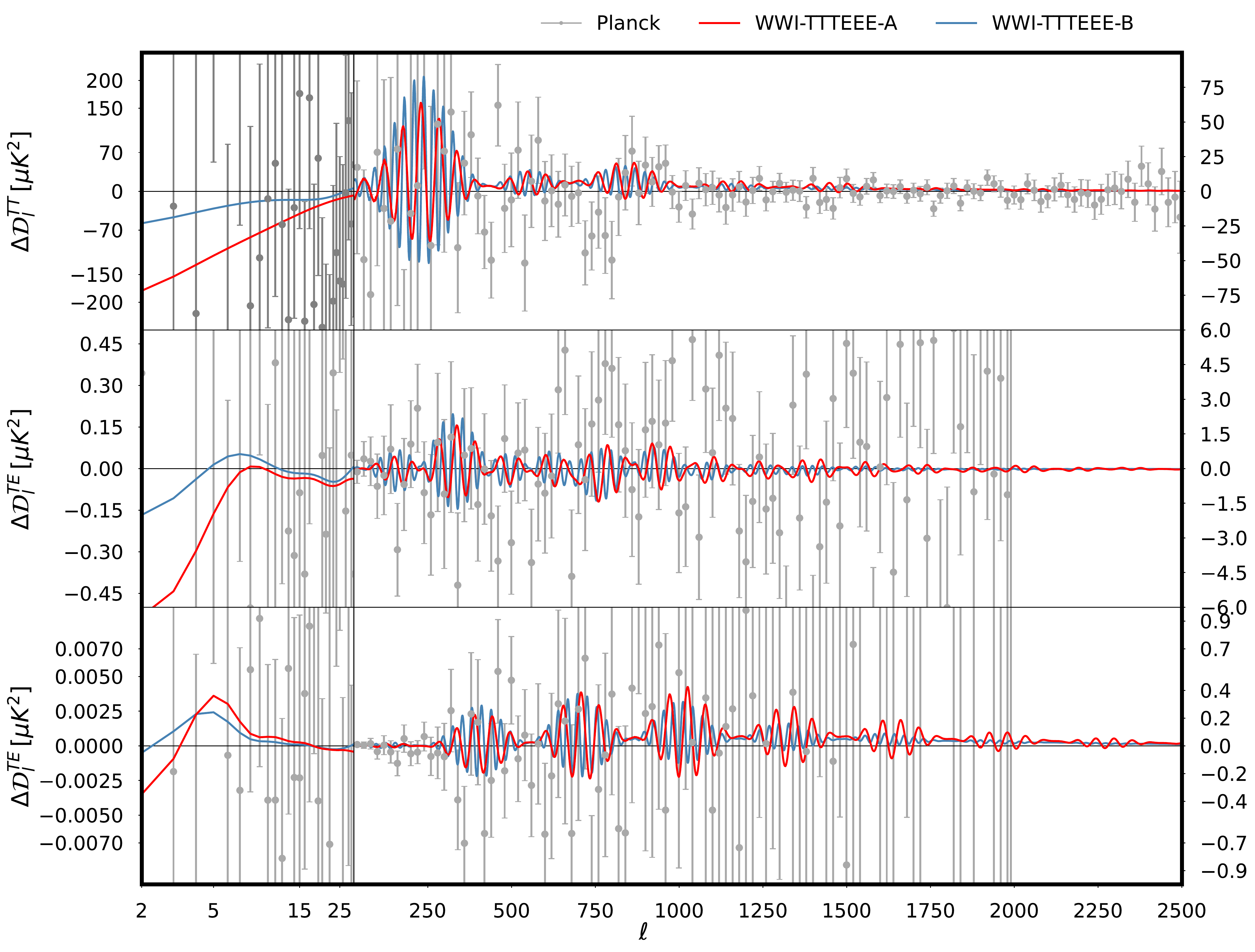}
\caption{\footnotesize\label{fig:resid-WWI-TTTEEE-AC} Best fit angular power spectra residual to the power law baseline best fit. Here we plot the residuals for two local best fits to TTTEEE + lowT+lowE data, namely WWI-TTTEEE-A and WWI-TTTEEE-B (PPS plotted in~\autoref{fig:PPS-WWI-TTTEEE}).} 
\end{figure*}

\begin{figure*}[!htb]
\centering
\advance\leftskip-1.5cm
\includegraphics[width=\columnwidth]{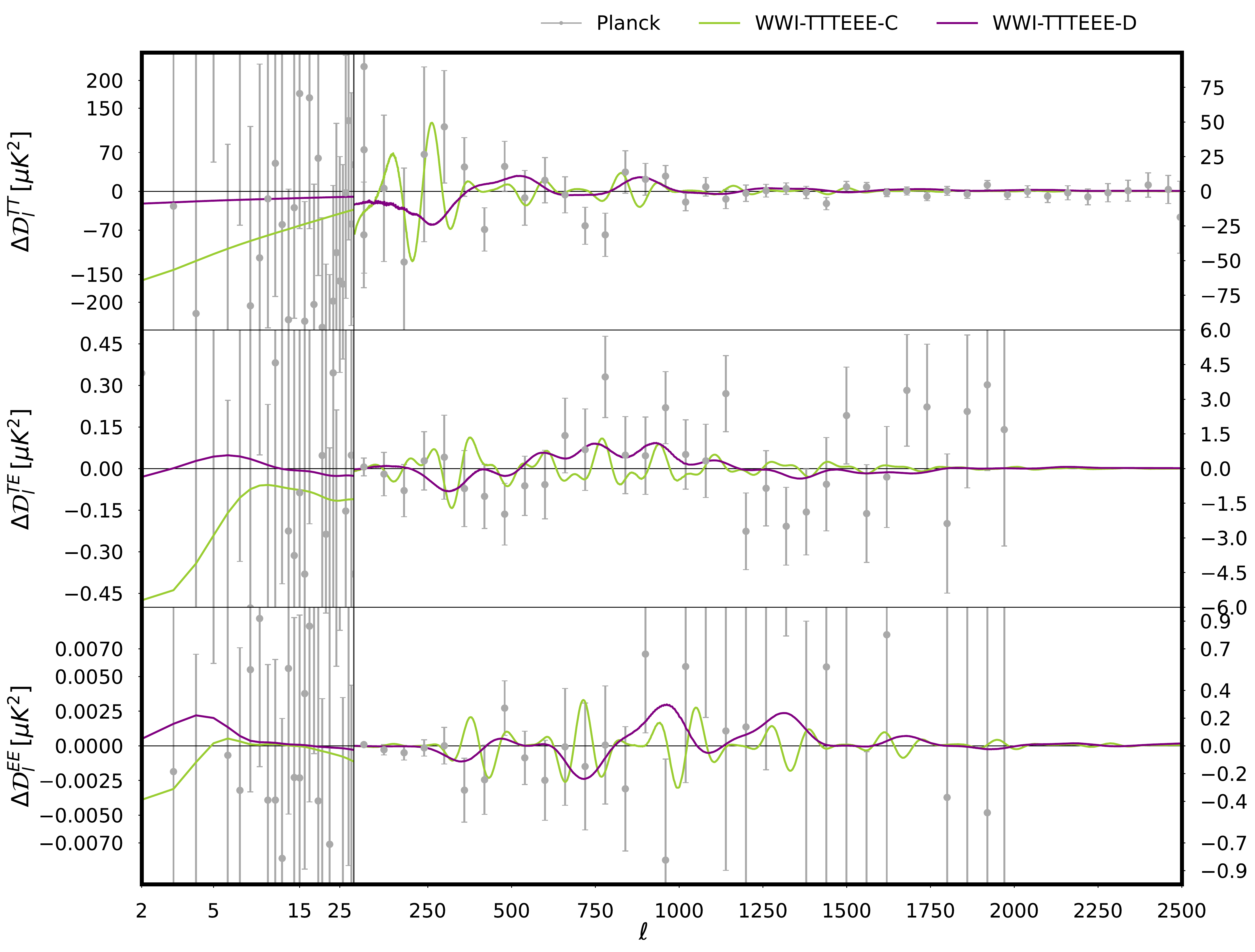}
\caption{\footnotesize\label{fig:resid-WWI-TTTEEE-BD} Angular power spectra residual to the power law baseline best fit. Here we plot the residuals for two local best fits to TTTEEE + lowT+lowE data, namely WWI-TTTEEE-C and WWI-TTTEEE-D (PPS plotted in~\autoref{fig:PPS-WWI-TTTEEE})..} 
\end{figure*}

\begin{figure*}[!htb]
\centering
\includegraphics[width=\columnwidth]{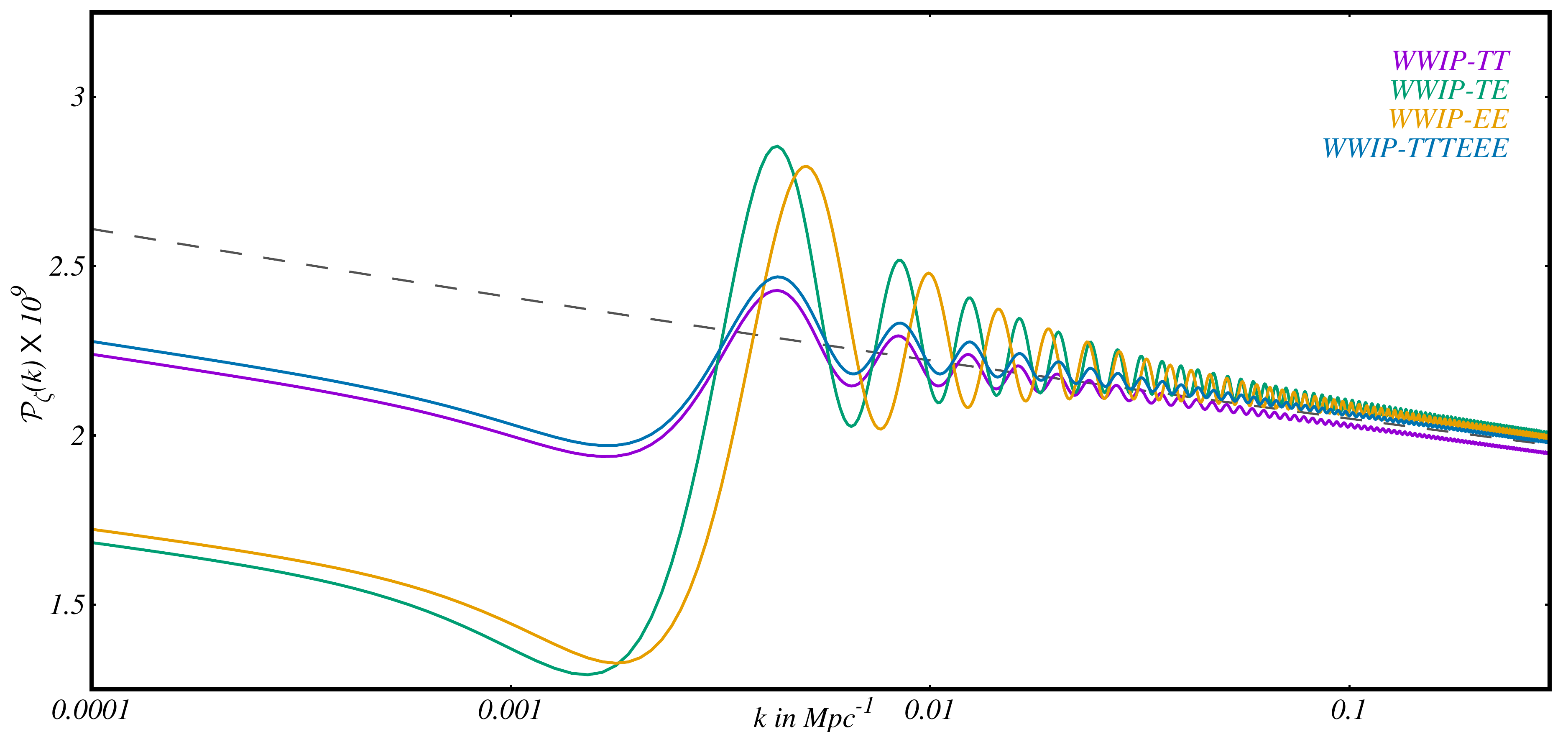}
\caption{\footnotesize\label{fig:PPS-WWIP} Primordial power spectra that are best fits to different dataset combinations in WWIP potential. Note that here we plot only one representative spectrum from the pool of best fits to each datasets since in~\autoref{fig:WWIP-Delchi2} we notice that most of the improvement is coming from large $\phi_T$ region.} 
\end{figure*}

\begin{figure*}[!htb]
\centering
\advance\leftskip-1.5cm
\includegraphics[width=\columnwidth]{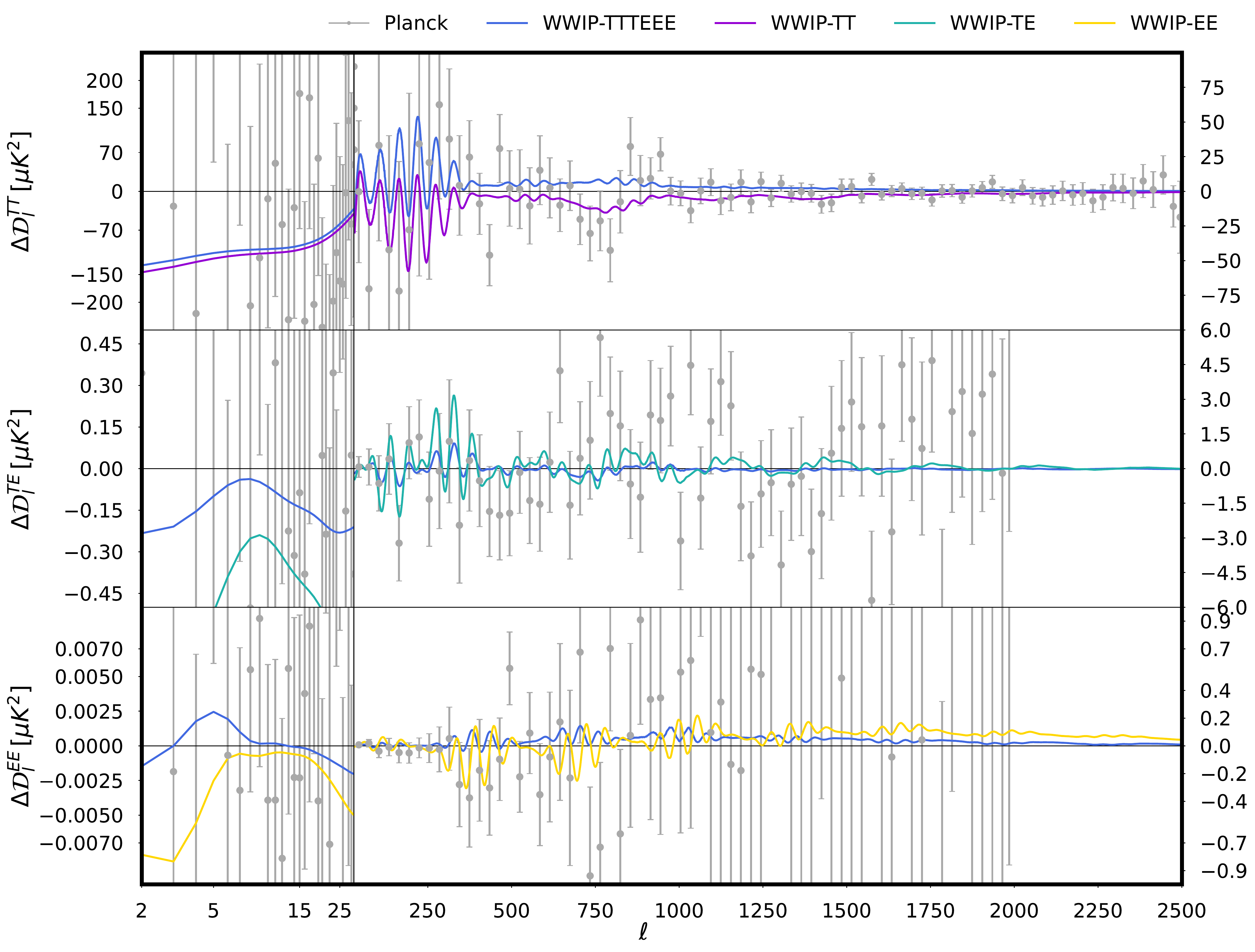}
\caption{\footnotesize\label{fig:resid-WWIP} Best fit angular power spectra residual to the power law baseline best fit. Here we plot the residuals for the best fits obtained from WWIP potential (PPS plotted in~\autoref{fig:PPS-WWIP}).}
\end{figure*}

The results for WWIP model are plotted in~\autoref{fig:WWIP} and the parameter mean values with uncertainties are provided in~\autoref{tab:WWIP}. Here we have two parameters that define the features, namely $\phi_0,\phi_T$. The marginalized posterior of $\phi_T$ peaks at large scales (higher $\phi_T$). EE+lowE and TE+lowE data show a peak around $\phi_0=0\Mpl$ and a plateau around $\phi_0=0.25\Mpl$. Combined dataset does not show the plateau as it is not supported by the temperature auto-correlation data. However, unlike TT+lowT+lowE where $\phi_0$ posterior sharply drops away from 0, the distribution obtained from the combined data is somewhat flat near the peak. However, we do not find any statistically significant support for the features. The evidence in~\autoref{tab:evidences} also shows weak preference for the baseline slow roll model. $\ln B$ in WWIP is comparatively higher {\it w.r.t.} WWI in TTTEEE+lowT+lowE. Due to two parameters less in the WWIP model, the marginal likelihood is less penalized compared to the WWI model in the combined dataset. The mean values on baryon density from temperature and polarization here too are in better agreement compared to slow roll potential as noticed in WWI potential.  

The samples with improvement in fit are plotted in~\autoref{fig:WWIP-Delchi2}. The improvements in $\chi^2$ are plotted against the position and amplitude parameter for the samples with features. As indicated by the marginalized posteriors, here we can notice most of the {\it favored} samples are located at large scales ($\phi_T\sim4.53\Mpl$). However compared to temperature data, the polarization data analysis prefers samples with sufficiently larger amplitude (warm colored samples). Larger uncertainties in the EE data allow large-amplitude wiggles in the primordial spectrum that are ruled out by temperature data. 

\subsection{Best fit power spectra}
Apart from parameter estimation and comparing marginal likelihoods, we also search for the best fits to each model and explore the possibility of the best fits to address temperature and polarization data individually and in combination. We use BOBYQA~\cite{BOBYQA} method for the minimization. The optimization is not very efficient in the large prior volume consisting of background, inflationary and nuisance parameters. Given the starting point of the best-fit search, the method usually converges to a local minima. To locate the local best fits we start the search for best fit from the parameter chains that represent the peaks in the $\Delta\chi^2$ plots presented in~\autoref{fig:WWI-Delchi2} and~\autoref{fig:WWIP-Delchi2}. Note that in similar way we had obtained four local minima to the Planck 2015 data using WWI in~\cite{Hazra:2016fkm}. In~\cite{Braglia:2021ckn} three methods of best fit search have been presented. However given the smaller number of nuisance parameters in \texttt{CamSpec} 12.5HMcln likelihood compared to \texttt{PlikHM} likelihood we use the most conservative method where we vary the nuisance parameters including the $\chi^2$ from priors in the total $\chi^2$ (method III in~\cite{Braglia:2021ckn}).

Since the WI model just provides suppression at large scales, we do not discuss the best fits here. In~\autoref{fig:PPS-WWI-TEX} we plot the local best fits to the individual datasets in WWI model. In the top, middle and bottom panels primordial power spectra that are local best fits are plotted from TT+lowT+lowE, TE+lowE and EE+lowE respectively. Corresponding to these best fit primordial spectra, the angular power spectra residual to TTTEEE+lowT+lowE featureless best fit angular power spectra are plotted in~\autoref{fig:resid-WWI-TEX}. In each case we plot two power spectra that can be distinguished in terms of frequency, amplitude or position of oscillations in the primordial spectra. WWI-TT-A has a suppression at large scales followed by oscillations extending to the small scales (till 0.1~$\mathrm{Mpc^{-1}}$). WWI-TT-B on the other hand has a pronounced dip around $k=0.002~\mathrm{Mpc^{-1}}$ that improved the fit around $\ell=22-30$. Given sharper transition in WWI-TT-A compared to WWI-TT-B we notice the former primordial spectrum having oscillations with higher frequency than the latter. This allows the spectrum to provide better fit to the data around $\ell=200-300$ and smaller scales. WWI-TE-A and WWI-TE-B both have lower amplitude oscillations in the primordial power spectra. In WWI-TE-A, the features start appearing at small scales. WWI-TE-B on the other hand has a suppression at large scales and persistent oscillations continuing to smallest scales probed by Planck. WWI-EE-A an WWI-EE-B represent two best fits to the EE data. The former contains primordial features located between $k=0.001-0.08~\mathrm{Mpc}^{-1}$ and the latter exhibits non-local oscillations from $k=0.001~\mathrm{Mpc}^{-1}$ that extend to the smallest scales in the Planck window. The middle and bottom panels of~\autoref{fig:resid-WWI-TEX} which plot the residual TE and EE angular power spectrum from TE and EE best fits also show similar localized and non-local oscillations in the residuals. While individual fits to the temperature and polarization data are not significant compared to the combined datasets, it is important to note that the best fits to individual datasets do not share similar features, since the amplitude, frequency and the scale of the oscillations are different.      
The WWI best fits to the combined data are plotted in~\autoref{fig:PPS-WWI-TTTEEE}. We present 4 best fits (WWI-TTTEEE-[A-D]) that are different in characteristics of the features. The inflationary parameters for these four best fits are tabulated in~\autoref{tab:Bestfits}. The residual angular power spectra are plotted in~\autoref{fig:resid-WWI-TTTEEE-AC} and in~\autoref{fig:resid-WWI-TTTEEE-BD}. WWI-TTTEEE-A contains large scale suppression followed by oscillations extending to the smallest Planck-CMB scales. WWI-TTTEEE-B represents an envelope of oscillations that decays and merges to power law spectrum at $k=0.1~\mathrm{Mpc}^{-1}$ similar to WWI-TT-A. These two best fits to the full data provide nearly 8 and 7 improvements in fit compared to the baseline. The high-frequency oscillations at intermediate to small scales are reflected in the residual plots where we notice envelope of oscillations at different scales. For the best fit WWI-TTTEEE-A we notice oscillations persist till the smallest scales probed by Planck while for WWI-TTTEEE-B we find the oscillations decay, as expected from the PPS. The data residual to the baseline model are plotted in fine bins that seem to be addressed by the features around $\ell\sim100-400$ in TT and $\ell\sim100-500$ in TE.  WWI-TTTEEE-C and  WWI-TTTEEE-D best-fit spectra have oscillations with lower amplitude and frequency. WWI-TTTEEE-C represents the global best fit (9.25 improvement in fit compared to power law) from the model. Data in the residual plot are binned with larger bin widths. Residuals show that the global best-fit addresses the outliers in TT and TE datasets. The oscillations in the WWI-TTTEEE-D best fit is even wider and can also be noticed in the residual plots. Angular spectra to this particular PPS seem to address TE and EE outliers at multipoles less than 1000. However, due to larger errors, the improvement in fit is only 6.3. Overall the improvement in fit drops 3 in $\chi^2$ compared to Planck 2015 data (Plik likelihood)~\cite{Hazra:2016fkm}.

WWIP best-fit PPS for different datasets are plotted in~\autoref{fig:PPS-WWIP}. Since this model generates one type of feature, we get single local best-fit (that is also the global best-fit) to each data. Note that the extent of suppression at the largest scales that fits the low-multipole anomalies determines the amplitude of the oscillation feature at intermediate and small scales. The residuals are plotted in~\autoref{fig:resid-WWIP}. In the top panel we plot TT residuals for WWIP-TT and WWIP-TTTEEE best fits. In the middle panel we plot TE residuals for WWIP-TE and WWIP-TTTEEE and the panel at the bottom contains EE residuals for WWIP-EE and WWIP-TTTEEE best fits. We keep the same color convention used in~\autoref{fig:PPS-WWIP}. As expected, best fit to individual data fits the outliers of the corresponding data. For example WWIP-TT fits the oscillatory residuals around $\ell\sim100-400$ and the dip around $\ell\sim750$ and $1500$. WWIP-TE allows significantly larger amplitude oscillations in the PPS compared to WWIP-TT (with similar peak positions) and the residual plots captures its attempts to fit oscillatory TT residuals within $\ell=50-700$. Oscillations in WWIP-EE best fit PPS are phase shifted (the first peak being shifted to smaller scales) compared to WWIP-TE with the oscillations. Since they are phase-shifted, the same power spectrum can not address outliers in the TE and EE spectra. WWIP-EE best fit PPS addresses the residual data around $\ell\sim400$ and between $600-800$. WWIP-TTTEEE best fit PPS is very similar to WWIP-TT with certain shift in the overall amplitude. Unlike WWIP-TT, here the residual plot does not seem to address the TT residual data around $\ell\sim750$ and $1500$. It fits the TT oscillatory residuals around $\ell\sim100-400$ and TE residuals with lower amplitude around $\ell=50-700$. WWIP-TTTEEE provides an improvement in fit by 6.5 in $\chi^2$ to the full data compared to featureless model. The best fit parameters are $\left[\ln(10^{10}V_i,),\phi_0,\phi_{\rm T}\right]=\left[0.1973,0.091,4.528\right]$. Compared to~\cite{Hazra:2016fkm} we notice an overall 6 decrease in the $\Delta\chi^2$. 

There are two reasons for this decrease in both WWI and WWIP models. Firstly, with the WI potential we find that while the large scale scalar suppression is supported by the TT data, EE data from HFI prefer no suppression. Therefore the extent of fit to the data decreases. Secondly, we are using the \texttt{Clean CamSpec} likelihood instead of \texttt{Plik}. It has been shown in~\cite{Braglia:2021ckn} that the improvement in fit to the data is less in \texttt{Clean CamSpec} compared to \texttt{Plik2018}. Overall, this decrease definitely points to the need to explore features that are in agreement with both temperature and polarization data.
\subsection{Non-Gaussianity: bispectrum}
We compute the bispectra for the best fit potential parameters. We use BINGO~\cite{Hazra:2012BINGO,Sreenath:2014BINGO} to numerically evaluate the $f_{\rm NL}$ parameter as functions of wave-numbers ($k$). 
Note that the bispectrum amplitude $f_{\rm NL}(\bk_1,\bk_2,\bk_3)$ is defined as,
\begin{equation}
f_{\rm NL}(\bk_1,\bk_2,\bk_3)=-\frac{10}{3}\frac{\sqrt{2\pi}k_1^3k_2^3k_3^3{\cal B}_\zeta(\bk_1,\bk_2,\bk_3)}{k_1^3{\cal P}_\zeta(k_2){\cal P}_\zeta(k_3)+k_2^3{\cal P}_\zeta(k_3){\cal P}_\zeta(k_1)+k_3^3{\cal P}_\zeta(k_1){\cal P}_\zeta(k_2)},    
\end{equation}
with ${\cal B}_\zeta(\bk_1,\bk_2,\bk_3)$ denoting the bispectra and ${\cal P}_\zeta(k)$ denoting the power spectrum.

\begin{figure*}
\centering
\includegraphics[width=0.44\columnwidth]{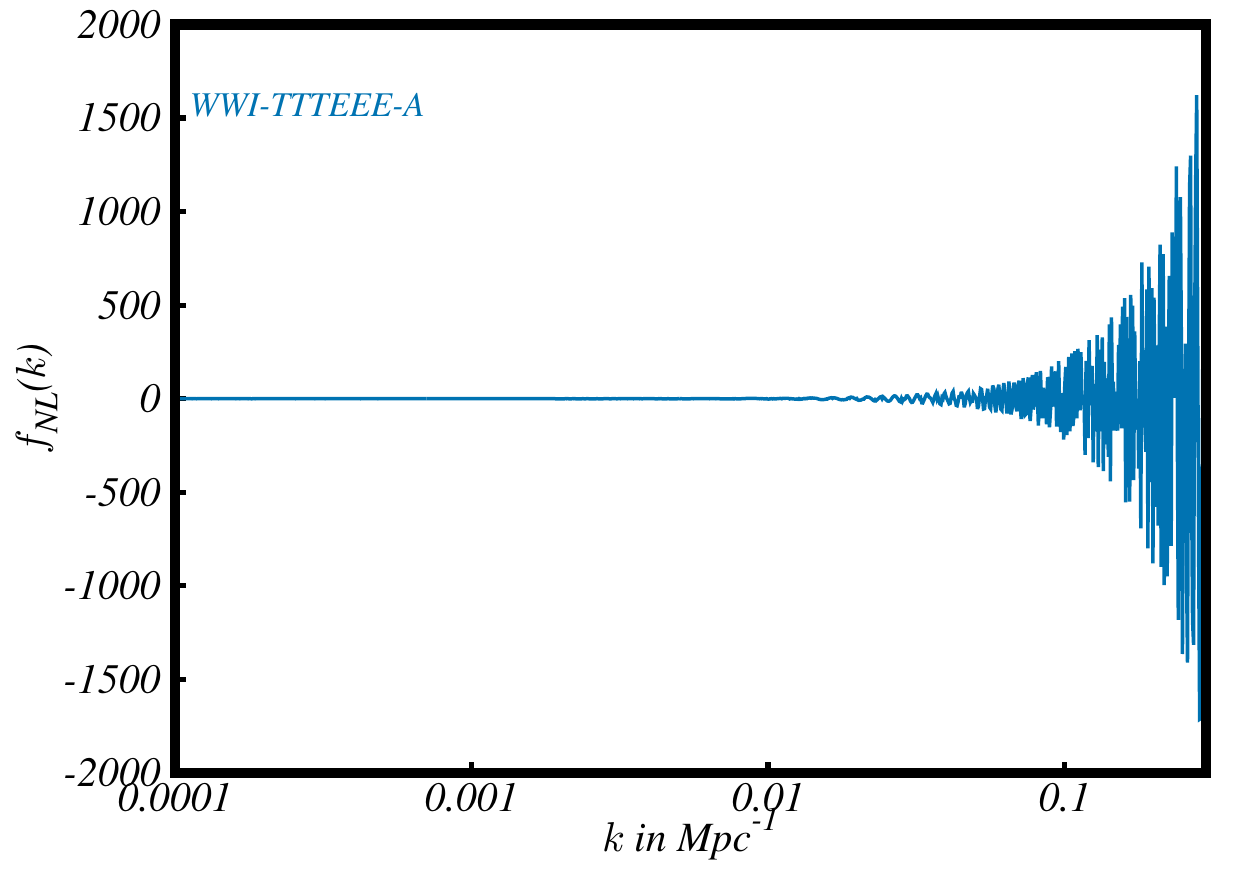}
\includegraphics[width=0.55\columnwidth]{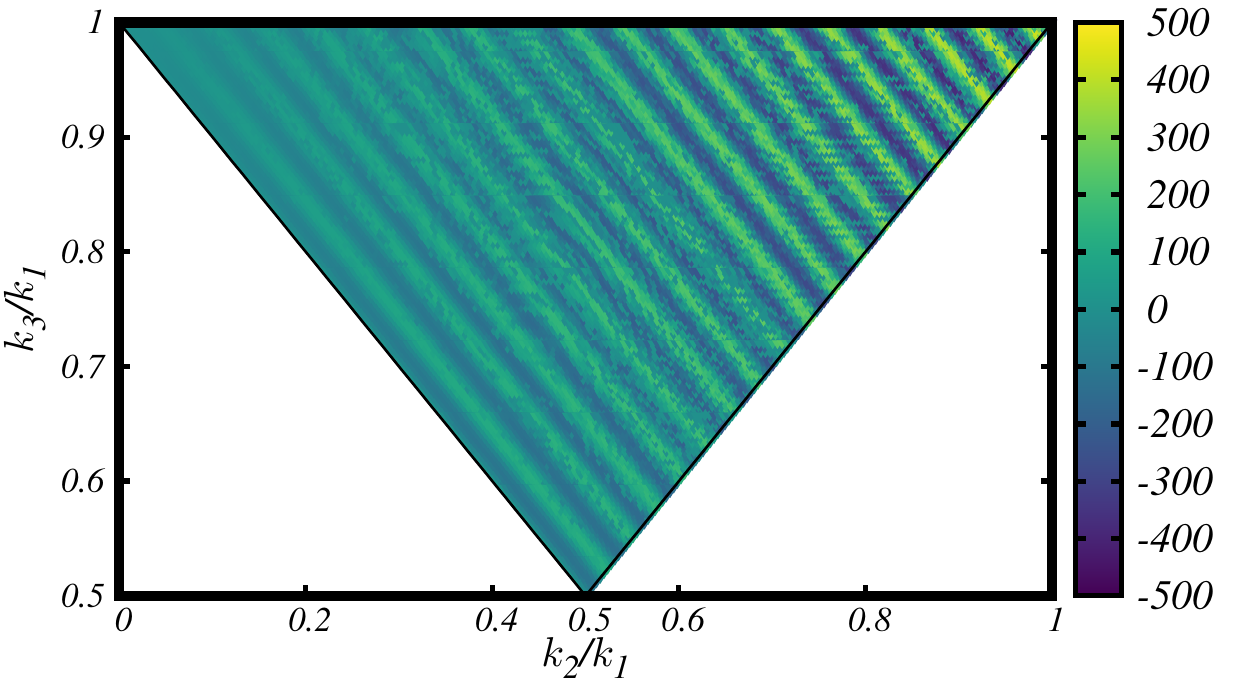}

\includegraphics[width=0.44\columnwidth]{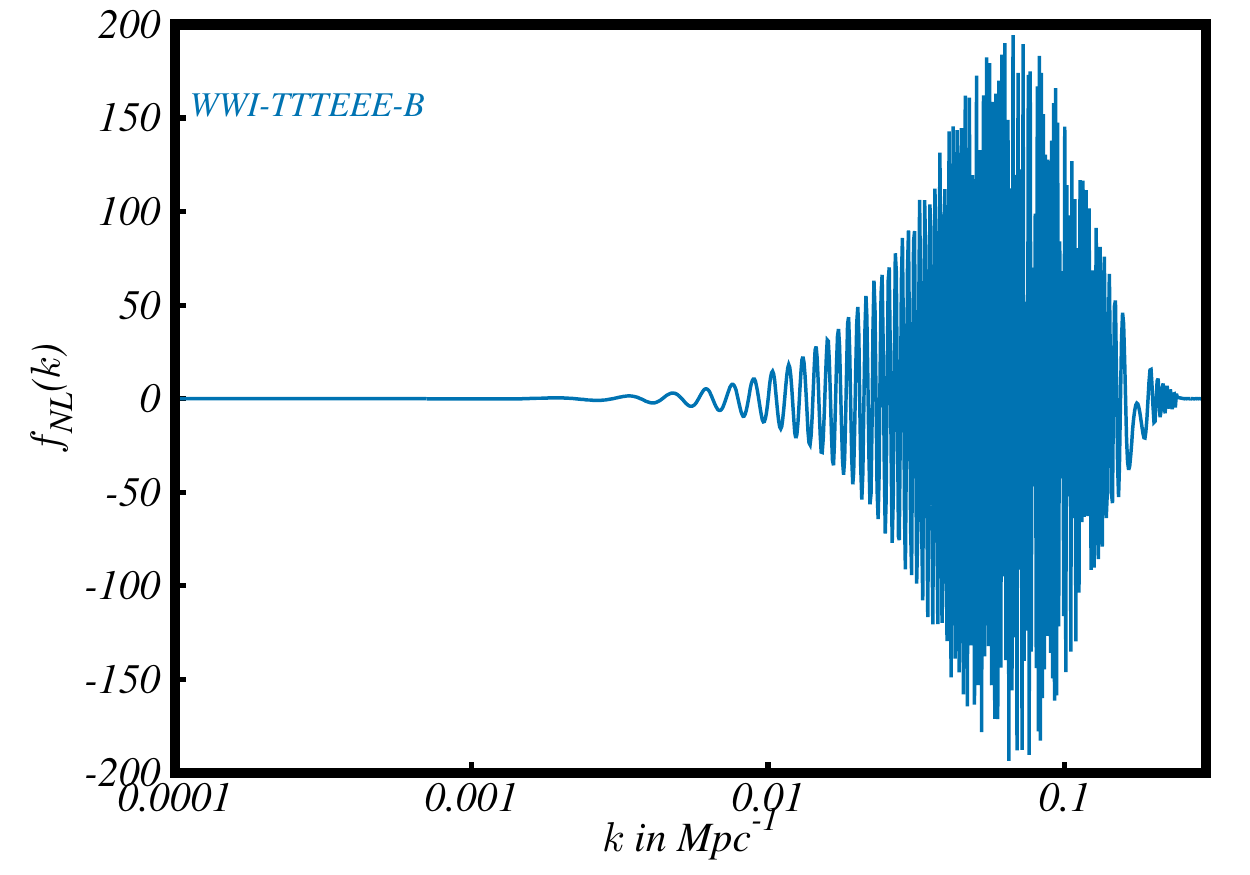}
\includegraphics[width=0.55\columnwidth]{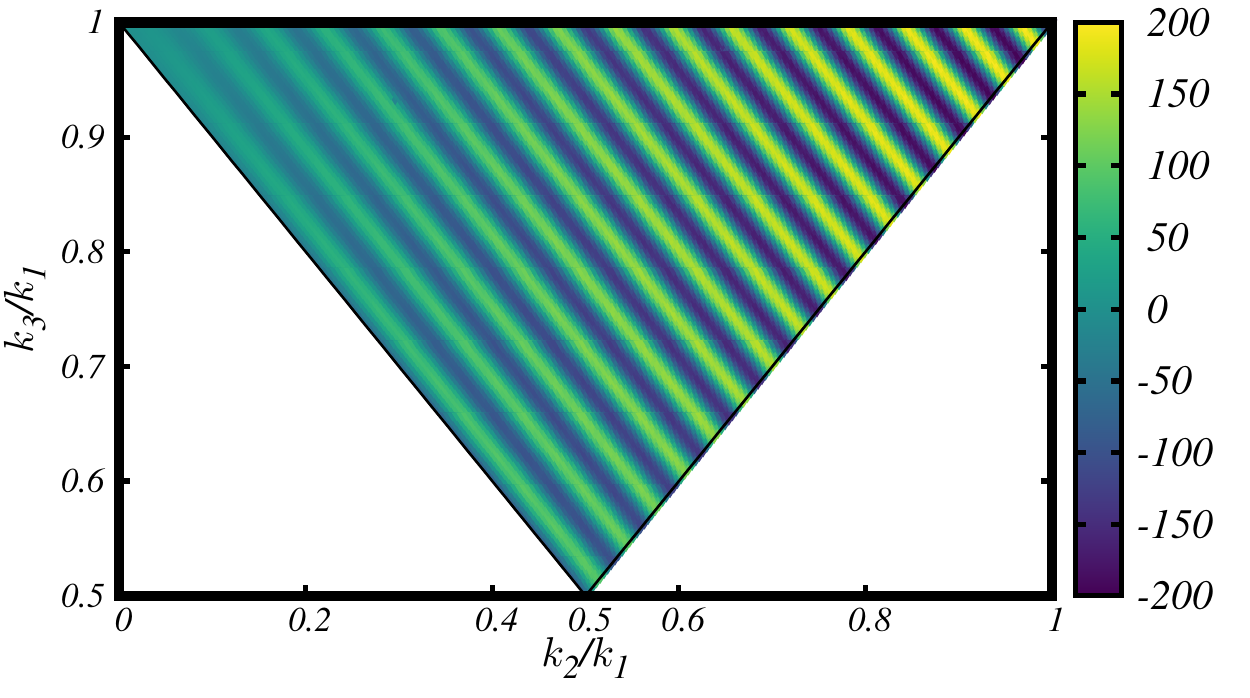}

\includegraphics[width=0.44\columnwidth]{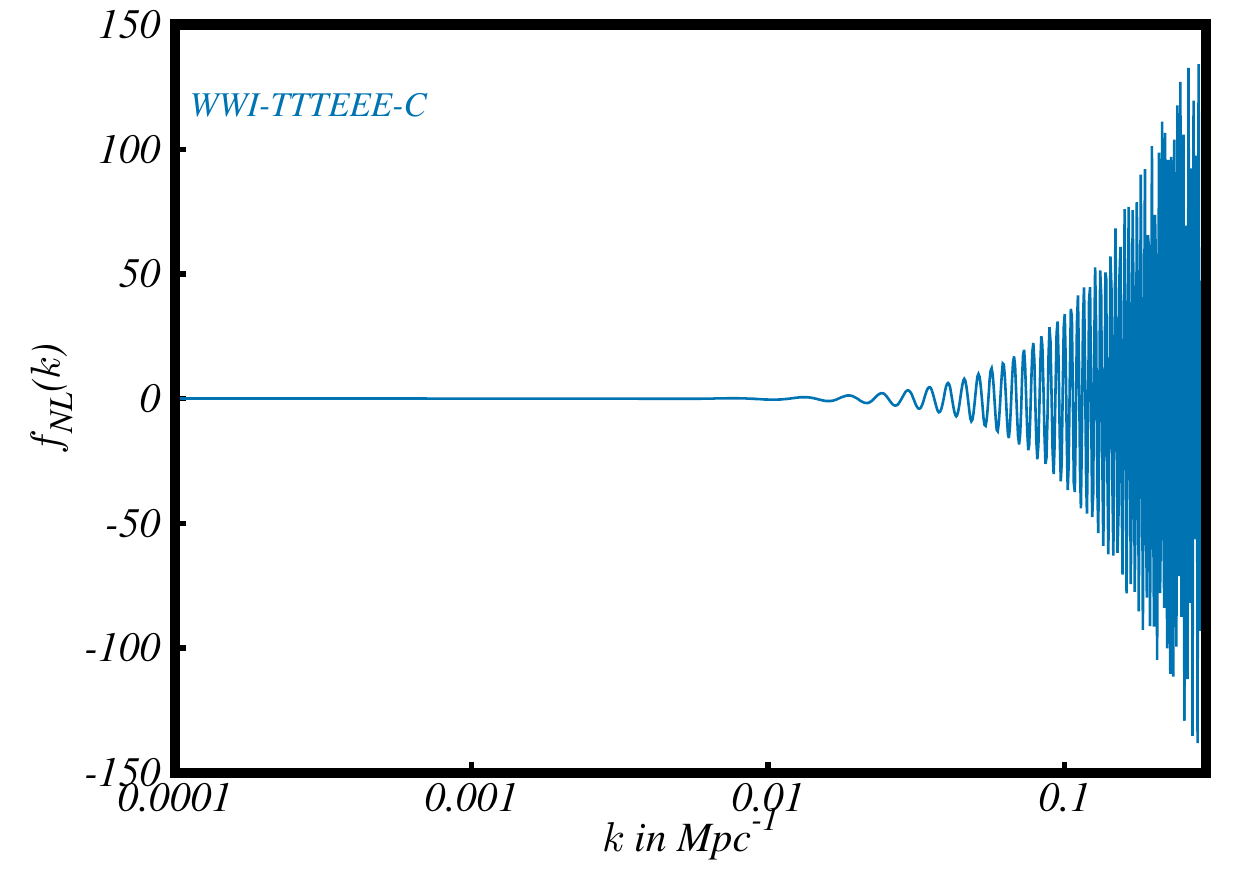}
\includegraphics[width=0.55\columnwidth]{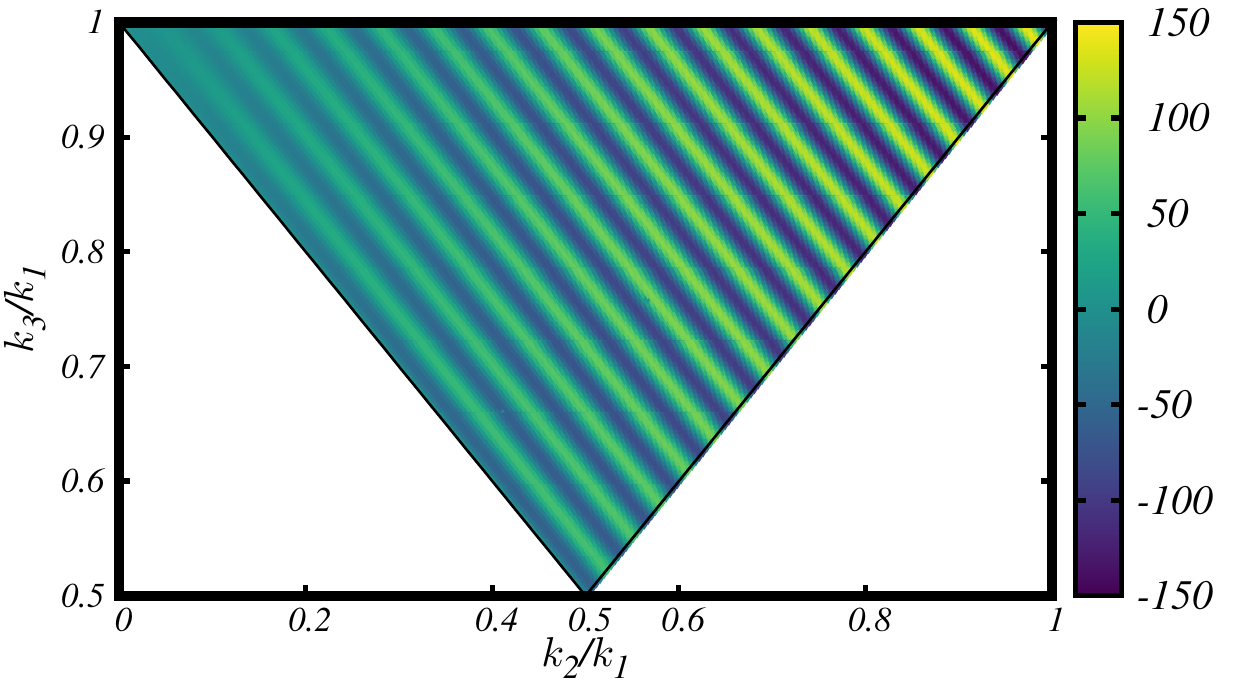}

\includegraphics[width=0.44\columnwidth]{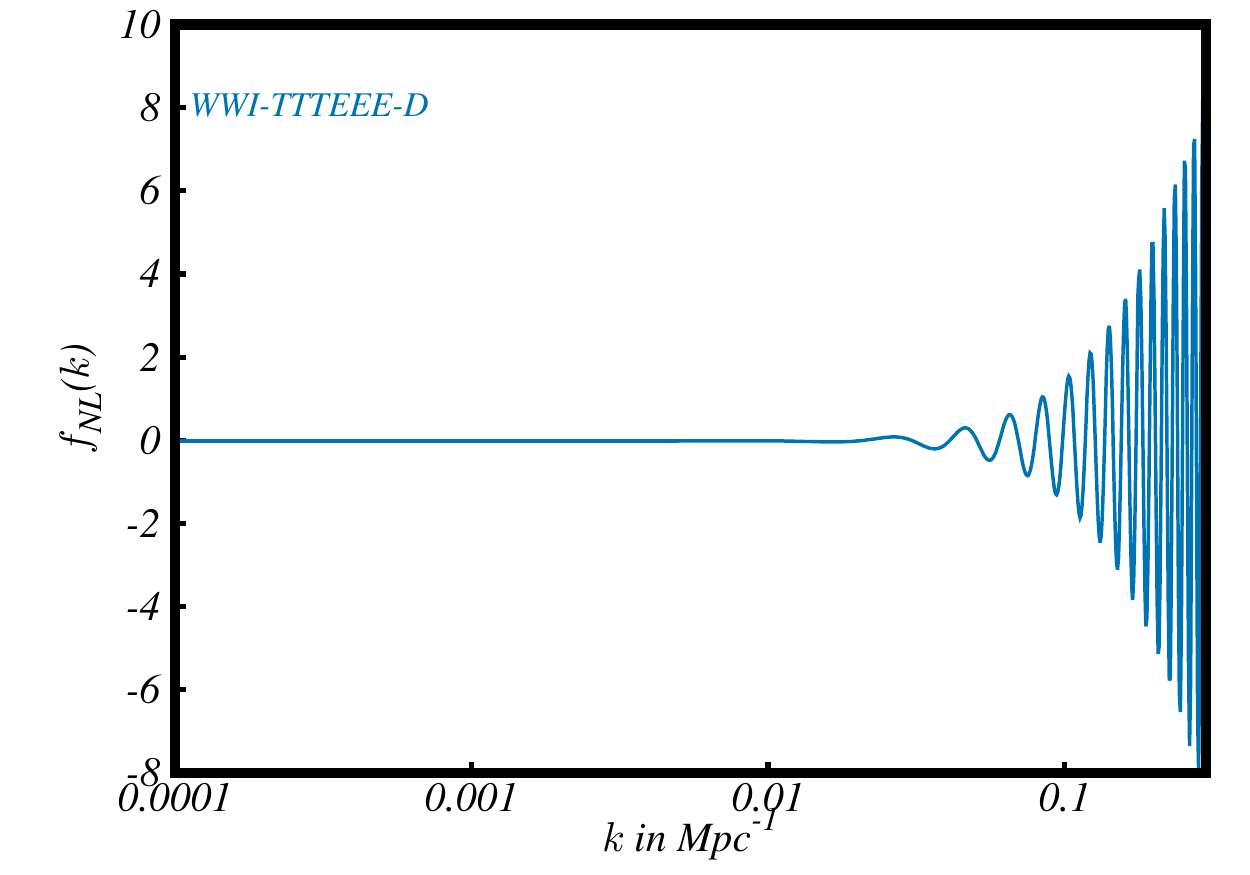}
\includegraphics[width=0.55\columnwidth]{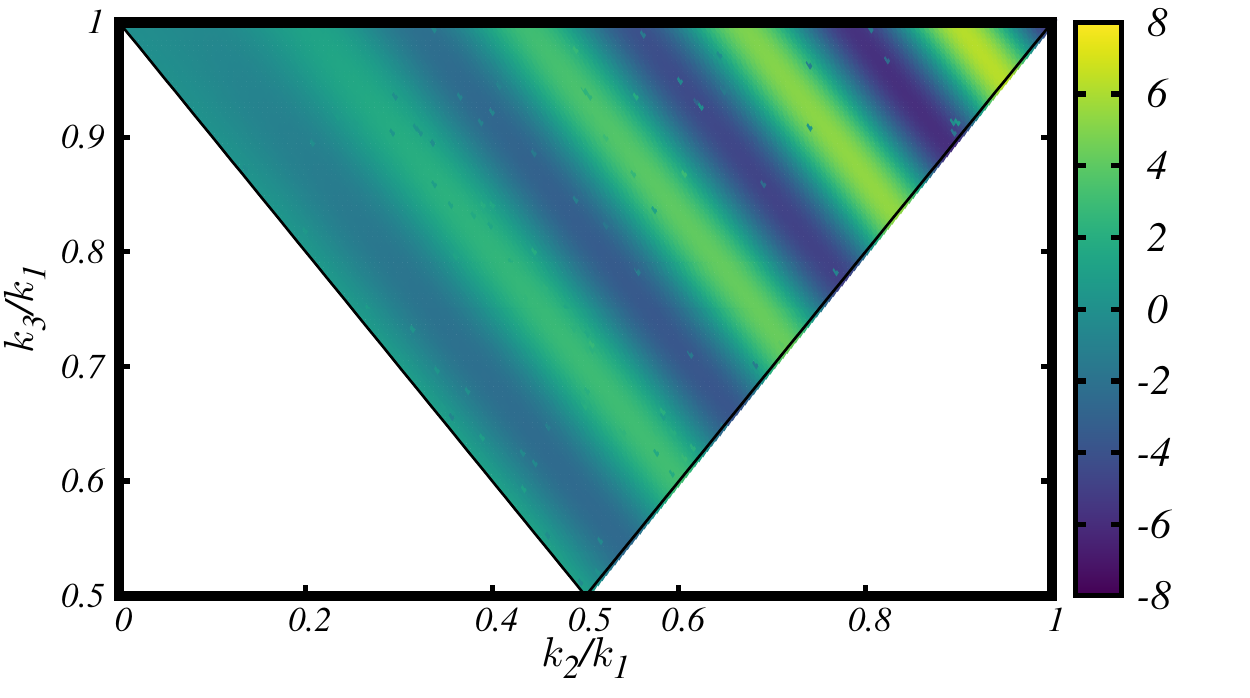}
\caption{\footnotesize\label{fig:WWI-NG} [Left] $f_{\rm NL}(k)$ computed for four best fits (PPS plotted in~\autoref{fig:PPS-WWI-TTTEEE}) to the Planck data in WWI potential in the equilateral limit. [Right] $f_{\rm NL}(\bk_1,\bk_2,\bk_3)$ plotted for the same best fits as on the left, but for arbitrary triangular configuration as a function of $k_2/k_1$ and $k_2/k_1$ where $k_1$ is chosen to be a particular small scale mode.} 
\end{figure*}

We compute all the terms contributing to the bispectra and combine them to calculate $f_{\rm NL}$ as discussed in~\cite{Martin:2011sn,Hazra:2012BINGO,Sreenath:2014BINGO}. In~\autoref{fig:WWI-NG} we plot the $f_{\rm NL}$ for the four WWI best fit potentials to the Planck data plotted in~\autoref{fig:PPS-WWI-TTTEEE}. We obtain oscillatory bispectra for all the best fits. Sharpness of the transition determines frequency of the oscillations in the power spectrum and that is reflected too in the bispectra. As we have discussed earlier, the discontinuity in the potential and its derivatives are modelled with a step function and its derivatives.  
When the step is sharp enough, we notice divergent $f_{\rm NL}$ in the equilateral limit. Therefore while the power spectrum displays oscillations of equal amplitude, the $f_{\rm NL}$ follows a power law divergence. In WWI-TTTEEE-B, the oscillations in the PPS decay at small scales as the step is not steep. The equilateral $f_{\rm NL}$ here grows till $0.08\Mpc{}^{-1}$ and then decays. Note that WWI-TTTEEE-D shows a growing $f_{\rm NL}$ till the smallest observed cosmological scales by Planck. It decays at smaller scales. Here we notice the divergence pattern at scales smaller $0.08\Mpc{}^{-1}$ (corresponding to WWI-TTTEEE-B). Although the sharpness and magnitude of the transition in potential for the best fit WWI-TTTEEE-B are more than WWI-TTTEEE-D, since the transition occurs at smaller scales in the latter we do not notice the decay of $f_{\rm NL}$ there. In the panel at the right we plot the density plots of $f_{\rm NL}(\bk_1,\bk_2,\bk_3)$ (represented by the colorbar). Here a small scale mode is fixed as $k_1$ and the $f_{\rm NL}$ is plotted as a function of $k_2/k_1$ and $k_3/k_1$.  

\begin{figure*}
\centering
\includegraphics[width=0.44\columnwidth]{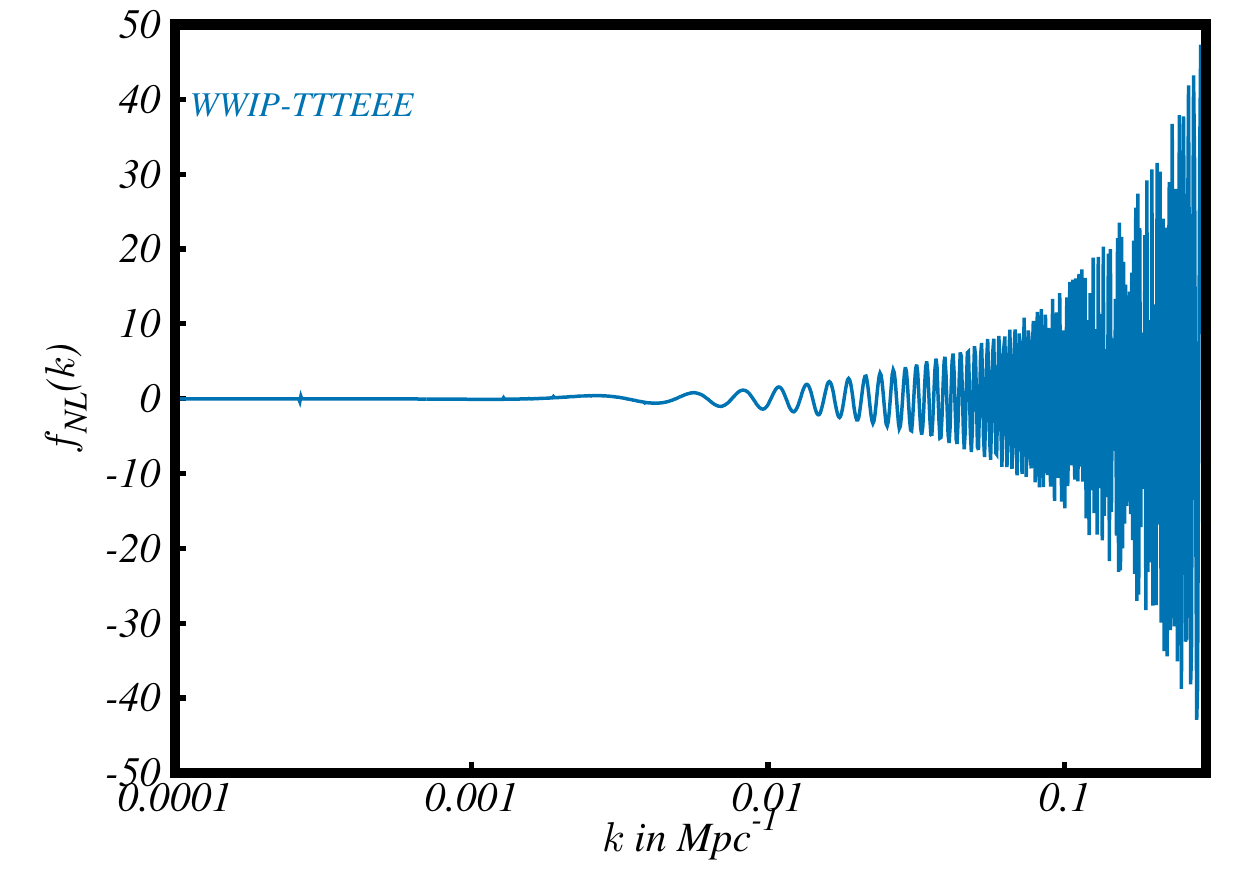}
\includegraphics[width=0.55\columnwidth]{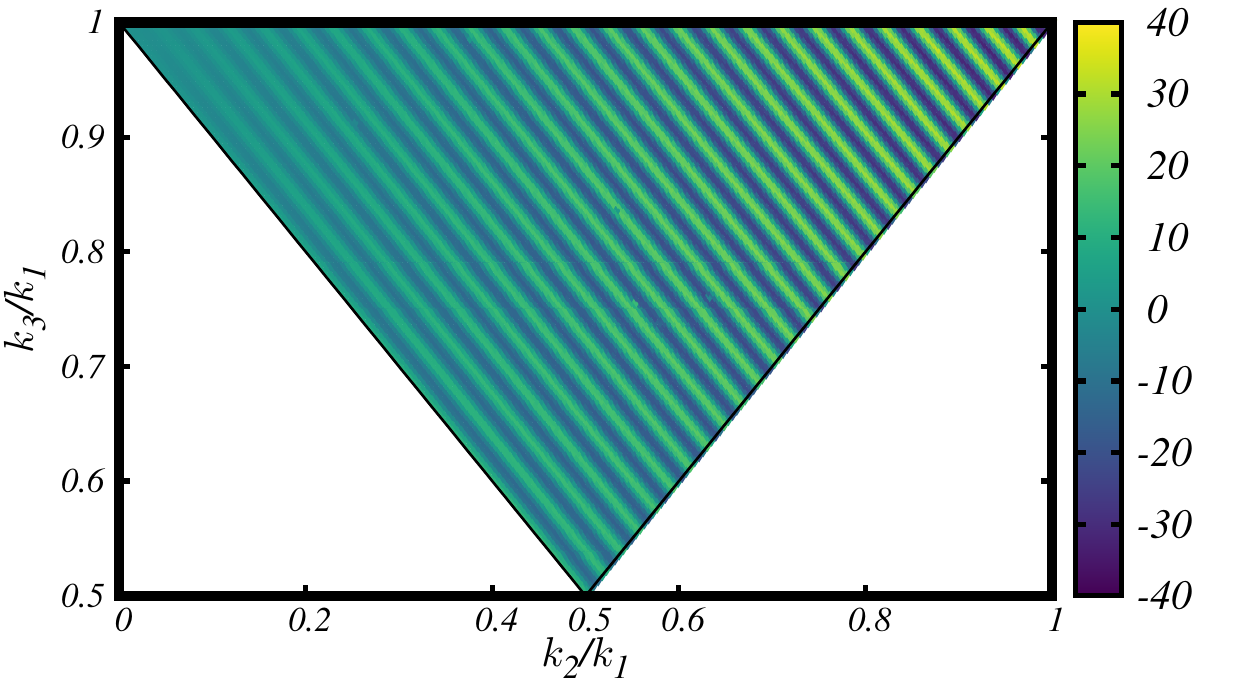}

\caption{\footnotesize\label{fig:WWIP-NG} [Left] $f_{\rm NL}(k)$ computed for WWIP-TTTEEE best fit (PPS plotted in~\autoref{fig:PPS-WWIP}) to the Planck data in the equilateral limit. [Right] $f_{\rm NL}(\bk_1,\bk_2,\bk_3)$ plotted for arbitrary triangular configuration as a function of $k_2/k_1$ and $k_2/k_1$ where $k_1$ is chosen to be a particular small scale mode.} 
\end{figure*}

In~\autoref{fig:WWIP-NG} for the best fit WWIP-TTTEEE we plot the $f_{\rm NL}(k)$ in the equilateral limit (left panel) and $f_{\rm NL}(\bk_1,\bk_2,\bk_3)$ with fixed $k_1$ (right panel). As has been discussed~\cite{Arroja:2011yu,Martin:2014kja} the $f_{\rm NL}(k)$ is divergent. Similar to WWI best fits here we also notice similar $f_{\rm NL}(\bk_1,\bk_2,\bk_3)$ in the plot at the right panel. Note that the Dirac delta function in the second derivative of the potential (the potential is continuous with discontinuous first derivative) is contributing to the divergent $f_{\rm NL}$. Unlike the WWI model, here the same power spectrum can provide different bispectra depending on the sharpness of the transition in the derivative of the potential. While for the power spectrum we do not get divergences from these discontinuities, divergences in the bispectrum will be constrained well from the bispectrum data that will add further constraints on transition width and possibly can rule our some of the best fit candidates.

In general the BINGO output for $f_{\rm NL}(\bk_1,\bk_2,\bk_3)$ with $\bk_1,\bk_2,\bk_3$ satisfying a triangle should be compared with the bispectrum from the CMB observation along with the power spectrum. Since such a joint likelihood for a generic power spectrum and bispectra is not available this estimation is beyond the scope of this paper.

\section{Discussions}~\label{DiscussionSection}
Using the baseline and feature variants of Wiggly Whipped Inflation we compare inflationary slow-roll dynamics and its departure with the data from Planck 2018 release. Planck binned power spectrum data for temperature, polarization anisotropy, lensing are used jointly with BICEP-Keck-Planck B-mode likelihood to constrain the baseline strict slow roll inflationary model. Once the baseline model is constrained with the full Planck+BK15 baseline, we explore the possible features of our models, appearing at different characteristics scales, and we compare the results for temperature and polarization (taken separately or in combination). Our main results are listed below:
\begin{itemize}
    \item Using the full Planck2018+BK15 baseline combination (binned \texttt{Plik} TTTEEE + lowT + lowE + lensing) we find, for Hilltop quartic inflation, the lower limit (at 95\%) on the field value at the end of inflation ($\mu$) is $13.4\Mpl$. 
    \item In Whipped Inflation, when TT+lowT data combination is used, a marginal (68\%) significance towards suppression is found with $\Delta\chi^2\sim3$ compared to the baseline. This significance decreases when lowE data is used in the complete data combination. A separate analysis with EE+lowE suggests no support for the suppression.
    For TT+lowT data we find positive but inconclusive preference for the WI model over the baseline with $\ln B=0.12$. TTTEEE+lowT+lowE data analysis does not show such preference as we find the marginal likelihood to be $\ln B=-1.8$. Though we considered some specific sub-class of possible potentials with features in our paper, our results reflect the following general tendency: taking CMB polarization data into account diminishes statistical importance of anomalies seen in the CMB temperature anisotropy for $\ell<30$. 
    
    \item When \texttt{Clean CamSpec} data is used to constrain the wiggles in the WWI (discontinuous potential) model, we find localized peaks in the posterior distribution of feature position. These peaks do not have overlaps between posteriors from different datasets. We find $\Delta\chi^2=9.25$ \textit{w.r.t.} baseline for \texttt{CamSpec} TTTEEE+lowT+lowE datasets. Here extra parameters are penalized more and we find $\ln B=-2.5$ \textit{w.r.t.} the baseline.
    
    \item WWIP model (with the discontinuity in the derivatives of a continuous potential) best fits for TT+lowT+lowE and EE+lowE show a mismatch in phases of the oscillations. Here for \texttt{CamSpec} TTTEEE+lowT+lowE analysis we find $\ln B=-1.6$. Best fit spectrum provides nearly 6.5 improvement in fit compared to the baseline. Although WWIP shows lower improvement in fit compared to the WWI, the Bayesian evidence here is better because lesser parameters for features decrease the prior volume.
    
    \item Overall we have noticed, compared to Planck 2015 data, a decrease in the improvement in fit to the data when HFI large scale polarization data is used and when \texttt{Clean CamSpec} with higher sky fraction and updated cleaning strategies are implemented.
    
    \item $f_{\rm NL}$ computed for the best fit candidates in both WWI and WWIP potentials show oscillatory bispectra. Provided a joint power spectrum and bispectra likelihood, in future joint estimation for feature models is possible.    
    \end{itemize}

To summarize, comparing the Bayesian evidences while we find a marginal favor of large scale suppression by the temperature anisotropy power spectrum, large-scale E-model polarization likelihood based on high frequency instrument cross spectrum does not support this suppression and in the combined data, the preference towards the
suppression becomes negligible. Oscillatory features that require more parameters are not favored compared to the baseline model although they provide upto 10 improvement in fit to the combined data. This is in agreement with other findings~\cite{Planck2018X,Braglia:2021ckn,Canas-Herrera:2020mme}. However, given the low signal-to-noise ratios in TE and EE datasets, we expect that a conclusive evidence for or against the features such as suppression and wiggles can be provided using other data from observations such as cosmic variance-limited polarization~\cite{LiteBIRD,CMBBHARAT,PICO} and large-scale-structure~\cite{DESI,Euclid,LSST,Hazra:2012LSS,Chen:2016LSS,Ballardini:2016LSS,LHuillier:2017lgm,Beutler:2019LSS,Ballardini:2019LSS,Debono20}.
\section*{Acknowledgements}
The authors acknowledge the use of computational resources at the Institute of Mathematical Science’s High Performance Computing facility (hpc.imsc.res.in) [Nandadevi] and at the  Istituto Nazionale di Fisica Nucleare's CNAF cluster. The authors would like to thank Matteo Braglia, Xingang Chen and Fabio Finelli for their comments on the manuscript. 
 DP acknowledges financial support by ASI  Grant 2016-24-H.0 and the agreement n. 2020-9-HH.0 ASI-UniRM2 “Partecipazione italiana alla fase A della missione LiteBIRD”. DP acknowledges the computing centre of Cineca and INAF, under the coordination of the “Accordo Quadro MoU per lo svolgimento di attivit\'a congiunta di ricerca Nuove frontiere in Astrofisica: HPC e Data Exploration di nuova generazione”, for the availability of computing resources and support with the project INA17-C5A42. AAS has been partially supported by the project number 0033-2019-0005 of the Russian Ministry of Science and Higher Education and by the Kazan Federal University Strategic Academic Leadership Program.

\bibliography{wwiref}

\providecommand{\href}[2]{#2}\begingroup\raggedright\begin{thebibliography}{100}

\bibitem{Starobinsky:1979ty}
A.~A. Starobinsky, \emph{{Spectrum of relict gravitational radiation and the
  early state of the universe}}, {\emph{JETP Lett.} {\bfseries 30} (1979) 682}.

\bibitem{Starobinsky:1980te}
A.~A. Starobinsky, \emph{{A New Type of Isotropic Cosmological Models Without
  Singularity}},
  \href{https://doi.org/10.1016/0370-2693(80)90670-X}{\emph{Phys. Lett. B}
  {\bfseries 91} (1980) 99}.

\bibitem{Guth}
A.~H. Guth, \emph{{The Inflationary Universe: A Possible Solution to the
  Horizon and Flatness Problems}},
  \href{https://doi.org/10.1103/PhysRevD.23.347}{\emph{Phys. Rev. D} {\bfseries
  23} (1981) 347}.

\bibitem{Sato}
K.~Sato, \emph{{First Order Phase Transition of a Vacuum and Expansion of the
  Universe}}, {\emph{Mon. Not. Roy. Astron. Soc.} {\bfseries 195} (1981) 467}.

\bibitem{Mukhanov}
V.~F. Mukhanov and G.~V. Chibisov, \emph{{Quantum Fluctuations and a
  Nonsingular Universe}}, {\emph{JETP Lett.} {\bfseries 33} (1981) 532}.

\bibitem{Linde:1981mu}
A.~D. Linde, \emph{{A New Inflationary Universe Scenario: A Possible Solution
  of the Horizon, Flatness, Homogeneity, Isotropy and Primordial Monopole
  Problems}}, \href{https://doi.org/10.1016/0370-2693(82)91219-9}{\emph{Phys.
  Lett. B} {\bfseries 108} (1982) 389}.

\bibitem{Albrecht:1982wi}
A.~Albrecht and P.~J. Steinhardt, \emph{{Cosmology for Grand Unified Theories
  with Radiatively Induced Symmetry Breaking}},
  \href{https://doi.org/10.1103/PhysRevLett.48.1220}{\emph{Phys. Rev. Lett.}
  {\bfseries 48} (1982) 1220}.

\bibitem{Hawking:1982cz}
S.~W. Hawking, \emph{{The Development of Irregularities in a Single Bubble
  Inflationary Universe}},
  \href{https://doi.org/10.1016/0370-2693(82)90373-2}{\emph{Phys. Lett. B}
  {\bfseries 115} (1982) 295}.

\bibitem{Starobinsky:1982ee}
A.~A. Starobinsky, \emph{{Dynamics of Phase Transition in the New Inflationary
  Universe Scenario and Generation of Perturbations}},
  \href{https://doi.org/10.1016/0370-2693(82)90541-X}{\emph{Phys. Lett. B}
  {\bfseries 117} (1982) 175}.

\bibitem{Guth:1982ec}
A.~H. Guth and S.~Y. Pi, \emph{{Fluctuations in the New Inflationary
  Universe}}, \href{https://doi.org/10.1103/PhysRevLett.49.1110}{\emph{Phys.
  Rev. Lett.} {\bfseries 49} (1982) 1110}.

\bibitem{Starobinsky:1983zz}
A.~A. Starobinsky, \emph{{The Perturbation Spectrum Evolving from a Nonsingular
  Initially De-Sitter Cosmology and the Microwave Background Anisotropy}},
  {\emph{Sov. Astron. Lett.} {\bfseries 9} (1983) 302}.

\bibitem{Linde}
A.~D. Linde, \emph{{Chaotic Inflation}},
  \href{https://doi.org/10.1016/0370-2693(83)90837-7}{\emph{Phys. Lett. B}
  {\bfseries 129} (1983) 177}.

\bibitem{COBE}
{\scshape COBE} collaboration, \emph{{Structure in the COBE differential
  microwave radiometer first year maps}},
  \href{https://doi.org/10.1086/186504}{\emph{Astrophys. J. Lett.} {\bfseries
  396} (1992) L1}.

\bibitem{COBEnormalization}
M.~J. White and E.~F. Bunn, \emph{{The COBE normalization of CMB
  anisotropies}}, \href{https://doi.org/10.1086/176158}{\emph{Astrophys. J.}
  {\bfseries 450} (1995) 477}
  [\href{https://arxiv.org/abs/astro-ph/9503054}{{\ttfamily
  astro-ph/9503054}}].

\bibitem{Hinshaw_2013}
G.~Hinshaw, D.~Larson, E.~Komatsu, D.~N. Spergel, C.~L. Bennett, J.~Dunkley
  et~al., \emph{Nine-year wilkinson microwave anisotropy probe ( wmap )
  observations: Cosmological parameter results},
  \href{https://doi.org/10.1088/0067-0049/208/2/19}{\emph{The Astrophysical
  Journal Supplement Series} {\bfseries 208} (2013) 19}.

\bibitem{Planck:2018Overview}
{\scshape Planck} collaboration, \emph{{Planck 2018 results. I. Overview and
  the cosmological legacy of Planck}},
  \href{https://doi.org/10.1051/0004-6361/201833880}{\emph{Astron. Astrophys.}
  {\bfseries 641} (2020) A1}
  [\href{https://arxiv.org/abs/1807.06205}{{\ttfamily 1807.06205}}].

\bibitem{Planck:2018Params}
{\scshape Planck} collaboration, \emph{{Planck 2018 results. VI. Cosmological
  parameters}},
  \href{https://doi.org/10.1051/0004-6361/201833910}{\emph{Astron. Astrophys.}
  {\bfseries 641} (2020) A6}
  [\href{https://arxiv.org/abs/1807.06209}{{\ttfamily 1807.06209}}].

\bibitem{BK15}
{\scshape BICEP2, Keck Array} collaboration, \emph{{BICEP2 / Keck Array x:
  Constraints on Primordial Gravitational Waves using Planck, WMAP, and New
  BICEP2/Keck Observations through the 2015 Season}},
  \href{https://doi.org/10.1103/PhysRevLett.121.221301}{\emph{Phys. Rev. Lett.}
  {\bfseries 121} (2018) 221301}
  [\href{https://arxiv.org/abs/1810.05216}{{\ttfamily 1810.05216}}].

\bibitem{Planck:2018HFI}
{\scshape Planck} collaboration, \emph{{Planck 2018 results. III. High
  Frequency Instrument data processing and frequency maps}},
  \href{https://doi.org/10.1051/0004-6361/201832909}{\emph{Astron. Astrophys.}
  {\bfseries 641} (2020) A3}
  [\href{https://arxiv.org/abs/1807.06207}{{\ttfamily 1807.06207}}].

\bibitem{Planck:2015Params}
{\scshape Planck} collaboration, \emph{{Planck 2015 results. XIII. Cosmological
  parameters}},
  \href{https://doi.org/10.1051/0004-6361/201525830}{\emph{Astron. Astrophys.}
  {\bfseries 594} (2016) A13}
  [\href{https://arxiv.org/abs/1502.01589}{{\ttfamily 1502.01589}}].

\bibitem{Planck2018X}
{\scshape Planck} collaboration, \emph{{Planck 2018 results. X. Constraints on
  inflation}}, \href{https://doi.org/10.1051/0004-6361/201833887}{\emph{Astron.
  Astrophys.} {\bfseries 641} (2020) A10}
  [\href{https://arxiv.org/abs/1807.06211}{{\ttfamily 1807.06211}}].

\bibitem{Hazra:recon13broad}
D.~K. Hazra, A.~Shafieloo and G.~F. Smoot, \emph{{Reconstruction of broad
  features in the primordial spectrum and inflaton potential from Planck}},
  \href{https://doi.org/10.1088/1475-7516/2013/12/035}{\emph{JCAP} {\bfseries
  12} (2013) 035} [\href{https://arxiv.org/abs/1310.3038}{{\ttfamily
  1310.3038}}].

\bibitem{Hazra:reconP13}
D.~K. Hazra, A.~Shafieloo and T.~Souradeep, \emph{{Primordial power spectrum
  from Planck}},
  \href{https://doi.org/10.1088/1475-7516/2014/11/011}{\emph{JCAP} {\bfseries
  11} (2014) 011} [\href{https://arxiv.org/abs/1406.4827}{{\ttfamily
  1406.4827}}].

\bibitem{Hunt:2015iua}
P.~Hunt and S.~Sarkar, \emph{{Search for features in the spectrum of primordial
  perturbations using Planck and other datasets}},
  \href{https://doi.org/10.1088/1475-7516/2015/12/052}{\emph{JCAP} {\bfseries
  12} (2015) 052} [\href{https://arxiv.org/abs/1510.03338}{{\ttfamily
  1510.03338}}].

\bibitem{Obied:2018qdr}
G.~Obied, C.~Dvorkin, C.~Heinrich, W.~Hu and V.~Miranda, \emph{{Inflationary
  versus reionization features from $Planck$ 2015 data}},
  \href{https://doi.org/10.1103/PhysRevD.98.043518}{\emph{Phys. Rev. D}
  {\bfseries 98} (2018) 043518}
  [\href{https://arxiv.org/abs/1803.01858}{{\ttfamily 1803.01858}}].

\bibitem{Hannestad:recon00}
S.~Hannestad, \emph{{Reconstructing the inflationary power spectrum from CMBR
  data}}, \href{https://doi.org/10.1103/PhysRevD.63.043009}{\emph{Phys. Rev. D}
  {\bfseries 63} (2001) 043009}
  [\href{https://arxiv.org/abs/astro-ph/0009296}{{\ttfamily
  astro-ph/0009296}}].

\bibitem{Shafieloo:recon03}
A.~Shafieloo and T.~Souradeep, \emph{{Primordial power spectrum from WMAP}},
  \href{https://doi.org/10.1103/PhysRevD.70.043523}{\emph{Phys. Rev. D}
  {\bfseries 70} (2004) 043523}
  [\href{https://arxiv.org/abs/astro-ph/0312174}{{\ttfamily
  astro-ph/0312174}}].

\bibitem{Mukherjee:recon03}
P.~Mukherjee and Y.~Wang, \emph{{Model-independent reconstruction of the
  primordial power spectrum from WMAP data}},
  \href{https://doi.org/10.1086/379161}{\emph{Astrophys. J.} {\bfseries 599}
  (2003) 1} [\href{https://arxiv.org/abs/astro-ph/0303211}{{\ttfamily
  astro-ph/0303211}}].

\bibitem{Bridle:recon03}
S.~L. Bridle, A.~M. Lewis, J.~Weller and G.~Efstathiou, \emph{{Reconstructing
  the primordial power spectrum}},
  \href{https://doi.org/10.1046/j.1365-8711.2003.06807.x}{\emph{Mon. Not. Roy.
  Astron. Soc.} {\bfseries 342} (2003) L72}
  [\href{https://arxiv.org/abs/astro-ph/0302306}{{\ttfamily
  astro-ph/0302306}}].

\bibitem{Kogo:recon05}
N.~Kogo, M.~Sasaki and J.~Yokoyama, \emph{{Constraining cosmological parameters
  by the cosmic inversion method}},
  \href{https://doi.org/10.1143/PTP.114.555}{\emph{Prog. Theor. Phys.}
  {\bfseries 114} (2005) 555}
  [\href{https://arxiv.org/abs/astro-ph/0504471}{{\ttfamily
  astro-ph/0504471}}].

\bibitem{Leach:recon05}
S.~M. Leach, \emph{{Measuring the primordial power spectrum: Principal
  component analysis of the cosmic microwave background}},
  \href{https://doi.org/10.1111/j.1365-2966.2006.10842.x}{\emph{Mon. Not. Roy.
  Astron. Soc.} {\bfseries 372} (2006) 646}
  [\href{https://arxiv.org/abs/astro-ph/0506390}{{\ttfamily
  astro-ph/0506390}}].

\bibitem{Tocchini-Valentini:recon05}
D.~Tocchini-Valentini, Y.~Hoffman and J.~Silk, \emph{{Non-parametric
  reconstruction of the primordial power spectrum at horizon scales from wmap
  data}}, \href{https://doi.org/10.1111/j.1365-2966.2006.10031.x}{\emph{Mon.
  Not. Roy. Astron. Soc.} {\bfseries 367} (2006) 1095}
  [\href{https://arxiv.org/abs/astro-ph/0509478}{{\ttfamily
  astro-ph/0509478}}].

\bibitem{Shafieloo:recon07}
A.~Shafieloo and T.~Souradeep, \emph{{Estimation of Primordial Spectrum with
  post-WMAP 3 year data}},
  \href{https://doi.org/10.1103/PhysRevD.78.023511}{\emph{Phys. Rev. D}
  {\bfseries 78} (2008) 023511}
  [\href{https://arxiv.org/abs/0709.1944}{{\ttfamily 0709.1944}}].

\bibitem{Paykari:recon09}
P.~Paykari and A.~H. Jaffe, \emph{{Optimal Binning of the Primordial Power
  Spectrum}}, \href{https://doi.org/10.1088/0004-637X/711/1/1}{\emph{Astrophys.
  J.} {\bfseries 711} (2010) 1}
  [\href{https://arxiv.org/abs/0902.4399}{{\ttfamily 0902.4399}}].

\bibitem{Nicholson:recon09}
G.~Nicholson and C.~R. Contaldi, \emph{{Reconstruction of the Primordial Power
  Spectrum using Temperature and Polarisation Data from Multiple Experiments}},
  \href{https://doi.org/10.1088/1475-7516/2009/07/011}{\emph{JCAP} {\bfseries
  07} (2009) 011} [\href{https://arxiv.org/abs/0903.1106}{{\ttfamily
  0903.1106}}].

\bibitem{Gauthier:recon12}
C.~Gauthier and M.~Bucher, \emph{{Reconstructing the primordial power spectrum
  from the CMB}},
  \href{https://doi.org/10.1088/1475-7516/2012/10/050}{\emph{JCAP} {\bfseries
  10} (2012) 050} [\href{https://arxiv.org/abs/1209.2147}{{\ttfamily
  1209.2147}}].

\bibitem{Vazquez:recon12}
J.~A. Vazquez, M.~Bridges, M.~P. Hobson and A.~N. Lasenby, \emph{{Model
  selection applied to reconstruction of the Primordial Power Spectrum}},
  \href{https://doi.org/10.1088/1475-7516/2012/06/006}{\emph{JCAP} {\bfseries
  06} (2012) 006} [\href{https://arxiv.org/abs/1203.1252}{{\ttfamily
  1203.1252}}].

\bibitem{Hazra:recon13}
D.~K. Hazra, A.~Shafieloo and T.~Souradeep, \emph{{Primordial power spectrum: a
  complete analysis with the WMAP nine-year data}},
  \href{https://doi.org/10.1088/1475-7516/2013/07/031}{\emph{JCAP} {\bfseries
  07} (2013) 031} [\href{https://arxiv.org/abs/1303.4143}{{\ttfamily
  1303.4143}}].

\bibitem{Hunt:recon13}
P.~Hunt and S.~Sarkar, \emph{{Reconstruction of the primordial power spectrum
  of curvature perturbations using multiple data sets}},
  \href{https://doi.org/10.1088/1475-7516/2014/01/025}{\emph{JCAP} {\bfseries
  01} (2014) 025} [\href{https://arxiv.org/abs/1308.2317}{{\ttfamily
  1308.2317}}].

\bibitem{Starobinsky:Kink}
A.~A. Starobinsky, \emph{{Spectrum of adiabatic perturbations in the universe
  when there are singularities in the inflation potential}}, {\emph{JETP Lett.}
  {\bfseries 55} (1992) 489}.

\bibitem{Ivanov:1994pa}
P.~Ivanov, P.~Naselsky and I.~Novikov, \emph{{Inflation and primordial black
  holes as dark matter}},
  \href{https://doi.org/10.1103/PhysRevD.50.7173}{\emph{Phys. Rev. D}
  {\bfseries 50} (1994) 7173}.

\bibitem{Adams:Step0}
J.~A. Adams, B.~Cresswell and R.~Easther, \emph{{Inflationary perturbations
  from a potential with a step}},
  \href{https://doi.org/10.1103/PhysRevD.64.123514}{\emph{Phys. Rev. D}
  {\bfseries 64} (2001) 123514}
  [\href{https://arxiv.org/abs/astro-ph/0102236}{{\ttfamily
  astro-ph/0102236}}].

\bibitem{Covi:Step1}
L.~Covi, J.~Hamann, A.~Melchiorri, A.~Slosar and I.~Sorbera, \emph{{Inflation
  and WMAP three year data: Features have a Future!}},
  \href{https://doi.org/10.1103/PhysRevD.74.083509}{\emph{Phys. Rev. D}
  {\bfseries 74} (2006) 083509}
  [\href{https://arxiv.org/abs/astro-ph/0606452}{{\ttfamily
  astro-ph/0606452}}].

\bibitem{Ashoorioon:Step2}
A.~Ashoorioon and A.~Krause, \emph{{Power Spectrum and Signatures for Cascade
  Inflation}},  \href{https://arxiv.org/abs/hep-th/0607001}{{\ttfamily
  hep-th/0607001}}.

\bibitem{Joy:2007na}
M.~Joy, V.~Sahni and A.~A. Starobinsky, \emph{{A New Universal Local Feature in
  the Inflationary Perturbation Spectrum}},
  \href{https://doi.org/10.1103/PhysRevD.77.023514}{\emph{Phys. Rev. D}
  {\bfseries 77} (2008) 023514}
  [\href{https://arxiv.org/abs/0711.1585}{{\ttfamily 0711.1585}}].

\bibitem{Joy:2008qd}
M.~Joy, A.~Shafieloo, V.~Sahni and A.~A. Starobinsky, \emph{{Is a step in the
  primordial spectral index favored by CMB data ?}},
  \href{https://doi.org/10.1088/1475-7516/2009/06/028}{\emph{JCAP} {\bfseries
  06} (2009) 028} [\href{https://arxiv.org/abs/0807.3334}{{\ttfamily
  0807.3334}}].

\bibitem{Hazra:Step3}
D.~K. Hazra, M.~Aich, R.~K. Jain, L.~Sriramkumar and T.~Souradeep,
  \emph{{Primordial features due to a step in the inflaton potential}},
  \href{https://doi.org/10.1088/1475-7516/2010/10/008}{\emph{JCAP} {\bfseries
  10} (2010) 008} [\href{https://arxiv.org/abs/1005.2175}{{\ttfamily
  1005.2175}}].

\bibitem{Miranda:Step4}
V.~Miranda, W.~Hu and P.~Adshead, \emph{{Warp Features in DBI Inflation}},
  \href{https://doi.org/10.1103/PhysRevD.86.063529}{\emph{Phys. Rev. D}
  {\bfseries 86} (2012) 063529}
  [\href{https://arxiv.org/abs/1207.2186}{{\ttfamily 1207.2186}}].

\bibitem{Benetti:Step5}
M.~Benetti, \emph{{Updating constraints on inflationary features in the
  primordial power spectrum with the Planck data}},
  \href{https://doi.org/10.1103/PhysRevD.88.087302}{\emph{Phys. Rev. D}
  {\bfseries 88} (2013) 087302}
  [\href{https://arxiv.org/abs/1308.6406}{{\ttfamily 1308.6406}}].

\bibitem{GallegoCadavid:Step6}
A.~Gallego~Cadavid and A.~E. Romano, \emph{{Effects of discontinuities of the
  derivatives of the inflaton potential}},
  \href{https://doi.org/10.1140/epjc/s10052-015-3733-x}{\emph{Eur. Phys. J. C}
  {\bfseries 75} (2015) 589} [\href{https://arxiv.org/abs/1404.2985}{{\ttfamily
  1404.2985}}].

\bibitem{Chluba:Step7}
J.~Chluba, J.~Hamann and S.~P. Patil, \emph{{Features and New Physical Scales
  in Primordial Observables: Theory and Observation}},
  \href{https://doi.org/10.1142/S0218271815300232}{\emph{Int. J. Mod. Phys. D}
  {\bfseries 24} (2015) 1530023}
  [\href{https://arxiv.org/abs/1505.01834}{{\ttfamily 1505.01834}}].

\bibitem{Bousso:Step8}
R.~Bousso, D.~Harlow and L.~Senatore, \emph{{Inflation After False Vacuum
  Decay: New Evidence from BICEP2}},
  \href{https://doi.org/10.1088/1475-7516/2014/12/019}{\emph{JCAP} {\bfseries
  12} (2014) 019} [\href{https://arxiv.org/abs/1404.2278}{{\ttfamily
  1404.2278}}].

\bibitem{Allahverdi:PI}
R.~Allahverdi, K.~Enqvist, J.~Garcia-Bellido and A.~Mazumdar, \emph{{Gauge
  invariant MSSM inflaton}},
  \href{https://doi.org/10.1103/PhysRevLett.97.191304}{\emph{Phys. Rev. Lett.}
  {\bfseries 97} (2006) 191304}
  [\href{https://arxiv.org/abs/hep-ph/0605035}{{\ttfamily hep-ph/0605035}}].

\bibitem{Jain:PI}
R.~K. Jain, P.~Chingangbam, J.-O. Gong, L.~Sriramkumar and T.~Souradeep,
  \emph{{Punctuated inflation and the low CMB multipoles}},
  \href{https://doi.org/10.1088/1475-7516/2009/01/009}{\emph{JCAP} {\bfseries
  01} (2009) 009} [\href{https://arxiv.org/abs/0809.3915}{{\ttfamily
  0809.3915}}].

\bibitem{Chen:Osc00}
X.~Chen, R.~Easther and E.~A. Lim, \emph{{Generation and Characterization of
  Large Non-Gaussianities in Single Field Inflation}},
  \href{https://doi.org/10.1088/1475-7516/2008/04/010}{\emph{JCAP} {\bfseries
  04} (2008) 010} [\href{https://arxiv.org/abs/0801.3295}{{\ttfamily
  0801.3295}}].

\bibitem{McAllister:Osc0}
L.~McAllister, E.~Silverstein and A.~Westphal, \emph{{Gravity Waves and Linear
  Inflation from Axion Monodromy}},
  \href{https://doi.org/10.1103/PhysRevD.82.046003}{\emph{Phys. Rev. D}
  {\bfseries 82} (2010) 046003}
  [\href{https://arxiv.org/abs/0808.0706}{{\ttfamily 0808.0706}}].

\bibitem{Flauger:Osc1}
R.~Flauger, L.~McAllister, E.~Pajer, A.~Westphal and G.~Xu, \emph{{Oscillations
  in the CMB from Axion Monodromy Inflation}},
  \href{https://doi.org/10.1088/1475-7516/2010/06/009}{\emph{JCAP} {\bfseries
  06} (2010) 009} [\href{https://arxiv.org/abs/0907.2916}{{\ttfamily
  0907.2916}}].

\bibitem{Pahud:Osc2}
C.~Pahud, M.~Kamionkowski and A.~R. Liddle, \emph{{Oscillations in the inflaton
  potential?}}, \href{https://doi.org/10.1103/PhysRevD.79.083503}{\emph{Phys.
  Rev. D} {\bfseries 79} (2009) 083503}
  [\href{https://arxiv.org/abs/0807.0322}{{\ttfamily 0807.0322}}].

\bibitem{Flauger:2010Osc}
R.~Flauger and E.~Pajer, \emph{{Resonant Non-Gaussianity}},
  \href{https://doi.org/10.1088/1475-7516/2011/01/017}{\emph{JCAP} {\bfseries
  01} (2011) 017} [\href{https://arxiv.org/abs/1002.0833}{{\ttfamily
  1002.0833}}].

\bibitem{Chen:2010Osc}
X.~Chen, \emph{{Folded Resonant Non-Gaussianity in General Single Field
  Inflation}}, \href{https://doi.org/10.1088/1475-7516/2010/12/003}{\emph{JCAP}
  {\bfseries 12} (2010) 003} [\href{https://arxiv.org/abs/1008.2485}{{\ttfamily
  1008.2485}}].

\bibitem{Aich:Osc3}
M.~Aich, D.~K. Hazra, L.~Sriramkumar and T.~Souradeep, \emph{{Oscillations in
  the inflaton potential: Complete numerical treatment and comparison with the
  recent and forthcoming CMB datasets}},
  \href{https://doi.org/10.1103/PhysRevD.87.083526}{\emph{Phys. Rev. D}
  {\bfseries 87} (2013) 083526}
  [\href{https://arxiv.org/abs/1106.2798}{{\ttfamily 1106.2798}}].

\bibitem{Peiris:Osc4}
H.~Peiris, R.~Easther and R.~Flauger, \emph{{Constraining Monodromy
  Inflation}}, \href{https://doi.org/10.1088/1475-7516/2013/09/018}{\emph{JCAP}
  {\bfseries 09} (2013) 018} [\href{https://arxiv.org/abs/1303.2616}{{\ttfamily
  1303.2616}}].

\bibitem{Meerburg:Osc5}
P.~D. Meerburg and D.~N. Spergel, \emph{{Searching for oscillations in the
  primordial power spectrum. II. Constraints from Planck data}},
  \href{https://doi.org/10.1103/PhysRevD.89.063537}{\emph{Phys. Rev. D}
  {\bfseries 89} (2014) 063537}
  [\href{https://arxiv.org/abs/1308.3705}{{\ttfamily 1308.3705}}].

\bibitem{Easther:Osc6}
R.~Easther and R.~Flauger, \emph{{Planck Constraints on Monodromy Inflation}},
  \href{https://doi.org/10.1088/1475-7516/2014/02/037}{\emph{JCAP} {\bfseries
  02} (2014) 037} [\href{https://arxiv.org/abs/1308.3736}{{\ttfamily
  1308.3736}}].

\bibitem{Motohashi:Osc7}
H.~Motohashi and W.~Hu, \emph{{Running from Features: Optimized Evaluation of
  Inflationary Power Spectra}},
  \href{https://doi.org/10.1103/PhysRevD.92.043501}{\emph{Phys. Rev. D}
  {\bfseries 92} (2015) 043501}
  [\href{https://arxiv.org/abs/1503.04810}{{\ttfamily 1503.04810}}].

\bibitem{Miranda:Osc8}
V.~Miranda, W.~Hu, C.~He and H.~Motohashi, \emph{{Nonlinear Excitations in
  Inflationary Power Spectra}},
  \href{https://doi.org/10.1103/PhysRevD.93.023504}{\emph{Phys. Rev. D}
  {\bfseries 93} (2016) 023504}
  [\href{https://arxiv.org/abs/1510.07580}{{\ttfamily 1510.07580}}].

\bibitem{Cremonini:2fm1}
S.~Cremonini, Z.~Lalak and K.~Turzynski, \emph{{Strongly Coupled Perturbations
  in Two-Field Inflationary Models}},
  \href{https://doi.org/10.1088/1475-7516/2011/03/016}{\emph{JCAP} {\bfseries
  1103} (2011) 016} [\href{https://arxiv.org/abs/1010.3021}{{\ttfamily
  1010.3021}}].

\bibitem{Achucarro:2fm2}
A.~Achucarro, J.-O. Gong, S.~Hardeman, G.~A. Palma and S.~P. Patil,
  \emph{{Features of heavy physics in the CMB power spectrum}},
  \href{https://doi.org/10.1088/1475-7516/2011/01/030}{\emph{JCAP} {\bfseries
  01} (2011) 030} [\href{https://arxiv.org/abs/1010.3693}{{\ttfamily
  1010.3693}}].

\bibitem{Braglia:2fm7}
M.~Braglia, D.~K. Hazra, L.~Sriramkumar and F.~Finelli, \emph{{Generating
  primordial features at large scales in two field models of inflation}},
  \href{https://doi.org/10.1088/1475-7516/2020/08/025}{\emph{JCAP} {\bfseries
  08} (2020) 025} [\href{https://arxiv.org/abs/2004.00672}{{\ttfamily
  2004.00672}}].

\bibitem{Braglia:2fm8}
M.~Braglia, D.~K. Hazra, F.~Finelli, G.~F. Smoot, L.~Sriramkumar and A.~A.
  Starobinsky, \emph{{Generating PBHs and small-scale GWs in two-field models
  of inflation}},
  \href{https://doi.org/10.1088/1475-7516/2020/08/001}{\emph{JCAP} {\bfseries
  08} (2020) 001} [\href{https://arxiv.org/abs/2005.02895}{{\ttfamily
  2005.02895}}].

\bibitem{Chen:2fm3}
X.~Chen, \emph{Primordial features as evidence for inflation},
  \href{https://doi.org/10.1088/1475-7516/2012/01/038}{\emph{JCAP} {\bfseries
  01} (2012) 038} [\href{https://arxiv.org/abs/1104.1323}{{\ttfamily
  1104.1323}}].

\bibitem{Chen:2fm4}
X.~Chen and C.~Ringeval, \emph{{Searching for Standard Clocks in the Primordial
  Universe}}, \href{https://doi.org/10.1088/1475-7516/2012/08/014}{\emph{JCAP}
  {\bfseries 08} (2012) 014} [\href{https://arxiv.org/abs/1205.6085}{{\ttfamily
  1205.6085}}].

\bibitem{Chen:2fm5}
X.~Chen, M.~H. Namjoo and Y.~Wang, \emph{{Models of the Primordial Standard
  Clock}}, \href{https://doi.org/10.1088/1475-7516/2015/02/027}{\emph{JCAP}
  {\bfseries 1502} (2015) 027}
  [\href{https://arxiv.org/abs/1411.2349}{{\ttfamily 1411.2349}}].

\bibitem{Chen:2fm6}
X.~Chen and M.~H. Namjoo, \emph{{Standard Clock in Primordial Density
  Perturbations and Cosmic Microwave Background}},
  \href{https://doi.org/10.1016/j.physletb.2014.11.002}{\emph{Phys. Lett. B}
  {\bfseries 739} (2014) 285}
  [\href{https://arxiv.org/abs/1404.1536}{{\ttfamily 1404.1536}}].

\bibitem{Braglia:2fm9}
M.~Braglia, X.~Chen and D.~K. Hazra, \emph{{Probing Primordial Features with
  the Stochastic Gravitational Wave Background}},
  \href{https://doi.org/10.1088/1475-7516/2021/03/005}{\emph{JCAP} {\bfseries
  03} (2021) 005} [\href{https://arxiv.org/abs/2012.05821}{{\ttfamily
  2012.05821}}].

\bibitem{Braglia:2021ckn}
M.~Braglia, X.~Chen and D.~K. Hazra, \emph{{Comparing multi-field primordial
  feature models with the Planck data}},
  \href{https://doi.org/10.1088/1475-7516/2021/06/005}{\emph{JCAP} {\bfseries
  06} (2021) 005} [\href{https://arxiv.org/abs/2103.03025}{{\ttfamily
  2103.03025}}].

\bibitem{Braglia:2fm10}
M.~Braglia, X.~Chen and D.~K. Hazra, \emph{{Uncovering the History of Cosmic
  Inflation from Anomalies in Cosmic Microwave Background Spectra}},
  \href{https://arxiv.org/abs/2106.07546}{{\ttfamily 2106.07546}}.

\bibitem{Braglia:long}
M.~Braglia, X.~Chen and D.~K. Hazra, \emph{{Primordial Standard Clock Models
  and CMB Residual Anomalies}},
  \href{https://arxiv.org/abs/2108.10110}{{\ttfamily 2108.10110}}.

\bibitem{Canas-Herrera:2020mme}
G.~Ca\~nas Herrera, J.~Torrado and A.~Ach\'ucarro, \emph{{Bayesian
  reconstruction of the inflaton\textquoteright{}s speed of sound using CMB
  data}}, \href{https://doi.org/10.1103/PhysRevD.103.123531}{\emph{Phys. Rev.
  D} {\bfseries 103} (2021) 12}
  [\href{https://arxiv.org/abs/2012.04640}{{\ttfamily 2012.04640}}].

\bibitem{WMAP:1Peiris}
{\scshape WMAP} collaboration, \emph{{First year Wilkinson Microwave Anisotropy
  Probe (WMAP) observations: Implications for inflation}},
  \href{https://doi.org/10.1086/377228}{\emph{Astrophys. J. Suppl.} {\bfseries
  148} (2003) 213} [\href{https://arxiv.org/abs/astro-ph/0302225}{{\ttfamily
  astro-ph/0302225}}].

\bibitem{Hazra:2016fkm}
D.~K. Hazra, A.~Shafieloo, G.~F. Smoot and A.~A. Starobinsky, \emph{{Primordial
  features and Planck polarization}},
  \href{https://doi.org/10.1088/1475-7516/2016/09/009}{\emph{JCAP} {\bfseries
  1609} (2016) 009} [\href{https://arxiv.org/abs/1605.02106}{{\ttfamily
  1605.02106}}].

\bibitem{Hazra:2014jka}
D.~K. Hazra, A.~Shafieloo, G.~F. Smoot and A.~A. Starobinsky, \emph{{Inflation
  with Whip-Shaped Suppressed Scalar Power Spectra}},
  \href{https://doi.org/10.1103/PhysRevLett.113.071301}{\emph{Phys. Rev. Lett.}
  {\bfseries 113} (2014) 071301}
  [\href{https://arxiv.org/abs/1404.0360}{{\ttfamily 1404.0360}}].

\bibitem{Hazra:2014goa}
D.~K. Hazra, A.~Shafieloo, G.~F. Smoot and A.~A. Starobinsky, \emph{{Wiggly
  Whipped Inflation}},
  \href{https://doi.org/10.1088/1475-7516/2014/08/048}{\emph{JCAP} {\bfseries
  1408} (2014) 048} [\href{https://arxiv.org/abs/1405.2012}{{\ttfamily
  1405.2012}}].

\bibitem{Hazra:2017joc}
D.~K. Hazra, D.~Paoletti, M.~Ballardini, F.~Finelli, A.~Shafieloo, G.~F. Smoot
  et~al., \emph{{Probing features in inflaton potential and reionization
  history with future CMB space observations}},
  \href{https://doi.org/10.1088/1475-7516/2018/02/017}{\emph{JCAP} {\bfseries
  1802} (2018) 017} [\href{https://arxiv.org/abs/1710.01205}{{\ttfamily
  1710.01205}}].

\bibitem{LHuillier:2017lgm}
B.~L'Huillier, A.~Shafieloo, D.~K. Hazra, G.~F. Smoot and A.~A. Starobinsky,
  \emph{{Probing features in the primordial perturbation spectrum with
  large-scale structure data}},
  \href{https://doi.org/10.1093/mnras/sty745}{\emph{Mon. Not. Roy. Astron.
  Soc.} {\bfseries 477} (2018) 2503}
  [\href{https://arxiv.org/abs/1710.10987}{{\ttfamily 1710.10987}}].

\bibitem{Debono20}
I.~Debono, D.~K. Hazra, A.~Shafieloo, G.~F. Smoot and A.~A. Starobinsky,
  \emph{{Constraints on features in the inflationary potential from future
  Euclid data}}, \href{https://doi.org/10.1093/mnras/staa1765}{\emph{Mon. Not.
  Roy. Astron. Soc.} {\bfseries 496} (2020) 3448}
  [\href{https://arxiv.org/abs/2003.05262}{{\ttfamily 2003.05262}}].

\bibitem{Planck:2019NG}
{\scshape Planck} collaboration, \emph{{Planck 2018 results. IX. Constraints on
  primordial non-Gaussianity}},
  \href{https://doi.org/10.1051/0004-6361/201935891}{\emph{Astron. Astrophys.}
  {\bfseries 641} (2020) A9}
  [\href{https://arxiv.org/abs/1905.05697}{{\ttfamily 1905.05697}}].

\bibitem{Maldacena:2002vr}
J.~M. Maldacena, \emph{{Non-Gaussian features of primordial fluctuations in
  single field inflationary models}},
  \href{https://doi.org/10.1088/1126-6708/2003/05/013}{\emph{JHEP} {\bfseries
  05} (2003) 013} [\href{https://arxiv.org/abs/astro-ph/0210603}{{\ttfamily
  astro-ph/0210603}}].

\bibitem{Seery:2005wm}
D.~Seery and J.~E. Lidsey, \emph{{Primordial non-Gaussianities in single field
  inflation}}, \href{https://doi.org/10.1088/1475-7516/2005/06/003}{\emph{JCAP}
  {\bfseries 06} (2005) 003}
  [\href{https://arxiv.org/abs/astro-ph/0503692}{{\ttfamily
  astro-ph/0503692}}].

\bibitem{Chen:2006xjb}
X.~Chen, R.~Easther and E.~A. Lim, \emph{{Large Non-Gaussianities in Single
  Field Inflation}},
  \href{https://doi.org/10.1088/1475-7516/2007/06/023}{\emph{JCAP} {\bfseries
  06} (2007) 023} [\href{https://arxiv.org/abs/astro-ph/0611645}{{\ttfamily
  astro-ph/0611645}}].

\bibitem{Chen:2010xka}
X.~Chen, \emph{{Primordial Non-Gaussianities from Inflation Models}},
  \href{https://doi.org/10.1155/2010/638979}{\emph{Adv. Astron.} {\bfseries
  2010} (2010) 638979} [\href{https://arxiv.org/abs/1002.1416}{{\ttfamily
  1002.1416}}].

\bibitem{Martin:2011sn}
J.~Martin and L.~Sriramkumar, \emph{{The scalar bi-spectrum in the Starobinsky
  model: The equilateral case}},
  \href{https://doi.org/10.1088/1475-7516/2012/01/008}{\emph{JCAP} {\bfseries
  01} (2012) 008} [\href{https://arxiv.org/abs/1109.5838}{{\ttfamily
  1109.5838}}].

\bibitem{Hazra:2012BINGO}
D.~K. Hazra, L.~Sriramkumar and J.~Martin, \emph{{BINGO: A code for the
  efficient computation of the scalar bi-spectrum}},
  \href{https://doi.org/10.1088/1475-7516/2013/05/026}{\emph{JCAP} {\bfseries
  05} (2013) 026} [\href{https://arxiv.org/abs/1201.0926}{{\ttfamily
  1201.0926}}].

\bibitem{Hazra:2012kq}
D.~K. Hazra, J.~Martin and L.~Sriramkumar, \emph{{The scalar bi-spectrum during
  preheating in single field inflationary models}},
  \href{https://doi.org/10.1103/PhysRevD.86.063523}{\emph{Phys. Rev. D}
  {\bfseries 86} (2012) 063523}
  [\href{https://arxiv.org/abs/1206.0442}{{\ttfamily 1206.0442}}].

\bibitem{Adshead:2013zfa}
P.~Adshead, W.~Hu and V.~Miranda, \emph{{Bispectrum in Single-Field Inflation
  Beyond Slow-Roll}},
  \href{https://doi.org/10.1103/PhysRevD.88.023507}{\emph{Phys. Rev. D}
  {\bfseries 88} (2013) 023507}
  [\href{https://arxiv.org/abs/1303.7004}{{\ttfamily 1303.7004}}].

\bibitem{Achucarro:2013cva}
A.~Ach\'ucarro, V.~Atal, P.~Ortiz and J.~Torrado, \emph{{Localized correlated
  features in the CMB power spectrum and primordial bispectrum from a transient
  reduction in the speed of sound}},
  \href{https://doi.org/10.1103/PhysRevD.89.103006}{\emph{Phys. Rev. D}
  {\bfseries 89} (2014) 103006}
  [\href{https://arxiv.org/abs/1311.2552}{{\ttfamily 1311.2552}}].

\bibitem{Sreenath:2014BINGO}
V.~Sreenath, D.~K. Hazra and L.~Sriramkumar, \emph{{On the scalar consistency
  relation away from slow roll}},
  \href{https://doi.org/10.1088/1475-7516/2015/02/029}{\emph{JCAP} {\bfseries
  02} (2015) 029} [\href{https://arxiv.org/abs/1410.0252}{{\ttfamily
  1410.0252}}].

\bibitem{Martin:2014kja}
J.~Martin, L.~Sriramkumar and D.~K. Hazra, \emph{{Sharp inflaton potentials and
  bi-spectra: Effects of smoothening the discontinuity}},
  \href{https://doi.org/10.1088/1475-7516/2014/09/039}{\emph{JCAP} {\bfseries
  09} (2014) 039} [\href{https://arxiv.org/abs/1404.6093}{{\ttfamily
  1404.6093}}].

\bibitem{Achucarro:2014msa}
A.~Achucarro, V.~Atal, B.~Hu, P.~Ortiz and J.~Torrado, \emph{{Inflation with
  moderately sharp features in the speed of sound: Generalized slow roll and
  in-in formalism for power spectrum and bispectrum}},
  \href{https://doi.org/10.1103/PhysRevD.90.023511}{\emph{Phys. Rev. D}
  {\bfseries 90} (2014) 023511}
  [\href{https://arxiv.org/abs/1404.7522}{{\ttfamily 1404.7522}}].

\bibitem{Fergusson:2014hya}
J.~R. Fergusson, H.~F. Gruetjen, E.~P.~S. Shellard and M.~Liguori,
  \emph{{Combining power spectrum and bispectrum measurements to detect
  oscillatory features}},
  \href{https://doi.org/10.1103/PhysRevD.91.023502}{\emph{Phys. Rev. D}
  {\bfseries 91} (2015) 023502}
  [\href{https://arxiv.org/abs/1410.5114}{{\ttfamily 1410.5114}}].

\bibitem{Meerburg:2015owa}
P.~D. Meerburg, M.~M\"unchmeyer and B.~Wandelt, \emph{{Joint resonant CMB power
  spectrum and bispectrum estimation}},
  \href{https://doi.org/10.1103/PhysRevD.93.043536}{\emph{Phys. Rev. D}
  {\bfseries 93} (2016) 043536}
  [\href{https://arxiv.org/abs/1510.01756}{{\ttfamily 1510.01756}}].

\bibitem{Appleby:2015bpw}
S.~Appleby, J.-O. Gong, D.~K. Hazra, A.~Shafieloo and S.~Sypsas, \emph{{Direct
  search for features in the primordial bispectrum}},
  \href{https://doi.org/10.1016/j.physletb.2016.07.004}{\emph{Phys. Lett. B}
  {\bfseries 760} (2016) 297}
  [\href{https://arxiv.org/abs/1512.08977}{{\ttfamily 1512.08977}}].

\bibitem{Dias:2016rjq}
M.~Dias, J.~Frazer, D.~J. Mulryne and D.~Seery, \emph{{Numerical evaluation of
  the bispectrum in multiple field inflation\textemdash{}the transport approach
  with code}}, \href{https://doi.org/10.1088/1475-7516/2016/12/033}{\emph{JCAP}
  {\bfseries 12} (2016) 033}
  [\href{https://arxiv.org/abs/1609.00379}{{\ttfamily 1609.00379}}].

\bibitem{Planck13:inflation}
{\scshape Planck} collaboration, \emph{{Planck 2013 results. XXII. Constraints
  on inflation}},
  \href{https://doi.org/10.1051/0004-6361/201321569}{\emph{Astron. Astrophys.}
  {\bfseries 571} (2014) A22}
  [\href{https://arxiv.org/abs/1303.5082}{{\ttfamily 1303.5082}}].

\bibitem{Euclid}
G.~D. {Racca}, R.~{Laureijs}, L.~{Stagnaro}, J.-C. {Salvignol}, J.~{Lorenzo
  Alvarez}, G.~{Saavedra Criado} et~al., \emph{{The Euclid mission design}},
  in \emph{Space Telescopes and Instrumentation 2016: Optical, Infrared, and
  Millimeter Wave}, H.~A. {MacEwen}, G.~G. {Fazio}, M.~{Lystrup}, N.~{Batalha},
  N.~{Siegler} and E.~C. {Tong}, eds., vol.~9904 of \emph{Society of
  Photo-Optical Instrumentation Engineers (SPIE) Conference Series}, p.~99040O,
  July, 2016, \href{https://doi.org/10.1117/12.2230762}{DOI}
  [\href{https://arxiv.org/abs/1610.05508}{{\ttfamily 1610.05508}}].

\bibitem{LSST}
{\scshape LSST} collaboration, \emph{{LSST: from Science Drivers to Reference
  Design and Anticipated Data Products}},
  \href{https://doi.org/10.3847/1538-4357/ab042c}{\emph{Astrophys. J.}
  {\bfseries 873} (2019) 111}
  [\href{https://arxiv.org/abs/0805.2366}{{\ttfamily 0805.2366}}].

\bibitem{Boubekeur:2005Hilltop}
L.~Boubekeur and D.~H. Lyth, \emph{{Hilltop inflation}},
  \href{https://doi.org/10.1088/1475-7516/2005/07/010}{\emph{JCAP} {\bfseries
  07} (2005) 010} [\href{https://arxiv.org/abs/hep-ph/0502047}{{\ttfamily
  hep-ph/0502047}}].

\bibitem{Efstathiou:2006ak}
G.~Efstathiou and S.~Chongchitnan, \emph{{The search for primordial tensor
  modes}}, \href{https://doi.org/10.1143/PTPS.163.204}{\emph{Prog. Theor. Phys.
  Suppl.} {\bfseries 163} (2006) 204}
  [\href{https://arxiv.org/abs/astro-ph/0603118}{{\ttfamily
  astro-ph/0603118}}].

\bibitem{Contaldi:2003KD}
C.~R. Contaldi, M.~Peloso, L.~Kofman and A.~D. Linde, \emph{{Suppressing the
  lower multipoles in the CMB anisotropies}},
  \href{https://doi.org/10.1088/1475-7516/2003/07/002}{\emph{JCAP} {\bfseries
  07} (2003) 002} [\href{https://arxiv.org/abs/astro-ph/0303636}{{\ttfamily
  astro-ph/0303636}}].

\bibitem{Hergt:2018KD}
L.~T. Hergt, W.~J. Handley, M.~P. Hobson and A.~N. Lasenby, \emph{{Constraining
  the kinetically dominated Universe}},
  \href{https://doi.org/10.1103/PhysRevD.100.023501}{\emph{Phys. Rev. D}
  {\bfseries 100} (2019) 023501}
  [\href{https://arxiv.org/abs/1809.07737}{{\ttfamily 1809.07737}}].

\bibitem{Ragavendra:2020KD}
H.~V. Ragavendra, D.~Chowdhury and L.~Sriramkumar, \emph{{Suppression of scalar
  power on large scales and associated bispectra}},
  \href{https://arxiv.org/abs/2003.01099}{{\ttfamily 2003.01099}}.

\bibitem{Kallosh:2013alpha1}
R.~Kallosh and A.~Linde, \emph{{Universality Class in Conformal Inflation}},
  \href{https://doi.org/10.1088/1475-7516/2013/07/002}{\emph{JCAP} {\bfseries
  07} (2013) 002} [\href{https://arxiv.org/abs/1306.5220}{{\ttfamily
  1306.5220}}].

\bibitem{Kallosh:2013alpha2}
R.~Kallosh and A.~Linde, \emph{{Multi-field Conformal Cosmological
  Attractors}},
  \href{https://doi.org/10.1088/1475-7516/2013/12/006}{\emph{JCAP} {\bfseries
  12} (2013) 006} [\href{https://arxiv.org/abs/1309.2015}{{\ttfamily
  1309.2015}}].

\bibitem{Kallosh:2013alpha3}
R.~Kallosh, A.~Linde and D.~Roest, \emph{{Superconformal Inflationary
  $\alpha$-Attractors}},
  \href{https://doi.org/10.1007/JHEP11(2013)198}{\emph{JHEP} {\bfseries 11}
  (2013) 198} [\href{https://arxiv.org/abs/1311.0472}{{\ttfamily 1311.0472}}].

\bibitem{Lewis:1999CAMB}
A.~Lewis, A.~Challinor and A.~Lasenby, \emph{{Efficient computation of CMB
  anisotropies in closed FRW models}},
  \href{https://doi.org/10.1086/309179}{\emph{Astrophys. J.} {\bfseries 538}
  (2000) 473} [\href{https://arxiv.org/abs/astro-ph/9911177}{{\ttfamily
  astro-ph/9911177}}].

\bibitem{Planck2018V}
{\scshape Planck} collaboration, \emph{{Planck 2018 results. V. CMB power
  spectra and likelihoods}},
  \href{https://doi.org/10.1051/0004-6361/201936386}{\emph{Astron. Astrophys.}
  {\bfseries 641} (2020) A5}
  [\href{https://arxiv.org/abs/1907.12875}{{\ttfamily 1907.12875}}].

\bibitem{PlanckVIII}
{\scshape Planck} collaboration, \emph{{Planck 2018 results. VIII.
  Gravitational lensing}},
  \href{https://doi.org/10.1051/0004-6361/201833886}{\emph{Astron. Astrophys.}
  {\bfseries 641} (2020) A8}
  [\href{https://arxiv.org/abs/1807.06210}{{\ttfamily 1807.06210}}].

\bibitem{EG20}
G.~Efstathiou and S.~Gratton, \emph{{A Detailed Description of the CamSpec
  Likelihood Pipeline and a Reanalysis of the Planck High Frequency Maps}},
  \href{https://arxiv.org/abs/1910.00483}{{\ttfamily 1910.00483}}.

\bibitem{Handley2015a}
W.~J. {Handley}, M.~P. {Hobson} and A.~N. {Lasenby}, \emph{{POLYCHORD:
  next-generation nested sampling}},
  \href{https://doi.org/10.1093/mnras/stv1911}{\emph{Mon. Not. Roy. Astron.
  Soc.} {\bfseries 453} (2015) 4384}
  [\href{https://arxiv.org/abs/1506.00171}{{\ttfamily 1506.00171}}].

\bibitem{Handley:2015fda}
W.~J. Handley, M.~P. Hobson and A.~N. Lasenby, \emph{{PolyChord: nested
  sampling for cosmology}},
  \href{https://doi.org/10.1093/mnrasl/slv047}{\emph{Mon. Not. Roy. Astron.
  Soc.} {\bfseries 450} (2015) L61}
  [\href{https://arxiv.org/abs/1502.01856}{{\ttfamily 1502.01856}}].

\bibitem{Lewis:2002COSMOMC}
A.~Lewis and S.~Bridle, \emph{{Cosmological parameters from CMB and other data:
  A Monte Carlo approach}},
  \href{https://doi.org/10.1103/PhysRevD.66.103511}{\emph{Phys. Rev. D}
  {\bfseries 66} (2002) 103511}
  [\href{https://arxiv.org/abs/astro-ph/0205436}{{\ttfamily
  astro-ph/0205436}}].

\bibitem{Lewis:2019GETDIST}
A.~Lewis, \emph{{GetDist: a Python package for analysing Monte Carlo samples}},
   \href{https://arxiv.org/abs/1910.13970}{{\ttfamily 1910.13970}}.

\bibitem{Jeffreys:1961}
H.~Jeffreys, \emph{Theory of Probability}. Oxford University Press: Oxford,
  1961.

\bibitem{Trotta:2007}
R.~{Trotta}, \emph{{Applications of Bayesian model selection to cosmological
  parameters}},
  \href{https://doi.org/10.1111/j.1365-2966.2007.11738.x}{\emph{\mnras}
  {\bfseries 378} (2007) 72}
  [\href{https://arxiv.org/abs/arXiv:astro-ph/0504022}{{\ttfamily
  arXiv:astro-ph/0504022}}].

\bibitem{BOBYQA}
M.~J. Powell, \emph{The bobyqa algorithm for bound constrained optimization
  without derivatives}, .

\bibitem{Arroja:2011yu}
F.~Arroja, A.~E. Romano and M.~Sasaki, \emph{{Large and strong scale dependent
  bispectrum in single field inflation from a sharp feature in the mass}},
  \href{https://doi.org/10.1103/PhysRevD.84.123503}{\emph{Phys. Rev. D}
  {\bfseries 84} (2011) 123503}
  [\href{https://arxiv.org/abs/1106.5384}{{\ttfamily 1106.5384}}].

\bibitem{LiteBIRD}
M.~Hazumi, P.~A. Ade, A.~Adler, E.~Allys, K.~Arnold, D.~Auguste et~al.,
  \emph{Litebird satellite: Jaxa’s new strategic l-class mission for all-sky
  surveys of cosmic microwave background polarization},
  \href{https://doi.org/10.1117/12.2563050}{\emph{Space Telescopes and
  Instrumentation 2020: Optical, Infrared, and Millimeter Wave} (2020) }.

\bibitem{CMBBHARAT}
``{CMB} {Bharat} – {Assessing} the prospects for frontier {CMB} space
  experiments from {India}.''

\bibitem{PICO}
K.~Young, M.~Alvarez, N.~Battaglia, J.~Bock, J.~Borrill, D.~Chuss et~al.,
  \emph{Optical design of pico, a concept for a space mission to probe
  inflation and cosmic origins},  2018.

\bibitem{DESI}
{\scshape DESI} collaboration, \emph{{The DESI Experiment Part I:
  Science,Targeting, and Survey Design}},
  \href{https://arxiv.org/abs/1611.00036}{{\ttfamily 1611.00036}}.

\bibitem{Hazra:2012LSS}
D.~K. Hazra, \emph{{Changes in the halo formation rates due to features in the
  primordial spectrum}},
  \href{https://doi.org/10.1088/1475-7516/2013/03/003}{\emph{JCAP} {\bfseries
  03} (2013) 003} [\href{https://arxiv.org/abs/1210.7170}{{\ttfamily
  1210.7170}}].

\bibitem{Chen:2016LSS}
X.~Chen, C.~Dvorkin, Z.~Huang, M.~H. Namjoo and L.~Verde, \emph{{The Future of
  Primordial Features with Large-Scale Structure Surveys}},
  \href{https://doi.org/10.1088/1475-7516/2016/11/014}{\emph{JCAP} {\bfseries
  11} (2016) 014} [\href{https://arxiv.org/abs/1605.09365}{{\ttfamily
  1605.09365}}].

\bibitem{Ballardini:2016LSS}
M.~Ballardini, F.~Finelli, C.~Fedeli and L.~Moscardini, \emph{{Probing
  primordial features with future galaxy surveys}},
  \href{https://doi.org/10.1088/1475-7516/2016/10/041}{\emph{JCAP} {\bfseries
  10} (2016) 041} [\href{https://arxiv.org/abs/1606.03747}{{\ttfamily
  1606.03747}}].

\bibitem{Beutler:2019LSS}
F.~Beutler, M.~Biagetti, D.~Green, A.~Slosar and B.~Wallisch, \emph{{Primordial
  Features from Linear to Nonlinear Scales}},
  \href{https://doi.org/10.1103/PhysRevResearch.1.033209}{\emph{Phys. Rev.
  Res.} {\bfseries 1} (2019) 033209}
  [\href{https://arxiv.org/abs/1906.08758}{{\ttfamily 1906.08758}}].

\bibitem{Ballardini:2019LSS}
M.~Ballardini, R.~Murgia, M.~Baldi, F.~Finelli and M.~Viel, \emph{{Non-linear
  damping of superimposed primordial oscillations on the matter power spectrum
  in galaxy surveys}},
  \href{https://doi.org/10.1088/1475-7516/2020/04/030}{\emph{JCAP} {\bfseries
  04} (2020) 030} [\href{https://arxiv.org/abs/1912.12499}{{\ttfamily
  1912.12499}}].

\end{thebibliography}\endgroup

\end{document}